\pdfoutput=1
\documentclass[11pt]{article}
\usepackage[letterpaper,margin=1in]{geometry}
\usepackage{times,setspace,amsmath,amssymb,amsthm,mathtools,graphicx,booktabs,longtable,array,natbib,microtype}
\usepackage[font=small,labelfont=bf]{caption}
\usepackage{xcolor,needspace,placeins}
\usepackage{xurl}
\usepackage[colorlinks=true,linkcolor=blue!45!black,citecolor=blue!45!black,urlcolor=blue!45!black,filecolor=blue!45!black]{hyperref}
\usepackage[nameinlink,capitalise]{cleveref}
\newcommand{\E}{\mathbb E}
\newcommand{\R}{\mathbb R}
\newcommand{\1}{\mathbf 1}
\newcommand{\dd}{\,\mathrm d}
\newtheorem{theorem}{Theorem}
\newtheorem{proposition}[theorem]{Proposition}

\newtheorem{corollary}[theorem]{Corollary}
\theoremstyle{definition}
\newtheorem{assumption}{Assumption}

\usepackage{fancyhdr}
\hypersetup{pdftitle={Trading Before a Committed Rebalance: Public Information and Optional Liquidity Supply},pdfauthor={Khizar Qureshi}}
\title{\bfseries Trading Before a Committed Rebalance\\[.3em]\large Public Information and Optional Liquidity Supply}
\author{Khizar Qureshi\thanks{Email: \href{mailto:kqureshi@mit.edu}{\nolinkurl{kqureshi@mit.edu}}.\newline Accepted to the INFORMS Conference on Financial Engineering. We are thankful to the participants and chairs for helpful commentary and ideas.}\\[.4em]\small MIT\\\small Cambridge, MA}
\date{Original version: July 10th, 2026\\Updated: November 25th, 2026}
\begin{document}
\makeatletter\immediate\write\@auxout{\gdef\string\@extra@binfo{}}\makeatother
% Only appendix sections are listed in the Internet Appendix contents.
\addtocontents{toc}{\protect\setcounter{tocdepth}{-2}}
\maketitle
\begin{abstract}
A public forecast of a committed portfolio adjustment changes both the institution's preparation and the inventory that other traders choose to supply. We characterize its effect on institutional expenditure in a Bayesian quantity game with avoidable activation and uniform-price terminal completion. For scalar targets, every refinement weakly lowers cost when participation is fixed. Optional supply can reverse that ranking: costly inventory may expand too rapidly, or helpful inventory may expand too slowly. The second mechanism persists with smooth cost densities, no pre-event impact, equal temporary charges, and participation near one-half. Favorable comparisons nevertheless hold for centered targets and centered Gaussian signals. For portfolios, we give an exact matrix criterion under which information weakly lowers cost relative to no information for every centered target, even with optional basket activation. Outside these comparisons, equal forecast accuracy need not imply equal execution value. We solve the finite-player equilibrium, give a sufficient condition for uniqueness among all Bayesian equilibria, and identify when individual strategic size reverses the continuum comparison. Fixed-prior experiments based on ETF price paths assess the role of observed target shapes. They characterize the model's information comparisons without estimating activation costs or institutional losses.
\end{abstract}
\noindent\textbf{Keywords:} committed portfolios, public information, optional liquidity, entry, terminal auctions, leveraged ETFs.

\noindent\textbf{Mathematics Subject Classification (2020):} 91G10, 91A15.

\section{Introduction}\label{sec:intro}
An institution that must complete a portfolio adjustment can use a public forecast to prepare its position. Other traders can use the same forecast to acquire inventory for the event. The institution must finish its trade. Prospective suppliers participate only when expected liquidation proceeds cover trading and activation costs. A better forecast therefore changes both the institution's preparation and the inventory available to it. The combined response can raise its execution bill.

We study this interaction in a simultaneous Bayesian quantity game. After a common public signal, the institution chooses its pre-event quantity. Prospective suppliers choose whether to activate an inventory allocation and how much to trade. All quantities follow a prescribed schedule. The institution then completes its target and suppliers liquidate at a uniform-price auction. Fixing the schedule permits an exact equilibrium calculation with private activation costs and explicit terminal cash payments. The institution acts before observing realized entry. Its target is external to affected execution prices.

The value of inventory and the value of information that induces it are distinct. Inventory can help at every target mean even when a forecast refinement raises expected cost. Suppose an inexpensive cohort of suppliers is nearly exhausted. An upward forecast revision then adds little inventory, while a downward revision can withdraw much more. The cost of that asymmetric response can exceed the institution's direct forecasting gain. Persistent price displacement creates a second mechanism: information can attract inventory whose effect on execution prices makes it costly to the institution.

The paper makes three contributions.

\paragraph{C1. Scalar information classification.}
We derive the posterior-mean cost function from the institution's cash account. Its curvature separates costly accelerating supply from helpful decelerating supply. The second mechanism can arise with smooth, positive cost densities, zero pre-event impact, equal temporary charges, and participation near any prescribed interior mass. A fixed two-cohort example operates near one-half participation. Three favorable results delimit these adverse comparisons. Information weakly lowers cost for every fixed roster and relative to no information for every centered scalar target. Centered Gaussian precision gains are beneficial throughout the risk-neutral model.

\paragraph{C2. Portfolio information classification.}
We give a necessary and sufficient matrix condition for information to weakly lower cost relative to no information for every centered vector target under optional basket activation. The activation distribution determines the target scale at which a given roster is reached. Execution matrices determine whether an adverse direction exists. Isotropic execution also preserves the centered Gaussian ranking. Outside these favorable comparisons, shared activation allows equally accurate signals about magnitude and composition to have opposite execution values under one unchanged prior.

\paragraph{C3. Finite strategic size and equilibrium selection.}
We solve the symmetric Bayesian equilibrium for each population size $N\geq2$ and give a sufficient condition under which every Bayesian equilibrium is symmetric. The condition covers the principal finite-population examples. Own price impact shifts the entry threshold, so a fixed refinement can be harmful in the continuum and beneficial at finite $N$. A joint limit identifies the relevant information scale. A second-order expansion determines the sign at the critical boundary. For small populations outside the uniqueness condition, asymmetric equilibria can yield different institutional costs.

Daily-reset leveraged funds motivate the committed-target problem. The empirical application uses observed benchmark paths to describe target distributions and evaluate feasible signal refinements under fixed priors. Each comparison holds the execution scenario unchanged. Chronological forecasts, reference-price checks, and dependence tests establish what these price-based exercises can support. They do not identify the activation distribution, strategic population, or historical execution bill.

Section~\ref{sec:institution} develops the institutional setting and literature. Sections~\ref{sec:model}--\ref{sec:information} derive equilibrium and the scalar classification. Sections~\ref{sec:portfolioinformation} and \ref{sec:finite} establish the portfolio and finite-player results. Sections~\ref{sec:numerical} and \ref{sec:data} give numerical implications and the price application. The appendix contains supporting proofs, risk and execution extensions, and the broader econometric, turnover, and factor analyses.

\section{Institutional motivation and related literature}\label{sec:institution}
\subsection{A target at the daily valuation time}
Daily leveraged and inverse ETFs state an objective relative to one benchmark calculation day, generally measured from one NAV determination to the next. Changes in benchmark value and fund wealth require the manager to restore the stated exposure. This creates a committed portfolio target, but does not require an exact cash-equity order at the closing print. Derivatives, investor flows, tracking tolerances, and prior trades determine the residual implementation. We model exact completion in fixed benchmark units as an idealization of that commitment.

Under a no-flow, no-fee one-period balance sheet, starting assets $A_0$ and exposure $\beta A_0$ become wealth $A_0(1+\beta R)$ and pre-reset exposure $\beta A_0(1+R)$. Restoring leverage therefore requires
\begin{equation}
 \Delta E=A_0\beta(\beta-1)R.\label{eq:reset}
\end{equation}
The coefficients are $12,6,2,0,2,6$ for leverage objectives $-3,-2,-1,1,2,3$, so a one-percent benchmark rise implies these numbers of cents of exposure adjustment per initial dollar. Carrying cost $C$ and net flow $\Phi$ under a terminal marking convention add $-\beta C+\beta\Phi$, while earlier flows need their own marking convention. The flow evidence in \citet{IvanovLenkey2018} makes that distinction material. A current AUM snapshot cannot reconstruct those historical quantities.

The target in our game is external to affected execution prices. With reference price $P_0$, the normalized dollar-exposure adjustment corresponds to $\Delta E/P_0$ fixed reference units. Actual terminal shares instead satisfy
\[
 \frac{P_0\Delta Q}{A_0}=\frac{\beta(\beta-1)R}{1+R},\qquad P_T=P_0(1+R).
\]
These are distinct target conventions. A conditional share forecast must average $R/(1+R)$ itself, rather than transform the conditional mean return. We use fixed reference units in the main application and examine the terminal-share convention in the appendix. Neither proxy identifies the residual order after earlier trades, flows, and derivative positions. The reset clock matters for returns as well: an ideal daily reset has log residual
\begin{equation}
 \log(1+\beta R_d)-\beta\log(1+R_d)
 =-\tfrac12\beta(\beta-1)R_d^2+O(R_d^3),\label{eq:discrete}
\end{equation}
whereas resetting at every intraday mark gives the sum of these residuals over intraday returns. The continuous-leverage identity of \citet{AvellanedaZhang2010} concerns the latter clock in a continuous-time limit. Appendix~\ref{app:identification} illustrates the distinction on matched paths. Current issuer documents support the daily mandate interpretation \citep{ProShares2026,Nasdaq2026}.

\subsection{Information, participation, and execution}
Public information can change participation and worsen financial outcomes. In \citet{Crego2020}, news encourages informed risk-averse investors to trade and worsens adverse selection. In \citet{DiamondVerrecchia1991}, disclosure can induce market-maker exit and reduce risk-bearing capacity. \citet{AdmatiPfleiderer1991} study how preannounced liquidity trades affect information asymmetry and coordination. Our participants are risk neutral and share target information. Private activation costs make inventory supply optional, while the committed institution adjusts its pretrade and pays the resulting execution bill.

The comparison uses \citet{Blackwell1953} and the posterior-mean geometry of \citet{GentzkowKamenica2016}. Persuasion provides a related framework \citep{KamenicaGentzkow2011,DworczakMartini2019}, but the institution has no disclosure instrument here. Ignoring a public forecast would not stop other traders from responding. Quadratic public-information objectives also admit matrix and spectral characterizations, as in \citet[Appendix D]{MiyashitaUi2025}. We derive the relevant matrix from institutional execution payments. Optional basket activation then creates the nonlinear comparison between information about magnitude and composition. The contribution is the resulting classification of institutional expenditure and its finite-size qualification. The underlying information-ordering methods are established.

Existing execution models allow decisions beyond quantity choice along a prescribed schedule. Stochastic targets and terminal constraints appear in \citet{DaiEtAl2023,BankVoss2018}, and endogenous timing and dropout in \citet{FuHagerHorst2026,FuHagerHorst2024}. Major-minor interaction and anticipation are studied by \citet{ChenEtAl2024,ChengEtAl2024,NeumanVoss2023}. Institutional trades reveal information in \citet{MoallemiEtAl2012}, while \citet{CampbellNutz2026,Strehle2018} analyze execution equilibria with rate, boundary, or transient-impact costs. The common signal and fixed schedule here isolate information value when activation remains avoidable and terminal settlement is uniform-price.

Table~\ref{tab:positioning} records the closest economic distinctions. Information and entry already interact in quantity competition \citep{DoKettering2025}, and prepared inventory is central to index execution in \citet{PegoraroEtAl2026}. \citet{Zhao2026} studies an affected-price leveraged-fund reset with capacity entry, auction mobilization, and finite speculators. In that paper's October 15 version, capacity in Appendix C.1.2 is committed before the open, while the fringe in Appendix B.9 enters within the auction. Here activation follows the common target forecast, and the target is external to affected execution prices. The distinction is the institutional expenditure induced by a common public forecast when activation remains avoidable.

The closest empirical comparison is \citet{BarbonEtAl2022}. Using historical assets and constituent weights, they connect leveraged-ETF and option rebalancing to end-of-day liquidity provision and subsequent dissipation. \citet{Brogger2021} finds smaller impact coefficients for larger and more predictable leveraged-volatility flows. These findings motivate favorable-information benchmarks alongside the adverse regions. Our price panel characterizes target uncertainty without identifying historical reset orders or institutional costs. Intraday predictability also has competing explanations \citep{GaoEtAl2018}. The model ranks the institutional payments induced by information. Directional ETF returns alone cannot identify that comparison.

\begin{table}[!htbp]\centering\footnotesize\setstretch{1.05}
\caption{Position relative to the closest literature. Reference names link to the cited articles or working-paper versions. The last column identifies the question or result added by the present paper.}\label{tab:positioning}
\begin{tabular}{@{}>{\raggedright\arraybackslash}p{.19\textwidth}>{\raggedright\arraybackslash}p{.35\textwidth}>{\raggedright\arraybackslash}p{.40\textwidth}@{}}\toprule
Reference & Established mechanism or object & Incremental contribution of this paper\\\midrule
\href{https://doi.org/10.1016/j.jfineco.2019.05.020}{Crego (2020)} & Public news encourages informed, risk-averse participation and can worsen adverse selection. & Common target information changes optional inventory supply among risk-neutral traders. Institutional cash cost can rise without private fundamental information.\\[.5em]
\href{https://doi.org/10.1111/j.1540-6261.1991.tb04620.x}{Diamond and Verrecchia (1991)} & Disclosure changes investor demand and market-maker risk-bearing capacity through exit. & A committed institution reoptimizes its pretrade. The posterior-mean cost function separates the value of prepared inventory from the information that induces it.\\[.5em]
\href{https://bce.yonsei.ac.kr/yeri/publications/report.do?articleNo=449442&attachNo=192344&mode=download}{Do and Kettering (2025)} & One firm has better demand information than its rival. Endogenous entry can reverse their profit ranking and make welfare nonmonotone in precision. & Common public experiments change the expenditure of a committed institution. The cash account yields a classification with favorable scalar, portfolio, and finite-player benchmarks.\\[.5em]
\href{https://ssrn.com/abstract=6772502}{Pegoraro, Sammon, and Shim (2026)} & Anticipatory traders provide liquidity around index reconstitution. Execution cost and tracking concerns shape index-linked trades. & The institution's quantity commitment is fixed. We compare public experiments when inventory activation follows the forecast, and identify both favorable and adverse regions.\\[.5em]
\href{https://xsci.com/preprint/doi/10.66977/xsci.2610.0014.pdf}{Zhao (2026)} & Price-dependent LETF resets create a feedback channel. Extensions include capacity entry, auction mobilization, and finite speculators. & The target is external to affected prices. A Blackwell comparison isolates forecast-induced activation, with an exact finite-population correction near entry.\\[.5em]
\href{https://www.abarbon.com/papers/liquidity-provision-to-leveraged-etfs-and-equity-options-rebalancing-flows}{Barbon et al. (2022)} & Historical rebalancing flows connect liquidity provision to end-of-day dynamics and subsequent dissipation. & The theoretical object is the institution's information value. Price evidence distinguishes mean forecasts from uncertainty calibration without identifying orders or structural activation.\\[.5em]
\href{https://arxiv.org/html/2312.09479v4}{Miyashita and Ui (2025)} & Public LQG information objectives admit matrix and spectral characterizations. & The institutional cash account determines the relevant matrix. Optional basket activation creates a nonlinear comparison beyond the fixed-roster quadratic.\\[.5em]
\href{https://doi.org/10.1287/opre.2021.2201}{Min, Maglaras, and Moallemi (2022)} & Portfolio liquidity provision and linear cross-impact couple execution across assets. & The execution account gives fixed-roster and centered optional-entry information criteria. Shared activation permits equally accurate signals to have opposite cash effects across target directions.\\\bottomrule
\end{tabular}
\end{table}
\nocite{MinEtAl2022}

\section{A game with pre-event trading and terminal completion}\label{sec:model}
\subsection{Information, timing, and admissible positions}
The pre-event interval is $[0,1]$. One public signal $Y$ is observed at time zero, and the terminal target $D$ is revealed at time one. Conditional moments are
\[
 \mu=\E[D\mid Y],\qquad v=\E[(D-\mu)^2\mid Y],
\]
where $D\in L^2$ and the signal experiment is independent of trading decisions. The institution and all prospective anticipators observe $Y$. Each anticipator also observes a private nonnegative entry cost $f$. This cost is drawn independently of the target and the entire family of compared public experiments. Its distribution has continuous CDF $F$ and $F(0)=0$. The limiting population has total mass one. Its private types follow $F$ in every public state. Strategies cannot use additional information before terminal revelation.

The charge $f$ is avoidable after observing the signal. Examples include the fixed cost of opening an event-specific inventory program or allocating an operational slot. Infrastructure or staffing costs paid before the signal belong to the fixed-roster comparison. Price and volume observations do not identify these activation costs.

The institution chooses a signed pre-event quantity $z\in\R$. A type $f$ either remains inactive or pays $f$ and chooses a signed inventory $q_f\in\R$. Pre-event trades use a common deterministic schedule $\phi\in L^2[0,1]$, with $\phi\geq0$ and $\int_0^1\phi_t\dd t=1$. The institution trades at rate $z\phi_t$, and an entrant at rate $q_f\phi_t$. Aggregate anticipator inventory is $m=\int q_f\1_{\{f\text{ enters}\}}\dd F(f)$, so the aggregate pre-event rate is $(z+m)\phi_t$. All positions are chosen simultaneously after the signal. Under Nash equilibrium, the institution holds the anticipator strategy fixed when it deviates. The institution is not a Stackelberg leader that commits first and internalizes subsequent re-entry decisions.

The common schedule may represent a prescribed execution window or an operational schedule whose size can be changed at a decision time. The game optimizes quantities along that schedule. Information comparisons hold the decision time and execution opportunity fixed. Waiting would shorten the horizon and change the schedule or integrated costs. Such a delay requires a different optimization problem. Once a trader enters, its entry status remains fixed until the event. Inventory follows $q_f\int_0^t\phi_s\dd s$ and becomes zero at terminal liquidation. Even if intermediate trading were allowed, flattening the position would not undo the entry decision.

\subsection{Transient impact and the terminal price}
Initially, take the unaffected benchmark price as the numeraire and hold it constant. Only execution-price deviations need be recorded. The pre-event impact process satisfies
\begin{equation}
 \dot I_t=-\rho I_t+\kappa(z+m)\phi_t,\qquad I_0=0,\label{eq:impact}
\end{equation}
with $\rho\geq0$ and $\kappa\geq0$. Define the average execution loading and the terminal carryover loading by
\begin{align}
 K&=\kappa\int_0^1\phi_t\int_0^t e^{-\rho(t-s)}\phi_s\dd s\dd t,\label{eq:K}\\
 H&=\kappa\int_0^1e^{-\rho(1-s)}\phi_s\dd s.\label{eq:H}
\end{align}
The common pre-event price component paid per unit is $K(z+m)$. The displacement still present just before the auction is $H(z+m)$. The institution also pays a temporary execution charge $\eta_0z^2/2$, and each entrant pays $\eta q_f^2/2$, where $\eta_0,\eta>0$. These coefficients include the schedule's squared-rate integral and use the same normalized cash units as price impact.

At the event, the institutional order is $J=D-z$, the aggregate anticipator order is $-m$, and net outside demand is $n=D-z-m$. An outside liquidity supplier has the linear inverse supply rule
\begin{equation}
 p_A=H(z+m)+\lambda n,\qquad \lambda>0,\label{eq:auction}
\end{equation}
and every event trade executes at this uniform price relative to the unaffected benchmark. The institution pays $(D-z)p_A$. Anticipators collectively receive $mp_A$, and the outside supplier receives $np_A$. The institution finishes with $z+J=D$. Each anticipator finishes flat, and the supplier absorbs the residual inventory. Any remaining supplier inventory is valued at the unaffected benchmark immediately after the accounting horizon. This convention completes the cash account without introducing a second market-clearing problem.

Equation~\eqref{eq:auction} specifies an elastic supply schedule. It is not derived from a strategic dealer equilibrium. Net demand can have either sign, and the model imposes no auction capacity constraint or additional post-auction manipulation opportunity. The institution internalizes its price effect because its order is large. An individual trader in the continuum takes the uniform auction price as given. The finite-player version below restores each minor trader's own price effect at its population scale.

\begin{assumption}\label{ass:primitive}
The execution coefficients satisfy $\lambda>0$, $0\leq H<\lambda$, $K\geq0$, and $\eta_0,\eta>0$. Private costs have a continuous CDF $F$ on $[0,\infty)$ with $F(0)=0$, are independent of the target and the entire family of compared public signals taken jointly, and the target has finite second moment.
\end{assumption}
The restriction on $H$ is sufficient rather than necessary. It compares inherited displacement per pre-event unit with the auction price response per unit of net terminal demand. Here $\lambda$ is inverse depth, so a larger value means less terminal liquidity for a given price change. The restriction is checked for every reported numerical scenario. The associated coefficient $b=\lambda+K-H$ is positive. This also excludes profitable zero-target round trips within the stipulated strategy class.

\subsection{Cash objectives and their curvature}
Conditional expected institutional expenditure relative to the unaffected target value is
\begin{align}
 C_0(z,m;\mu,v)
 &=Kz(z+m)+\frac{\eta_0}2z^2
  +\E[(D-z)\{H(z+m)+\lambda(D-z-m)\}\mid Y]\notag\\
 &=\frac A2z^2+bzm-d\mu z-(\lambda-H)\mu m+\lambda(\mu^2+v),\label{eq:major}
\end{align}
where
\begin{equation}
 b=\lambda+K-H,\qquad A=\eta_0+2b,\qquad d=2\lambda-H.\label{eq:abd}
\end{equation}
An entrant's conditional expected profit before its fixed cost is
\begin{equation}
 \pi(q;z,m,\mu)=q\{\lambda\mu-b(z+m)\}-\frac\eta2q^2.\label{eq:minor}
\end{equation}
These expressions follow by subtracting pre-event purchases and temporary charges from terminal liquidation receipts. The variance term $\lambda v$ is retained in \eqref{eq:major} because it changes when information changes. It does not affect the pre-event best response under risk neutrality.

Both optimization problems are globally well posed in their own controls. The major's objective has curvature $A>0$, and the entrant's profit has curvature $-\eta<0$. These conditions do not require a positive joint signal-inventory state matrix. Completing the square in the control bounds the linear signal-profit term required by the cash account. Every equilibrium position is bounded by a constant times $|\mu|$, so finite second moments suffice for ex ante integrability.

An unaffected square-integrable price martingale $S$ can be restored in a filtration $(\mathcal F_t)$ with $\sigma(Y)\subseteq\mathcal F_0$. It must remain a martingale in that enlarged filtration, and the positions must retain the stated deterministic schedule. Its conditional expected contribution $z\int\phi_t(S_t-S_1)\dd t$ to institutional shortfall is zero. The corresponding inventory-profit term uses the opposite price difference and also has zero conditional expectation. The target may be correlated with the terminal martingale innovation without changing this calculation. The institution's terminal fundamental payment cancels against the value of its mandated final inventory. The analysis therefore concerns the execution component of the target rather than a fundamental-return alpha.

\section{Equilibrium and cost incidence}\label{sec:eq}
Define
\begin{equation}
 s=\lambda-\frac{bd}{A}=\frac{\lambda\eta_0+bH}{A}>0,
 \qquad c=b\left(1-\frac bA\right)=\frac{b(\eta_0+b)}A>0.\label{eq:sc}
\end{equation}
The institutional best response to $m$ is $(d\mu-bm)/A$. Substituting it into the entrant's first-order condition gives $\eta q=s\mu-cm$. This is an algebraic solution of simultaneous best responses. It does not give the institution an earlier commitment opportunity.

\Needspace{12\baselineskip}
\begin{theorem}[Unique participation equilibrium]\label{thm:eq}
Under Assumption~\ref{ass:primitive}, for every conditional mean $\mu$ there is a unique equilibrium entered mass $a\in[0,1]$, aggregate inventory $m$, and institutional quantity $z$, up to entry decisions by indifferent zero-measure types. The common entrant inventory $q$ denotes a hypothetical entrant's optimum when $a=0$. They satisfy
\begin{align}
 a&=F\left(\frac{\eta s^2\mu^2}{2(\eta+ca)^2}\right),\label{eq:mass}\\
 q&=\frac{s\mu}{\eta+ca},\qquad m=aq,\qquad
 z=\frac{d\mu-bm}{A}.\label{eq:quantities}
\end{align}
A type enters exactly when $f\leq\eta q^2/2$. The solution is continuous in $\mu$, its entered mass is nondecreasing in $|\mu|$, and its positions have finite second moments. The equilibrium is independent of $v$, although its institutional cost increases by $\lambda v$.
\end{theorem}
\begin{proof}
Strict own-control curvature gives unique conditional best responses. Every entrant has the same trading technology and hence the same optimal $q$, with maximized gross profit $\eta q^2/2$. Aggregation gives $m=aq$ and therefore \eqref{eq:quantities}, while the entry rule gives \eqref{eq:mass}. Let $h(a)=a-F(\eta s^2\mu^2/[2(\eta+ca)^2])$. Continuity of $F$ implies continuity of $h$, its endpoint values satisfy $h(0)\leq0$ and $h(1)\geq0$, and for $a_2>a_1$ the nonincreasing argument of $F$ gives $h(a_2)-h(a_1)\geq a_2-a_1>0$. There is therefore exactly one zero, and compactness of $[0,1]$, continuity of the equation, and uniqueness imply continuous dependence on $\mu$, while the same ordering establishes monotonicity in $|\mu|$. Finally, $|q|\leq s|\mu|/\eta$ and $|m|\leq|q|$, with an analogous bound on $z$, establish the moment claim and complete verification of both agents' global best responses.
\end{proof}

If $F$ has density $f_F$ at the entry cutoff and participation is interior, implicit differentiation for $u=|\mu|>0$ gives
\begin{equation}
 \frac{\dd a}{\dd u}
 =\frac{ f_F(g)\eta s^2u/(\eta+ca)^2}
 {1+f_F(g)\eta s^2u^2c/(\eta+ca)^3},
 \qquad g=\frac{\eta s^2u^2}{2(\eta+ca)^2}.\label{eq:derivative}
\end{equation}
This derivative is nonnegative. It is strictly positive only where the density is positive and the participation constraint does not bind. Equation~\eqref{eq:mass} characterizes the equilibrium implicitly for a general distribution and does not give a closed-form solution for $a$. Monotonicity supports globally reliable bisection without a contraction estimate for the entry response.

\subsection{When inventory provision raises institutional cost}
Let $z_0=d\mu/A$ be the institution's optimal pre-event position in the absence of anticipators. The corresponding cost is $C_0^{\varnothing}=\lambda(\mu^2+v)-d^2\mu^2/(2A)$. Define
\begin{equation}
 e=\frac{db}{A}-(\lambda-H)
   =\frac{Hb-(\lambda-H)\eta_0}{A},\qquad k=\frac{b^2}{2A}.\label{eq:ek}
\end{equation}
\begin{proposition}[Institutional cost decomposition]\label{prop:cost}
For any signed anticipator inventory $m$, the institution's optimized cost satisfies
\begin{equation}
 \min_z C_0(z,m;\mu,v)-C_0^{\varnothing}
    =e\mu m-km^2.\label{eq:costdiff}
\end{equation}
If its pre-event quantity is frozen at $z_0$, the cost change is $e\mu m$, and adapting that quantity saves $km^2$. At equilibrium, $\mu m\geq0$, so entry weakly lowers institutional cost whenever $e\leq0$. If $e>0$ and $\mu\neq0$, entry raises cost exactly when $0<m/\mu<e/k$.
\end{proposition}
\begin{proof}
Completing the square in \eqref{eq:major} gives the minimum
\[
 \lambda(\mu^2+v)-(\lambda-H)\mu m
      -\frac{(d\mu-bm)^2}{2A}.
\]
Subtracting $C_0^{\varnothing}$ proves \eqref{eq:costdiff}. Evaluation at $z_0$ gives the frozen-position expression, and subtraction yields $b^2m^2/(2A)$. The equilibrium sign follows from \eqref{eq:quantities}, after which the stated conditions are the signs of $e(m/\mu)-k(m/\mu)^2$.
\end{proof}

The coefficient $e$ measures the cost effect of the first unit of anticipator inventory at the institution's unperturbed optimum. When $Hb>(\lambda-H)\eta_0$, the price displacement carried into the auction exceeds the initial benefit of additional auction supply. Adjusting its own pretrade gives the institution a quadratic saving. At sufficiently small participation, however, that saving cannot offset a positive first-order cost effect. When $H=0$, $e=-\lambda\eta_0/A<0$, and anticipator inventory is beneficial throughout this benchmark. This sign determines the onset mechanism. Section~\ref{sec:information} shows that information can also be harmful when all anticipator inventory is beneficial.

Appendix~\ref{app:heterogeneouscosts} allows correlated fixed and marginal inventory costs while retaining a one-dimensional equilibrium equation. Appendix~\ref{app:riskextension} introduces conditional execution risk through a mean--variance objective. In that extension, posterior variance affects supply and can reverse the favorable Gaussian precision ranking. The baseline risk-neutral results remain unchanged.

\section{Information and optional inventory supply}\label{sec:information}
\subsection{The cost function of a posterior mean}
Let $m(\mu)$ denote equilibrium inventory from Theorem~\ref{thm:eq}, and define
\begin{equation}
 \psi(\mu)=-\frac{d^2}{2A}\mu^2+e\mu m(\mu)-k m(\mu)^2.
 \label{eq:psi}
\end{equation}
Throughout, implementation costs are independent of the target and all public signals in the comparison, considered jointly. Their conditional distribution therefore remains $F$. For any such public information $\mathcal H$, the institution's ex ante optimized expenditure is
\begin{equation}
 \mathcal C(\mathcal H)=\lambda\E[D^2]+\E[\psi(\E[D\mid\mathcal H])].
 \label{eq:representation}
\end{equation}
The first term is unchanged across experiments under the same target law. The second combines the institution's direct forecasting benefit with the cost of the inventory response. The value of information therefore depends on the shape of the equilibrium cost function.

\begin{theorem}[Information characterization]\label{thm:geometry}
Under Assumption~\ref{ass:primitive}, \eqref{eq:representation} holds for every square-integrable target. On an interval $I$, every feasible Blackwell refinement with posterior means in $I$ weakly lowers cost if $\psi$ is concave on $I$. Conversely, if $\psi$ is not concave on $I$, there exists a two-point target law supported in $I$ for which revelation raises cost relative to no information. Thus concavity is equivalent to a beneficial information ranking across all target laws supported in $I$.

Where $m$ is twice differentiable,
\begin{equation}
 \psi''=-\frac{d^2}{A}+2em'-2k(m')^2+(e\mu-2km)m''.
 \label{eq:curvature}
\end{equation}
At a point $\tau$ where $m$ is continuous with one-sided derivatives, the derivative jump is
\begin{equation}
 J_\tau:=\psi'_+(\tau)-\psi'_-(\tau)
       =[e\tau-2km(\tau)]\,[m'_+(\tau)-m'_-(\tau)].
 \label{eq:jump}
\end{equation}
For a positive-signal interior region with continuously differentiable cost density $f_F$, put $g=\eta q^2/2$, $B_F=F(g)+2g f_F(g)$, and $R_F=\eta+cB_F$. Then
\begin{equation}
 m'=\frac{sB_F}{R_F},\qquad
 m''=\frac{s^2\eta^2q\{3f_F(g)+2g f_F'(g)\}}{R_F^3}.
 \label{eq:primitivecurvature}
\end{equation}
\end{theorem}
\begin{proof}
Substitute Proposition~\ref{prop:cost} and the conditional variance identity into expected cost to obtain \eqref{eq:representation}. A Blackwell refinement admits posterior means $M_0,M_1$ satisfying $\E[M_1\mid\mathcal H_0]=M_0$, so conditional Jensen proves the concavity implication. If concavity fails, choose $x,y\in I$ and $\theta\in(0,1)$ such that $\theta\psi(x)+(1-\theta)\psi(y)>\psi(\theta x+(1-\theta)y)$. A target equal to $x$ or $y$ with these probabilities and its full revelation implement precisely that strict inequality, with the same target law in both experiments. The moment bounds in Theorem~\ref{thm:eq} justify all expectations. Differentiating \eqref{eq:psi} gives \eqref{eq:curvature} and \eqref{eq:jump}. Finally, $m=qF(g)$ and $\mu=q[\eta+cF(g)]/s$ imply $\dd m/\dd q=B_F$ and $\dd\mu/\dd q=R_F/s>0$. A second differentiation, using $B_F'=\eta q(3f_F+2g f_F')$, proves \eqref{eq:primitivecurvature}.
\end{proof}

To interpret \eqref{eq:curvature}, separate marginal inventory incidence from the direct forecasting gain. For $\mu>0$, define
\begin{equation}
 G(\mu)=e\mu-2km(\mu),\qquad
 Q(h)=\frac{d^2}{A}-2eh+2kh^2.
 \label{eq:marginalincidence}
\end{equation}
Here $G$ is the marginal effect of supplied inventory on optimized institutional cost, holding the target mean fixed. For every $h\geq0$, $Q(h)>0$. If $e\leq0$, each term in $Q$ is nonnegative and the first is positive. If $e>0$, the identity $e=db/A-(\lambda-H)<db/A$ and $2k=b^2/A$ give $Q(h)\geq d^2/A-e^2/(2k)>0$. Consequently,
\[
 \psi''(\mu)=G(\mu)m''(\mu)-Q(m'(\mu)).
\]
Within a smooth region, information is locally harmful precisely when $Gm''>Q(m')$. With marginally costly inventory, harm requires sufficiently accelerating supply. With marginally helpful inventory, it requires sufficiently decelerating supply. Opposite signs of $G$ and $m''$ imply strict concavity. At a slope discontinuity, the corresponding condition is $J_\tau=G(\tau)\Delta m'(\tau)$. The total inventory effect, $e\mu m-km^2$, has a different threshold. For $e>0$ and $m>0$, its sign changes at $m/\mu=e/k$, whereas marginal incidence changes at $e/(2k)$. Reflection gives the negative-mean case.

The converse allows the target law to vary across counterexamples. It does not permit arbitrary posterior-mean distributions under a fixed prior. For a given nonbinary prior, feasible posterior means satisfy the additional restrictions characterized by \citet{GentzkowKamenica2016}. The ordering argument uses standard information geometry. The execution coefficients and inventory derivatives determine the financial comparison through the shape of $\psi$.

If a roster of mass $a$ is fixed before information arrives, with implementation costs already sunk, inventory is $m=t(a)\mu$, where $t(a)=as/(\eta+ca)$. In this case
\begin{equation}
 \psi_a(\mu)=\left[-(\lambda-H)t(a)-\frac{\{d-bt(a)\}^2}{2A}\right]\mu^2.
 \label{eq:infofixed}
\end{equation}
The bracket is strictly negative. Every Blackwell refinement therefore lowers cost for each fixed roster, strictly whenever it increases the second moment of the posterior mean. The institution reoptimizes its quantity whether information arrives before or after activation. Only information arriving before avoidable activation can also change who participates.

\paragraph{An exact cash decomposition.}
The cost of a finite refinement can be decomposed without a local approximation. For nested posterior means $M_0,M_1$, write $m_i=m(M_i)$, $\Delta M=M_1-M_0$, and $\Delta m=m_1-m_0$. Equation~\eqref{eq:psi} gives
\begin{equation}
 \Delta\mathcal C
 =-\frac{d^2}{2A}\E[(\Delta M)^2]
 +\E[(eM_1-2km_0)\Delta m]
 -k\E[(\Delta m)^2].\label{eq:cashdecomposition}
\end{equation}
Indeed, $\E[M_1^2-M_0^2]=\E[(\Delta M)^2]$ and $\E[m_0\Delta M]=0$. Expanding $M_1m_1-M_0m_0=M_1\Delta m+m_0\Delta M$ and $m_1^2-m_0^2=2m_0\Delta m+(\Delta m)^2$ proves the identity, including refinements that cross participation boundaries.

The original cash objective gives each term an interpretation. Hold suppliers' inventory at $m_0$ and let the institution adjust its pretrade to $M_1$. The resulting saving is the first term. Now change inventory to $m_1$, keeping that pretrade fixed. The change in expenditure is the middle term. Finally, let the institution reoptimize against $m_1$. The resulting cost change is $-k\E[(\Delta m)^2]$, since $k=b^2/(2A)$. Information is harmful exactly when the middle cost increase exceeds the two savings together. These steps are accounting counterfactuals, not separate asymmetric-information equilibria. Appendix~\ref{app:cashcomponents} evaluates them in the smooth example.

\subsection{Information conditional on a trading environment}\label{sec:state}
A public range or activity measure may reveal trading conditions as well as the target. To distinguish these channels, let $Z$ be observed before both target experiments. Let execution coefficients and the activation-cost law be measurable functions of $Z$. Suppose the primitive assumptions hold conditional on $Z$. Costs must be conditionally independent of the target and the entire compared signal family given $Z$. In the finite game, private costs must also be conditionally independent across allocations. With $\mathcal H_0\subseteq\mathcal H_1$ both containing $\sigma(Z)$, the same calculation gives
\begin{equation}
 \mathcal C(\mathcal H_i)=\E[\lambda_Z D^2]+
 \E[\psi_Z(\E[D\mid\mathcal H_i])].\label{eq:conditionalstate}
\end{equation}
For example, integrability follows from square-integrable $D$ and execution coefficients in a compact admissible set bounded away from zero denominators. The equilibrium and Jensen arguments then apply conditional on $Z$, after which expectations can be taken. This conditional formulation does not identify $F_Z$ from trading volume. Newly revealing $Z$, changing the conditional cost law, or delaying execution goes beyond a target-information comparison within a fixed environment.

Forecast accuracy and institutional information value are different comparisons. True nested conditional means $M_i=\E[D\mid\mathcal H_i]$ satisfy
\begin{equation}
 \E[(D-M_0)^2]-\E[(D-M_1)^2]=\E[(M_1-M_0)^2]\geq0.\label{eq:projection}
\end{equation}
The adverse effect in Theorem~\ref{thm:geometry} can therefore coexist with improved target prediction. Worse performance by a fitted regression reflects estimation or specification error, rather than harmful information in this sense. Appendix~\ref{app:forecastbound} bounds the error in a model-implied cost comparison when posterior means are estimated.

\subsection{Entry onset and saturation}
Assume a cost distribution supported on $[f_0,f_1]$, where $0<f_0<f_1<\infty$, with endpoint expansions
\begin{equation}
 F(f_0+x)=\ell_0x+O(x^2),\qquad
 1-F(f_1-x)=\ell_1x+O(x^2),\qquad \ell_0,\ell_1>0.
 \label{eq:endpoints}
\end{equation}
Define the onset and full-participation thresholds and the saturated inventory slope by
\begin{equation}
 \mu_0=\frac{\sqrt{2\eta f_0}}s,\qquad
 \mu_1=\frac{\eta+c}{s}\sqrt{\frac{2f_1}{\eta}},\qquad
 t_1=\frac{s}{\eta+c}.
 \label{eq:threshold}
\end{equation}
Entry starts at $\mu_0$, and the entire pool participates above $\mu_1$. Full participation exhausts the entry margin, but each active quantity remains unrestricted. Aggregate inventory continues to grow as $m=t_1\mu$. There is no hard position or balance-sheet cap.

\begin{theorem}[Two sources of harmful information]\label{thm:boundaries}
Under Assumption~\ref{ass:primitive} and \eqref{eq:endpoints}, set
\[
 L_0=\frac{2s f_0\ell_0}{\eta+2cf_0\ell_0},\qquad
 B_1=1+2f_1\ell_1,\qquad L_1=\frac{sB_1}{\eta+cB_1}>t_1.
\]
The jumps of $\psi'$ at the positive boundaries are
\begin{equation}
 J_0=e\mu_0L_0,\qquad
 J_1=\mu_1(e-2kt_1)(t_1-L_1).
 \label{eq:boundaryjumps}
\end{equation}
Consequently, onset has an upward jump if $e>0$, and saturation has an upward jump if $e<2kt_1$, including every $e\leq0$. At any boundary $\tau$ with $J_\tau>0$, fix $\delta>0$ and the target $D=\tau+\delta B$, where $B$ is a symmetric sign. A noisy observation of $B$ with accuracy $(1+\epsilon/\delta)/2$, for $0<\epsilon<\delta$, refines no information and changes institutional cost by
\begin{equation}
 \Delta\mathcal C=\frac{J_\tau}{2}\epsilon+O(\epsilon^2)>0
 \quad\text{for sufficiently small }\epsilon.
 \label{eq:boundaryinfo}
\end{equation}
Every admissible execution environment with \eqref{eq:endpoints} therefore admits a harmful refinement for a suitably located prior, as well as beneficial refinements in regions with constant participation.
\end{theorem}
\begin{proof}
At onset, $m=0$ below the boundary and expansion of the equilibrium equation gives $m(\mu_0+x)=L_0x+O(x^2)$. At saturation, $m(\mu_1)=t_1\mu_1$, the left slope is $L_1$ by the endpoint expansion, and the right slope is $t_1$. The endpoint expansions imply quadratic remainders for these inventory expansions, as shown in Appendix~\ref{app:information}. Formula~\eqref{eq:jump} gives the stated jumps. Bayes' rule gives posterior means $\tau\pm\epsilon$ with equal probability, so the cost change is the symmetric difference $[\psi(\tau+\epsilon)+\psi(\tau-\epsilon)]/2-\psi(\tau)$. Its linear term is $J_\tau\epsilon/2$. If $e>0$ the onset jump is positive, while if $e\leq0$ the saturation jump is positive because $k,t_1>0$ and $L_1>t_1$. A binary refinement confined to the nonempty no-entry region has the strictly negative change $-d^2\epsilon^2/(2A)$, which proves the final assertion.
\end{proof}

At onset, the first entrants create more inherited price displacement than they save in the institution's auction bill. Saturation has a different mechanism when $e\leq0$. Inventory lowers institutional cost at every signal, so a downward forecast revision withdraws helpful supply. An upward revision cannot attract more participants from the exhausted pool, although incumbents still increase their quantities. This asymmetry can make information harmful even when participation is helpful. Persistent pre-event impact is necessary for harmful onset in this model, but not for harmful information near saturation.

For an explicit example without pre-event impact, take $\lambda=\eta_0=\eta=1$, $K=H=0$, and uniform costs on $[0.02,0.04]$. Then $A=3$, $b=1$, $d=2$, $s=1/3$, $c=2/3$, $e=-1/3$, $k=1/6$, and $\mu_1=\sqrt2$. The inventory slopes are $5/13$ from the left and $1/5$ from the right, yielding
\begin{equation}
 J_1=\frac{24\sqrt2}{325},\qquad
 \Delta\mathcal C=\frac{12\sqrt2}{325}\epsilon+O(\epsilon^2).
 \label{eq:saturationexample}
\end{equation}
With participation fixed at one, the same experiment instead changes cost by $-0.74\epsilon^2$.

\subsection{A primitive classification between the boundaries}
\begin{proposition}[Uniform implementation costs]\label{prop:uniform}
For $F$ uniform on $[f_0,f_1]$, inventory has $m'>0$ and $m''>0$ throughout the positive-signal interior. If $e\leq0$, $\psi$ is strictly concave in every smooth participation region, its onset jumps are nonpositive, and its only failures of concavity occur at saturation. If $e>0$, positive smooth curvature can occur only where
\begin{equation}
 \frac{m(\mu)}{\mu}<\frac{e}{2k},\qquad \mu>0.
 \label{eq:interiornecessary}
\end{equation}
Negative signals obey the reflected classification.
\end{proposition}
\begin{proof}
The uniform density gives $3f_F+2gf_F'=3/(f_1-f_0)>0$ in \eqref{eq:primitivecurvature}. For $e\leq0$, every term in \eqref{eq:curvature} is nonpositive and its first term is strictly negative. Onset jumps have the sign of $e$. On constant-participation regions \eqref{eq:infofixed} gives strict concavity. For $e>0$, completing the square in the first three terms of \eqref{eq:curvature} bounds their sum above by $-d^2/A+e^2/(2k)<0$, since $0<e<db/A$ and $2k=b^2/A$. Positive curvature therefore requires $e\mu-2km>0$. Finally, the equilibrium has odd inventory $m$ and even cost $\psi$, which gives reflection across zero.
\end{proof}

Condition~\eqref{eq:interiornecessary} is stricter than $m/\mu<e/k$, the condition under which participation raises the level of cost. Participation can therefore be costly even where additional information is locally beneficial. The institution's direct forecasting gain can outweigh the additional supply response before inventory provision becomes beneficial in levels.

\subsection{When information remains beneficial}\label{sec:beneficial}
Information can also be ranked favorably without fixing participation. The alternative representation
\begin{equation}
 \psi(\mu)=-(\lambda-H)\mu m(\mu)-\frac{\{d\mu-bm(\mu)\}^2}{2A}
 \label{eq:psimonotone}
\end{equation}
gives the following benchmark.
\begin{proposition}[Centered targets and Gaussian precision]\label{prop:beneficial}
For every cost law in Assumption~\ref{ass:primitive}, $\psi$ is even, satisfies $\psi(0)=0$, and is strictly decreasing on $[0,\infty)$. Hence, for any square-integrable target with $\E D=0$, a public signal lowers institutional cost relative to no information, strictly if its posterior mean is nonzero with positive probability. If $D\sim N(0,\sigma_D^2)$ with $\sigma_D>0$, every increase in the precision of an independent Gaussian signal strictly lowers cost, including with optional participation.
\end{proposition}
\begin{proof}
For $\mu\geq0$, parameterize the equilibrium by $q\geq0$. Both $q$ and $m=qF(\eta q^2/2)$ are nondecreasing in $\mu$, and $q$ is strictly increasing because $\eta q+cm=s\mu$. For $\mu_2>\mu_1\geq0$, this identity gives
\[
 0\leq m(\mu_2)-m(\mu_1)\leq\frac{s}{c}(\mu_2-\mu_1),\qquad
 d-\frac{bs}{c}=\frac{A(\lambda-H)}{\eta_0+b}>0.
\]
Thus $d\mu-bm$ is positive and strictly increasing for $\mu>0$, while $\mu m$ is nonnegative and nondecreasing. Equation~\eqref{eq:psimonotone} proves the first assertion, without differentiability of $F$. Relative to a zero posterior mean, every nonzero posterior mean strictly reduces $\psi$. For a centered Gaussian prior, the posterior mean has distribution $\sigma_D\sqrt r\,Z$, where $Z\sim N(0,1)$ and $r\in[0,1]$ is the fraction of target variance explained. Coupling these distributions with the same $|Z|$ gives strict decrease as $r$ increases. The quadratic moment bound justifies taking expectations.
\end{proof}

The no-information benchmark differs from a refinement after the direction is known. Let $S,B$ be independent symmetric signs, fix $0<\epsilon<\delta<\tau$, and put $D=S(\tau+\delta B)$. This target is centered. Coarse information reveals $S$. A symmetric channel independent of $S$ reports $T\in\{-1,1\}$ with $\Pr(T=B)=(1+\epsilon/\delta)/2$. Fine information reveals $(S,T)$, giving
\[
 M_0=S\tau,\qquad M_1=S(\tau+\epsilon T),\qquad
 \Delta\mathcal C=\frac{\psi(\tau-\epsilon)+\psi(\tau+\epsilon)}2-\psi(\tau).
\]
Evenness of $\psi$ gives the last equality. Choosing $\tau$ in an adverse region and sufficiently small $\epsilon$ makes this refinement harmful, although learning $S$ initially is beneficial. This four-point prior is distinct from the centered Gaussian precision family. A signal that reveals magnitude while keeping the posterior mean zero gives equality under risk neutrality. Appendix~\ref{app:favorable} gives a complementary global example: with no pre-event impact and exponential activation costs, $\psi$ is strictly concave and every nontrivial posterior-mean refinement is beneficial.

\subsection{Harmful information in a smooth interior}\label{sec:smoothinterior}
Harmful information does not require a derivative jump. In \eqref{eq:marginalincidence}, smooth supply curvature can also outweigh the institution's direct forecasting benefit. The following construction uses accelerating supply with positive marginal incidence.

\begin{proposition}[Smooth-interior harm]\label{prop:smoothinterior}
Suppose the cost density is continuously differentiable near the equilibrium cutoffs on a neighborhood of $\mu_*>0$, and $\psi''(\mu_*)>0$. There exist $r_*>0$ and an open interval $I$ containing $\mu_*$ such that $\psi''\geq r_*$ on $I$. Let $X$ have a continuously differentiable density with compact support inside $I$, let $B$ be an independent symmetric sign, and fix $D=X+\delta B$, $\delta>0$. Coarse information reveals $X$. A noisy observation of $B$ with accuracy $(1+\epsilon/\delta)/2$, where $0<\epsilon<\delta$ and $\operatorname{supp}(X)+[-\epsilon,\epsilon]\subset I$, changes cost by
\begin{equation}
 \Delta\mathcal C
 =\E\!\left[\frac{\psi(X+\epsilon)+\psi(X-\epsilon)}2-\psi(X)\right]
 \geq\frac{r_*}{2}\epsilon^2>0.
 \label{eq:smoothharm}
\end{equation}
Such an economy exists with a smooth positive density on all positive costs and participation $a=1/4$. The strict inequality persists under sufficiently small admissible perturbations of the execution parameters, for a fixed experiment contained in a smaller interval.
\end{proposition}
\begin{proof}
The strictly increasing map $q\mapsto q[\eta+cF(\eta q^2/2)]/s$ has positive derivative. Its inverse and \eqref{eq:primitivecurvature} therefore give continuous $\psi''$ locally. Choose $I,r_*$ by continuity. Bayes' rule gives the two refined posterior means $X\pm\epsilon$. Integrating the bound on $\psi''$ twice proves \eqref{eq:smoothharm}. The unconditional target distribution is continuous, and all compared posterior means stay in the same smooth region.

For existence, use a uniform execution schedule and
\[
 (\lambda,\kappa,\rho,\eta_0,\eta)=(1,0.99,0,0.01,4).
\]
Let costs be lognormal with mean $0.021$ and variance $0.002^2/12$, so
\[
 \sigma_f^2=\log(1+1/1323),\qquad
 \ell_f=\log(0.021)-\sigma_f^2/2.
\]
Put $z_* =\Phi^{-1}(1/4)$, $g_*=\exp(\ell_f+\sigma_fz_*)$, $q_* =\sqrt{2g_*/\eta}$, and $\mu_*=q_*(\eta+c/4)/s$. Then $a=1/4$ exactly. Substitution in \eqref{eq:primitivecurvature} and \eqref{eq:curvature}, certified in Appendix~\ref{app:certificate}, gives
\[
 \mu_*\approx0.82506008,\qquad 4.4003<\psi''(\mu_*)<4.4005.
\]
There is neither a positive cost floor nor a finite full-participation threshold. The same local formulas are continuous in admissible execution parameters, so the strict curvature bound survives in a smaller fixed neighborhood, including for sufficiently small positive decay.
\end{proof}

Persistent price displacement, a low institutional temporary charge, a higher supplier charge, and concentrated activation costs produce the rapidly changing participation response in this construction. These are model assumptions, not estimates from ETF prices. The matched gamma law also has positive curvature at one-quarter participation, whereas the lognormal example has negative curvature at median participation. The result identifies an adverse region without implying an adverse ranking for every smooth cost law or every forecast.

A fixed adverse experiment remains adverse under sufficiently small uniform perturbations of the activation CDF. Appendix~\ref{sec:robustness} proves this stability result, classifies other cost-tail behavior, and integrates local effects across dispersed coarse forecasts. Appendix~\ref{app:informationclock} gives the corresponding posterior-martingale decomposition at a fixed execution opportunity.

\subsection{Helpful inventory with smooth interior harm}\label{sec:smoothcohorts}
Information can be harmful even when inventory is helpful at every signal and total participation remains well below one. Consider a low-cost supplier cohort followed by a more expensive cohort. When the first cohort is nearly exhausted, an upward forecast revision attracts few additional participants. A downward revision can still remove helpful suppliers. Smoothing the cohort distributions preserves this asymmetry without introducing an entry boundary.

\begin{proposition}[Smooth harm at any interior participation mass]\label{prop:smoothcohorts}
Fix an execution environment satisfying Assumption~\ref{ass:primitive} with $e\leq0$. For every $w\in(0,1)$ and every $h>0$, there is a cost distribution with a smooth positive density on $(0,\infty)$ and an open positive-mean interval $I$ such that
\[
 |a(\mu)-w|<h\quad\text{and}\quad\psi''(\mu)>0\qquad(\mu\in I).
\]
A binary target supported in $I$ therefore admits a harmful public refinement under its unchanged prior. Inventory nevertheless lowers institutional cost at every nonzero posterior mean.
\end{proposition}
The proof in Appendix~\ref{app:smoothcohorts} first exhausts a continuous low-cost cohort of mass $w$ and then smooths its positive costs multiplicatively. Before smoothing, the derivative jump is strictly positive even when $e=0$. The institutional readjustment term contributes $-2km$. Stability of a fixed information experiment preserves the positive gap after smoothing.

For an explicit example, set $\lambda=\eta_0=\eta=1$ and $K=H=0$. Let activation costs have an equal mixture of lognormal distributions with means $0.03$ and $0.30$, each with coefficient of variation $0.03$. At $\mu=1.013$, the entered mass is $0.49569$, aggregate inventory is $0.12580$, and
\[
 e=-\tfrac13,\qquad m''(1.013)=-8.50496,\qquad
 \psi''(1.013)=1.72430>0.
\]
Under the fixed prior $D=1.013\pm0.02$, a binary signal with accuracy $0.75$ gives posterior means $1.013\pm0.01$. It raises normalized expenditure by $0.00007589$, or $0.02552\%$ of the coarse cost $0.29734$. Direct terminal payments agree with the reduced cost representation. Smooth harm therefore occurs without persistent impact, unequal temporary charges, or participation near one. Its magnitude is modest and depends on concentrated cost cohorts. Appendix~\ref{app:smoothcohorts} reports sensitivity to dispersion and cohort weights under a fixed prior, including cases in which the sign reverses.

\begin{figure}[htbp]\centering
\includegraphics[width=.97\textwidth]{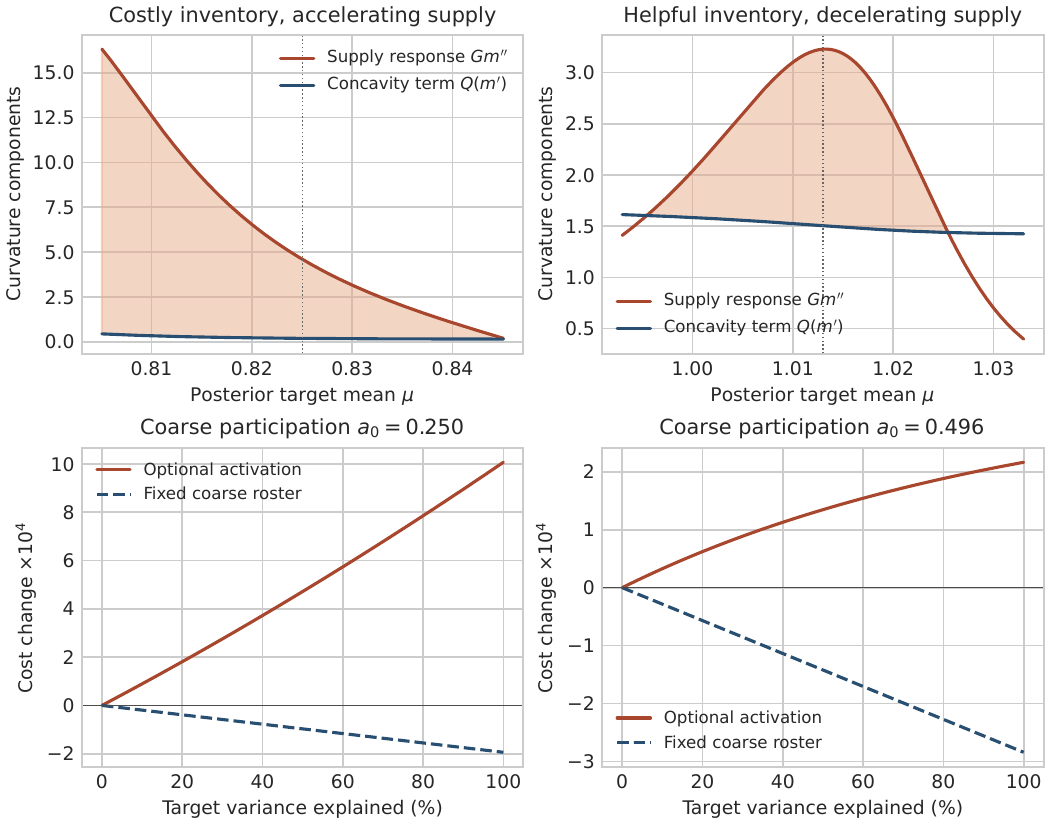}
\caption{Two smooth mechanisms for harmful information. The left column uses the lognormal construction of Proposition~\ref{prop:smoothinterior}, with coarse mean $\mu_*=0.82506008$. The right uses the two-cohort construction of Section~\ref{sec:smoothcohorts}, with coarse mean $1.013$. The upper panels compare the supply-curvature contribution $Gm''$ with $Q(m')$. Shading marks positive cost curvature. The lower panels hold each binary prior $D=\bar\mu\pm0.02$ fixed and vary signal accuracy. The horizontal axis is $100(\epsilon/0.02)^2$. Dashed curves fix participation at its own coarse-signal level. The cost scales differ across columns.}\label{fig:scalarclassification}
\end{figure}

\FloatBarrier
\section{The direction of portfolio information}\label{sec:portfolioinformation}
A supplier pays one private activation cost for a basket and then chooses quantities across $p\geq2$ assets. The vector game in Appendix~\ref{app:jointmodel} has an isotropic specialization with
\[
 \Lambda=\lambda I_p,\quad K=K_sI_p,\quad H=H_sI_p,\quad
 E_0=\eta_0I_p,\quad E=\eta I_p,
\]
where $(\lambda,K_s,H_s,\eta_0,\eta,F)$ satisfy Assumption~\ref{ass:primitive}. Equilibrium quantities are parallel to the posterior mean $\mu$. The optimized mean-dependent cost is $\Psi(\mu)=\psi(\|\mu\|)$, where the scalar function is defined in \eqref{eq:psi}. At a nonzero smooth state, put $r=\|\mu\|$ and $n=\mu/r$. Then
\begin{equation}
 \nabla^2\Psi(\mu)=\psi''(r)nn^\top
 +\frac{\psi'(r)}r(I_p-nn^\top).\label{eq:radialhessian}
\end{equation}
The radial direction changes basket magnitude. Orthogonal directions change composition to first order. The following experiments translate this local distinction into a feasible cash comparison under one prior.

\begin{corollary}[Equal forecast accuracy, opposite execution values]\label{cor:directional}
Let $n,v\in\R^p$ be orthonormal, and choose a smooth state $r>0$ with $\psi''(r)>0$. Fix $\delta>0$ and the target law
\begin{equation}
 D=rn+\delta(B_1n+B_2v),\label{eq:directionalprior}
\end{equation}
where $B_1,B_2$ are independent symmetric signs. For $0<\epsilon<\min\{\delta,r\}$, let $U_1,U_2$ be independent signs, independent of the target signs and activation costs, with $\Pr(U_j=1)=(1+\epsilon/\delta)/2$. Observing $T_1=B_1U_1$ alone or $T_2=B_2U_2$ alone gives posterior means $rn\pm\epsilon n$ or $rn\pm\epsilon v$, respectively, each pair equiprobable. Both refine no information and reduce ordinary squared forecast error by $\epsilon^2$ and $\Lambda$-weighted squared error by $\lambda\epsilon^2$. Their cost changes are
\begin{align}
 \Delta\mathcal C_{\rm magnitude}
 &=\tfrac12\{\psi(r+\epsilon)+\psi(r-\epsilon)\}-\psi(r)
 =\tfrac12\psi''(r)\epsilon^2+o(\epsilon^2),\label{eq:directionalgaps}\\
 \Delta\mathcal C_{\rm composition}
 &=\psi\!\left(\sqrt{r^2+\epsilon^2}\right)-\psi(r)<0.\nonumber
\end{align}
The magnitude signal raises cost for sufficiently small $\epsilon>0$. The composition signal lowers cost for every $0<\epsilon<\delta$.
\end{corollary}

\begin{figure}[htbp]\centering
\includegraphics[width=\textwidth]{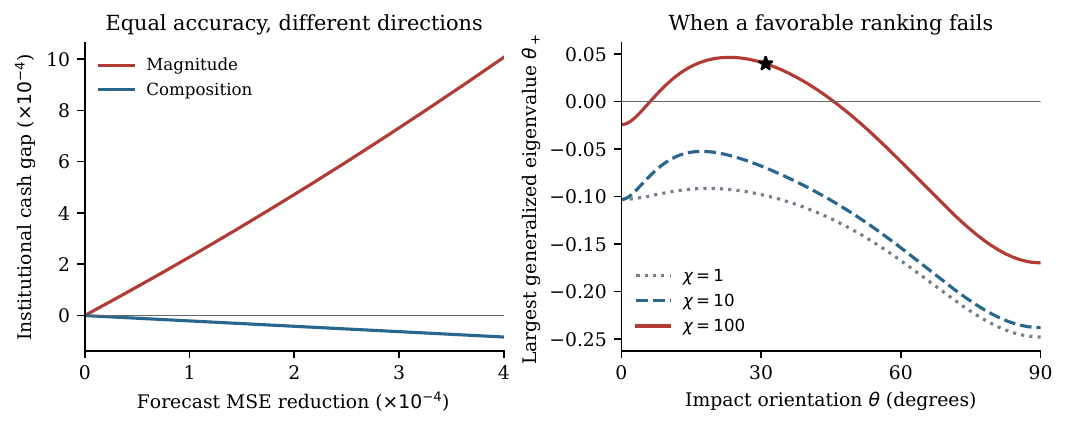}
\caption{Two portfolio mechanisms. Left: optional-entry magnitude and composition signals under the same isotropic four-point target law, with $\delta=.02$ and $\lambda=1$. The common coarse bill is $.351308300279$, including both shocks. The horizontal coordinate is the MSE reduction shared by the two signals. Right: the largest generalized eigenvalue of $(P_{1/2},\Lambda)$ across the execution technologies in \eqref{eq:spectralfamily}, for three institutional charge ratios $\chi$. Positive values permit harmful information at a fixed roster. The star marks the cross-impact counterexample. Costs are normalized model quantities.}\label{fig:portfoliomain}
\end{figure}

The proof is in Appendix~\ref{app:directional}. Both signals refine the same coarse experiment, but they need not be Blackwell ordered relative to each other. Figure~\ref{fig:portfoliomain} evaluates the comparison at the certified smooth state of Proposition~\ref{prop:smoothinterior}. Table~\ref{tab:directionalmain} gives numerical magnitudes. The signals remove the same amount of statistical and execution-weighted forecast error. Their different costs arise from the direction of the forecast revision.

\paragraph{When the scalar rankings survive.}
For general execution matrices, let $\Lambda$ denote terminal inverse depth, $H$ terminal carryover, and $E$ the suppliers' temporary charge. The remaining coefficient matrices are defined in \eqref{eq:vectorcoeff}. At a participation mass $a$, aggregate inventory is $T_a\mu$, where
\begin{align}
 T_a&=a(E+aC)^{-1}S,\nonumber\\
 P_a&=-\operatorname{Sym}\{(\Lambda-H)T_a\}
 -\tfrac12(L-BT_a)^\top A^{-1}(L-BT_a),\label{eq:vectorP}
\end{align}
and $\operatorname{Sym}(X)=(X+X^\top)/2$. The optimized mean-dependent cost is $\mu^\top P_a\mu$. Its first term measures the cost of inventory supplied in the target direction. Its second is the saving from the institution's optimal pretrade. Cross-impact can rotate supplied inventory away from the target and make the first term positive.

This matrix gives two exact rankings. For a fixed roster, every refinement weakly lowers cost across all square-integrable target laws if and only if $P_a\preceq0$ (Proposition~\ref{prop:vectorranking}). Now fix any one continuous cost law and allow optional basket activation. Every signal weakly improves on no information across all centered vector targets if and only if $P_a\preceq0$ at \emph{every} $a\in[0,1]$ (Proposition~\ref{prop:vectorcentered}). In the converse construction, the execution matrices determine an adverse direction. The cost law determines a target scale that implements its participation mass. This centered-prior criterion does not rank arbitrary pairs of informative signals.

Simultaneously diagonalizable execution matrices satisfy both favorable conditions. A strictly negative generalized-eigenvalue margin preserves these rankings in an open neighborhood as impact orientation and trading charges change. Appendix~\ref{app:portfoliospectral} measures this margin and maps its failure around the counterexample. The map uses $\lambda_{\max}(\Lambda^{-1/2}P_a\Lambda^{-1/2})$. This quantity measures the largest cost effect per unit of execution-weighted forecast improvement and is invariant to asset denomination.

Optional entry also preserves a Gaussian benchmark under isotropic execution. Every nested jointly Gaussian refinement for a centered Gaussian portfolio target weakly lowers cost, with a strict decrease when posterior-mean covariance changes (Proposition~\ref{prop:vectorGaussian}). More information increases the distribution of the forecast's radius, and $\psi$ decreases with that radius. The conclusion requires Gaussian experiments as well as a Gaussian target. It does not extend here to every diagonal but non-isotropic basket technology.

\section{Finite strategic populations}\label{sec:finite}
\subsection{The symmetric Bayesian equilibrium at finite size}
For $N$ potential inventory allocations, trader $i$ purchases $q_i/N$. It pays temporary charge $\eta q_i^2/(2N)$ and activation cost $f_i/N$, and liquidates the same position at the event. Costs are conditionally independent across traders and independent of $(D,Y)$. Equilibrium means Bayesian Nash equilibrium with independent private randomization conditional on the public signal. The population scaling preserves the total mass of potential allocations as $N$ changes. Normalized profit equals $N$ times an individual's cash profit. The institution chooses simultaneously with the traders, before observing realized entries. Its first-order condition therefore holds their strategies fixed.

\begin{theorem}[Exact symmetric finite-population equilibrium]\label{thm:exactfinite}
Under Assumption~\ref{ass:primitive}, for each $N\geq2$ there is a unique symmetric Bayesian equilibrium in entry and normalized quantities, up to zero-measure indifferent types. Put
\begin{equation}
 \eta_N=\eta+\frac{2b}{N},\qquad c_N=c-\frac bN>0.
 \label{eq:finitecoeff}
\end{equation}
Its entry probability, common entrant quantity, mean aggregate inventory, and institutional pretrade satisfy
\begin{align}
 a_N&=F\left(\frac{\eta_Ns^2\mu^2}{2(\eta_N+c_Na_N)^2}\right),&
 q_N&=\frac{s\mu}{\eta_N+c_Na_N},\label{eq:finiteexact}\\
 \bar m_N&=a_Nq_N,&z_N&=\frac{d\mu-b\bar m_N}{A}.\notag
\end{align}
All equilibrium positions have finite second moments. The realized aggregate has conditional variance $a_N(1-a_N)q_N^2/N$, and expected institutional cost is
\begin{equation}
 \mathcal C_N(\mathcal H)=\lambda\E[D^2]+\E[\psi_N(\E[D\mid\mathcal H])],\qquad
 \psi_N(\mu)=-\frac{d^2\mu^2}{2A}+e\mu\bar m_N-k\bar m_N^2.
 \label{eq:finitecost}
\end{equation}
\end{theorem}
\begin{proof}
Against symmetric rival entry and quantity $(a,q)$, an entering trader choosing normalized quantity $u$ obtains conditional expected normalized gross profit
\[
 u\{\lambda\mu-bz-b(1-1/N)aq\}-\frac{\eta_N}{2}u^2.
\]
Strict concavity makes the entered quantity independent of private activation cost, and the optimal gross profit is $\eta_Nq^2/2$. The institution's objective is affine in random aggregate inventory, which is independent of the target innovation conditional on $Y$, so its expected objective for every possible deviation is \eqref{eq:major} evaluated at $m=aq$. Its unique best response is therefore $z=(d\mu-baq)/A$. Substituting this into the minor's first-order condition yields $\eta_Nq=s\mu-c_Naq$. Since $c-b/2=b\eta_0/(2A)>0$, $c_N>0$ for all $N\geq2$, and the monotone-root proof of Theorem~\ref{thm:eq} applies to \eqref{eq:finiteexact}. This proves existence and uniqueness within the symmetric Bayesian class and verifies global deviations in both quantity and entry. The bounds $|q_N|\leq s|\mu|/\eta_N\leq s|\mu|/\eta$ give integrability. Independent Bernoulli entry gives the variance formula, and completing the institution's square after taking the conditional expectation gives \eqref{eq:finitecost}.
\end{proof}

The square in \eqref{eq:finitecost} is the square of expected inventory. The institution chooses before learning which traders enter. Using the expected square of realized inventory would give it additional information and describe a different game. Theorem~\ref{thm:exactfinite} establishes uniqueness within the symmetric class for every $N\geq2$. The following condition rules out asymmetric Bayesian equilibria as well.

\begin{proposition}[Uniqueness without imposed symmetry]\label{prop:finiteall}
In addition to Assumption~\ref{ass:primitive}, suppose $F$ is locally absolutely continuous and
\[
 \mathcal B_F:=\mathop{\rm ess\,sup}_{g>0}\{F(g)+2g f_F(g)\}<\infty.
\]
For every integer $N\geq2$ satisfying
\begin{equation}
 N\eta+2b>b\mathcal B_F,\label{eq:finiteallcondition}
\end{equation}
the equilibrium in Theorem~\ref{thm:exactfinite} is the unique Bayesian Nash equilibrium, up to indifferent types of probability zero.
\end{proposition}
\begin{proof}
In any equilibrium, strict concavity makes trader $i$'s entering quantity $q_i$ independent of its private fixed cost. Its entry probability is $a_i=F(\eta_Nq_i^2/2)$. With $\bar m=N^{-1}\sum_i a_iq_i$, its first-order condition is
\[
 \left[\eta_N-\frac bN F(\eta_Nq_i^2/2)\right]q_i
 =\lambda\mu-bz-b\bar m.
\]
The left side, viewed as a function of $q_i$, is locally absolutely continuous with derivative at least $(N\eta+2b-b\mathcal B_F)/N>0$ almost everywhere. It is therefore strictly increasing. All traders have the same quantity and entry probability, and Theorem~\ref{thm:exactfinite} gives uniqueness.
\end{proof}

For uniform costs on $[0.02,0.04]$ under the principal execution parameters, \eqref{eq:finiteallcondition} holds for $N\geq32$. For the smooth lognormal economy of Proposition~\ref{prop:smoothinterior}, it holds for $N\geq4$. The reported information reversals at those population sizes therefore do not depend on selecting a symmetric equilibrium. Appendix~\ref{app:finiteselection} derives these bounds and constructs strict asymmetric equilibria at $N=2$, where the sufficient condition fails.

\begin{corollary}[A nonuniform information limit]\label{cor:nonuniform}
Assume $f_0>0$, $F(f_0+x)=\ell_0x+O(x^2)$ with $\ell_0>0$, and $e>0$. Define the binary-refinement gap at the continuum entry boundary by
\[
 \Delta_N(\epsilon)=\tfrac12\{\psi_N(\mu_0+\epsilon)+\psi_N(\mu_0-\epsilon)\}-\psi_N(\mu_0).
\]
The target law is held fixed within each information comparison. The finite entry boundary and right inventory slope are
\begin{equation}
 \mu_{0,N}=\frac{\sqrt{2\eta_Nf_0}}s>\mu_0,\qquad
 L_N=\frac{2sf_0\ell_0}{\eta_N+2c_Nf_0\ell_0}.
 \label{eq:finitethreshold}
\end{equation}
A binary refinement centered at $\mu_{0,N}$ changes cost by $e\mu_{0,N}L_N\epsilon/2+O(\epsilon^2)>0$ locally. At the unchanged continuum center $\mu_0$, however, every fixed $N$ has
\begin{equation}
 \Delta_N(\epsilon)=-\frac{d^2}{2A}\epsilon^2<0
 \quad\text{if }0<\epsilon<\min\{\mu_0,\mu_{0,N}-\mu_0\}.
 \label{eq:finitequiet}
\end{equation}
For each fixed finite displacement the cost comparison converges to the continuum comparison as $N\to\infty$, but
\begin{equation}
 \lim_{N\to\infty}\lim_{\epsilon\downarrow0}\frac{\Delta_N(\epsilon)}{\epsilon}=0,
 \qquad
 \lim_{\epsilon\downarrow0}\lim_{N\to\infty}\frac{\Delta_N(\epsilon)}{\epsilon}
 =\frac{e\mu_0L_0}{2}>0.
 \label{eq:nonuniform}
\end{equation}
\end{corollary}
\begin{proof}
Apply the endpoint expansion with $(\eta_N,c_N)$ in place of $(\eta,c)$. Below $\mu_{0,N}$ neither finite branch enters, and \eqref{eq:finitecost} reduces to the no-entry quadratic, proving \eqref{eq:finitequiet}. For fixed $\mu$, any subsequential limit of the roots in the compact interval $[0,1]$ solves the continuum equation by continuity. Uniqueness gives convergence of the whole sequence, and the quantities and costs then converge. The two iterated limits follow from \eqref{eq:finitequiet} and Theorem~\ref{thm:boundaries}.
\end{proof}

Individual price impact raises the forecast threshold at which entry first becomes profitable. Near that threshold, a small shift can reverse the information comparison. An approximate equilibrium payoff bound cannot settle the sign of an information effect smaller than the bound. The exact equilibrium determines that sign without imposing a uniform position bound.

\begin{corollary}[Joint information and population scale]\label{cor:jointscale}
Under Corollary~\ref{cor:nonuniform}, define
\[
 \xi_0=\frac{b\mu_0}{\eta}>0,\qquad \gamma_0=\frac{e\mu_0L_0}{2}>0.
\]
Then $\mu_{0,N}=\mu_0+\xi_0/N+O(N^{-2})$. For the unchanged binary prior centered at $\mu_0$ and information displacement $\epsilon_N=x/N$, with fixed $x>0$,
\begin{equation}
 \lim_{N\to\infty}N\Delta_N(x/N)=\gamma_0(x-\xi_0)_+.
 \label{eq:jointscale}
\end{equation}
If $x<\xi_0$, the change is exactly $-d^2x^2/(2AN^2)$ for all sufficiently large $N$. If $x>\xi_0$, it is positive of order $N^{-1}$. At equality the first-order limit is zero. Appendix~\ref{app:finiteselection} resolves the next order: the gap is strictly negative and of order $N^{-2}$.
\end{corollary}
\begin{proof}
Expand $\mu_{0,N}=\mu_0\sqrt{1+2b/(\eta N)}$. The fixed endpoint expansion of $F$ and coefficients converging to strictly positive limits give, uniformly for small $h\geq0$, $\bar m_N(\mu_{0,N}+h)=L_Nh+O(h^2)$, with $L_N\to L_0$. For large $N$, the center and lower branch are inactive. The upper branch consequently has inventory
$L_N(x-\xi_0)_+/N+O(N^{-2})$, including an $O(N^{-2})$ bound at equality. Inserting this into \eqref{eq:finitecost} proves \eqref{eq:jointscale}. When $x<\xi_0$, both branches are below the finite threshold eventually, giving the exact quadratic expression.
\end{proof}

The relevant resolution near onset is $N\epsilon$. This limit explains how information interacts with strategic size. It does not give an exact sign rule for moderate $N$. Under the principal numerical parameters, $\xi_0\approx0.953145$ and $\gamma_0\approx0.0333783$. At $x=1$, $N\Delta_N(1/N)$ is negative at $N=100$ and positive at $N=1{,}000$, approaching $0.00156393$.

Appendix~\ref{sec:facility} studies a signed outside completion facility. It preserves the sign of initial inventory incidence under the stated coefficient restrictions and gives a separate check on the role of terminal execution technology.

\section{Numerical implications}\label{sec:numerical}
All numerical costs are normalized by $\lambda D_*^2$, where $D_*$ is a reference target quantity. Inventory is measured in units of $D_*$ and execution coefficients in units of $\lambda$. The numerical exercises use constructed economic environments. Each percentage change is relative to that scenario's coarse institutional bill. It is not a return on fund assets.

Table~\ref{tab:principalcomparisons} compares four scalar refinements at the same posterior-mean displacement, $\epsilon=.01$. Each row holds its own prior fixed. The onset environment has $(\lambda,\kappa,\rho,\eta_0,\eta)=(1,1,.1,.05,.05)$ and uniform activation costs on $[.02,.04]$. Its impact half-life is $6.93$ pre-event horizons. The saturation environment has zero pre-event impact. The one-cohort smooth example is the certified construction of Proposition~\ref{prop:smoothinterior}. The two-cohort example removes both persistent displacement and unequal temporary charges. In the latter three cases, the adverse effects are small relative to the constructed execution bills. An adverse region need not produce a large average loss.

\begin{table}[htbp]\centering\small
\caption{Fixed-prior scalar comparisons at $\epsilon=.01$. Coarse bills include target variance. The prior half-spread is $.04$ at onset, $.10$ at saturation, and $.02$ in both smooth examples. The last column is $100\Delta\mathcal C/\mathcal C_0$.}\label{tab:principalcomparisons}
\begin{tabular}{@{}lrrr@{}}\toprule
Scenario & Coarse bill & Cost change & Change, \%\\\midrule
Costly supply at onset & .005665 & +.00030485 & 5.3812\\
Helpful supply at full participation & .530000 & +.00044338 & .0837\\
Smooth costly supply & .350908 & +.00022785 & .0649\\
Smooth helpful supply near one-half & .297339 & +.00007589 & .0255\\\bottomrule
\end{tabular}
\end{table}

Finite strategic size can reverse the first row's sign. Under its unchanged continuum-centered prior, the information gap is $-.00001503$ for $N=100$ and $+.00014455$ for $N=200$, compared with $+.00030485$ in the continuum. Both finite cases satisfy the all-equilibrium uniqueness condition. At $N=100$, the entry threshold moves from $.089562$ to $.098634$. Recentering the prior at the latter threshold produces a positive gap of $.00033884$, but changes the target distribution. The finite-population comparison distinguishes this recentered exercise from the fixed-prior result.

Local curvature need not determine the value of full revelation. Under the fixed Gaussian prior centered at the full-participation boundary, small precision gains raise cost, whereas full revelation lowers it by $.00370761$. For centered Gaussian priors, information remains favorable throughout, as the theory requires. Appendix~\ref{app:numericalcomparisons} reports these precision paths, exact finite-threshold plots, schedule comparisons, and outside-completion results. Appendix~\ref{app:smoothscale} reports sensitivity to smooth cost distributions and execution parameters.

\FloatBarrier
\section{Price-informed target distributions}\label{sec:data}
\subsection{What the prices identify}
Daily-reset funds provide a concrete setting for the model's committed target. Prices describe its uncertainty, while residual orders, activation costs, and execution bills remain unobserved. We use chronological forecasts to assess target information and explicit public signals to evaluate the model's cost comparison under a fixed empirical target law.

The five-minute panel begins on November 9, 2021 for every benchmark and geared product. The common forecasting sample contains 717 normal sessions in 2023--2025 and 219 in 2026 for SPY, QQQ, IWM, and DIA. Features use completed bars in New York time. Under the primary interval-start convention, the noon and terminal marks have labels 11:55 and 15:55. Appendix~\ref{app:products} gives the session rules, past-only training windows, and consistent interval-end sensitivity. Some 2026 prices had already been examined during specification work, so that period provides a retrospective temporal evaluation.

At noon, the observed reset-period return $R_{0,t}$ is known. The remaining return $r_{t,T}$ determines the normalized no-flow target
\begin{equation}
 D=a_\beta\{R_{0,t}+(1+R_{0,t})r_{t,T}\},
 \qquad a_\beta=\beta(\beta-1).
 \label{eq:pricetarget}
\end{equation}
The observed component therefore remains informative about the target even when the remaining-return forecast is zero. The target is an exposure proxy in fixed reference units. Converting it to shares requires division by the terminal price before conditional averaging. Different leverage labels rescale the same benchmark paths and add no independent observations.

\begin{table}[htbp]\centering\footnotesize
\caption{Information in the price panel. The periods use identical dates across the four families. Remaining-return models use 252 prior observations. Interval-score changes are range-scaled minus raw bands, with lower scores preferred.}\label{tab:history}
\begin{tabular}{p{.45\textwidth}rr}
\toprule
Quantity & 2023--2025 & 2026 \\
\midrule
Zero-forecast remaining-return RMSE (bp) & 66.94 & 49.47 \\
Returns-plus-range RMSE / zero RMSE & 1.0159 & 1.0123 \\
Range-minus-raw interval score (bp) & $-42.59$ & $-26.70$ \\
HAC 95\% interval for score difference & $[-93.55,8.38]$ & $[-50.44,-2.96]$ \\
Holm-adjusted $p$, six score comparisons & 0.304 & 0.137 \\
\bottomrule
\end{tabular}
\end{table}

Every fitted mean model in the stated comparison has a higher point RMSE than the zero remaining-return forecast. Range information lowers the estimated interval score, although a few dates account for much of the gain and its statistical support depends on the inference method. The year 2025 accounts for $96.51\%$ of the reference period's net improvement. The five largest absolute date contributions account for $65.80\%$ and $61.51\%$ of the respective net gains. Five-session bootstrap intervals exclude zero, whereas the reference-period ten-session HAC interval includes it. The reference range-versus-RV comparison is more stable. Appendices~\ref{app:calibrationdetails} and \ref{app:econometrics} retain the full comparisons. The improvement concerns uncertainty calibration. Incremental mean prediction and additional activation in the risk-neutral model remain unestablished.

\subsection{A controlled information comparison}
For each period, we place equal mass on the date--family observations and normalize $D$ in \eqref{eq:pricetarget} by $0.01a_\beta$. We retain the observed return distribution without centering or trimming. Family identity is public. The coarse signal adds the sign of $R_{0,t}$, and the fine signal adds its absolute-magnitude category, using fixed boundaries at 25, 50, and 100 basis points. Both signals use information present at the same noon decision. Their posterior means are exact cell averages under this finite law, so $\E[M_1\mid Y_0]=M_0$ holds by construction. Both posterior means use the same finite law. The exercise evaluates a distributional scenario and makes no out-of-sample forecasting claim.

The prior, physical execution horizon, schedule, and cost law are fixed within each comparison. Table~\ref{tab:priceexperiment} compares two specified technologies: the persistent-impact onset parameters of Section~\ref{sec:numerical}, and $K=H=0$, $\lambda=\eta_0=\eta=1$. Both use uniform activation costs on $[0.02,0.04]$. The target scale multiplies all observations without recentering. A fixed-roster comparison holds participation at its coarse-signal average. The finite calculations satisfy the sufficient condition for uniqueness among all Bayesian equilibria.

\begin{table}[htbp]\centering\footnotesize
\caption{Noon sign-to-magnitude refinement under unchanged price-informed priors. Cash units are normalized by the reference target and $\lambda=1$. Negative changes benefit the institution. Both target scales belong to the declared sensitivity grid.}\label{tab:priceexperiment}
\begin{tabular}{llrrrrr}
\toprule
Period & Technology & Scale & Coarse bill & $\Delta\mathcal C$ & Fixed roster & $N=100$ \\
\midrule
2023--2025 & Persistent & 0.125 & 0.01648 & +0.00016 & -0.00189 & +0.00018 \\
2023--2025 & Persistent & 1 & 1.20155 & -0.02636 & -0.02759 & -0.02697 \\
2023--2025 & Zero & 0.125 & 0.01498 & -0.00276 & -0.00276 & -0.00276 \\
2023--2025 & Zero & 1 & 0.95043 & -0.20884 & -0.18016 & -0.20874 \\
2026 & Persistent & 0.125 & 0.01538 & +0.00022 & -0.00193 & +0.00032 \\
2026 & Persistent & 1 & 1.15179 & -0.02979 & -0.03006 & -0.03040 \\
2026 & Zero & 0.125 & 0.01302 & -0.00307 & -0.00307 & -0.00307 \\
2026 & Zero & 1 & 0.81873 & -0.23397 & -0.20237 & -0.23382 \\
\bottomrule
\end{tabular}

\end{table}

At unit target scale, the institution benefits under both technologies in both periods. In the persistent-impact case, the reference-period direct forecasting saving is $0.13064$, the inventory response adds $0.19019$, and institutional readjustment saves $0.08592$. Their sum is a cost reduction of $0.02636$. At scale $1/8$, the same return distribution places posterior means closer to entry. Information then raises cost under persistent impact. The cost increases are $0.00016$ and $0.00022$, about $0.99\%$ and $1.46\%$ of their constructed coarse bills. These are model expenditures under illustrative parameters, with no interpretation as estimated fund losses.

\begin{figure}[htbp]\centering
\includegraphics[width=.93\textwidth]{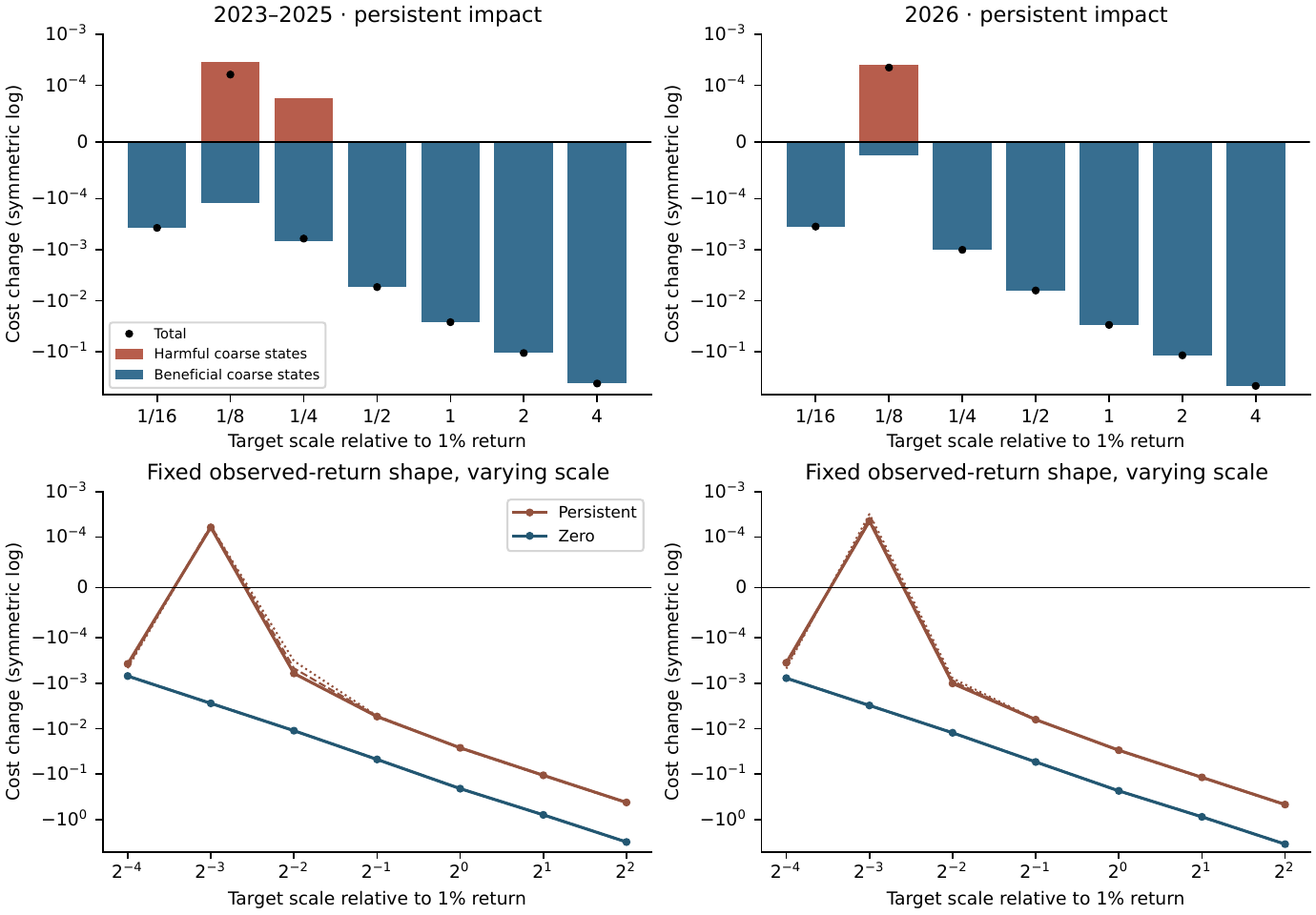}
\caption{Where the fixed-prior comparison changes sign. Upper panels aggregate beneficial and harmful information values within coarse family--sign states, then weight by their probabilities. Lower panels show total changes across the declared target scales. Solid curves are continuum equilibria, dotted curves use $N=100$, and dashed curves use $N=200$. The two technologies are those in Table~\ref{tab:priceexperiment}. The vertical scale is symmetric logarithmic with a linear region around zero.}\label{fig:priceexperiment}
\end{figure}

The broader grid varies impact persistence, temporary costs, cost dispersion, target scale, and population size. Only 4 of 336 continuum cases in 2023--2025 and 3 of 336 in 2026 have positive total cost changes for the observed-noon signal. A separate hypothetical terminal-sign refinement gives 3 positive cases in each period. These counts summarize the stated grid. They do not estimate the probability of adverse information in markets. Appendix~\ref{app:priceexperiments} gives the complete design, signed contributions, date-block sensitivity, and independent cash checks. Replacing the previous intraday mark by the matched daily reference, with or without cash-distribution adjustment, preserves the signs of all reported matched-panel scenario comparisons.

The remaining price evidence describes trading conditions and target uncertainty. Signed post-close movements are $-1.23$ basis points before 2026 and $-0.12$ in 2026, with pooled paired-clock intervals containing zero. Frozen correlation gives lower point scores for joint uncertainty than rolling correlation. The appendix retains these results, turnover decompositions, and factor attribution. They neither identify activation nor estimate structural cross-impact.

\FloatBarrier
\section{Conclusion}\label{sec:conclusion}
A public forecast changes how an institution prepares for its committed order and how much inventory other traders supply. The shape of that supply response determines whether the institution benefits. Costly accelerating supply and helpful decelerating supply can each outweigh the direct forecasting gain. The smooth two-cohort construction establishes the second mechanism without persistent displacement or participation near the full pool.

The favorable results define the limits of these adverse comparisons. Information weakly lowers cost for fixed scalar rosters and relative to no information for centered scalar targets. Centered Gaussian precision gains remain favorable under risk neutrality. For portfolios, an exact matrix condition extends the centered-target comparison to optional basket activation. Signal direction matters outside that class. Finite strategic size adds a further qualification. The uniqueness condition identifies comparisons that hold without selecting a symmetric equilibrium.

The price application characterizes target distributions and checks their measurement. Structural entry and institutional losses remain unidentified. In the risk-neutral model, information affects positions through the conditional target mean. The appendix's risk extension allows conditional variance to affect supply through inventory risk. All comparisons hold the physical execution opportunity fixed. Allowing traders to wait or change their schedule would add a strategic decision and requires a separate model.

\begingroup
\setstretch{1.05}

\endgroup

%% ===================================================================
%% Internet Appendix (separately paginated, as in the companion PDF)
%% ===================================================================
\clearpage
\pagestyle{plain}
\setlength{\headheight}{12pt}
\setcounter{page}{1}
\def\theHpage{IA.\arabic{page}}
\setcounter{footnote}{0}
% Appendix citations resolve to the appendix reference list (natbib/hyperref).
\makeatletter\gdef\@extra@b@citeb{@IA}\immediate\write\@auxout{\gdef\string\@extra@binfo{@IA}}\makeatother
\setcounter{table}{0}
\setcounter{figure}{0}
\setcounter{theorem}{0}
\setcounter{assumption}{0}
\setcounter{remark}{0}
\setcounter{equation}{0}
\renewcommand{\theHtheorem}{IA.\arabic{theorem}}
\renewcommand{\theHtable}{IA.\arabic{table}}
\renewcommand{\theHfigure}{IA.\arabic{figure}}
\renewcommand{\theHassumption}{IA.\arabic{assumption}}
\renewcommand{\theHremark}{IA.\arabic{remark}}
% Internet Appendix title block (same layout as the article class \maketitle).
\thispagestyle{plain}
\null\vskip 2em
\begin{center}
{\LARGE \bfseries Internet Appendix\\[.3em]\large Trading Before a Committed Rebalance\\Public Information and Optional Liquidity Supply \par}
\vskip 1.5em
{\large\lineskip .5em\begin{tabular}[t]{c}\end{tabular}\par}
\vskip 1em
{\large \par}
\end{center}
\par\vskip 1.5em
\setstretch{1.08}
\noindent This appendix contains supporting proofs, finite-population selection results, portfolio information criteria, and extensions for heterogeneous trading costs and inventory risk. The price application documents exact fixed-prior experiments and reconciles reset references using the existing daily records. The remaining sections retain the forecasting, econometric, turnover, and factor analyses. Returns and price changes are in basis points unless stated otherwise.
\setcounter{tocdepth}{1}
\begingroup\small\setstretch{1.0}
\makeatletter
\renewcommand{\l@section}[2]{\@dottedtocline{1}{0em}{1.8em}{#1}{#2}}
\makeatother
\tableofcontents
\addtocontents{toc}{\protect\setcounter{tocdepth}{1}}
\endgroup
\clearpage
\appendix
\numberwithin{equation}{section}
\renewcommand{\thetable}{IA.\arabic{table}}
\renewcommand{\thefigure}{IA.\arabic{figure}}
\renewcommand{\thetheorem}{IA.\arabic{theorem}}
\section{Smooth-interior certification and favorable information}\label{app:certificate}
\subsection{A reproducible curvature certificate}
For the lognormal construction in Proposition~\ref{prop:smoothinterior}, the reduced coefficients are
\[
 b=\frac{101}{200},\quad A=\frac{51}{50},\quad d=\frac{101}{100},\quad
 s=\frac{10199}{20400},\quad c=\frac{10403}{40800},\quad
 e=\frac{9997}{20400},\quad k=\frac{10201}{81600}.
\]
With $z=(\log g-\ell_f)/\sigma_f$, the density and its derivative are
\[
 f_F(g)=\frac{\varphi(z)}{\sigma_fg},\qquad
 f_F'(g)=-\frac{f_F(g)}{g}\left(1+\frac{z}{\sigma_f}\right).
\]
The quantile $z_*=\Phi^{-1}(1/4)$ lies in $[-0.674490,-0.674489]$. To certify the bracket, for $x=-z>0$ evaluate
\[
 \Phi(-x)=\frac12-\frac1{\sqrt{2\pi}}\sum_{j=0}^{12}
 \frac{(-1)^jx^{2j+1}}{2^jj!(2j+1)}+R(x),\qquad
 |R(x)|\leq\frac{x^{27}}{\sqrt{2\pi}\,2^{13}13!\,27}.
\]
The terms decrease at both bracket endpoints, so the alternating-series remainder applies. Outward interval evaluation at 50 decimal digits gives $\Phi(-0.674490)<0.249999921$ and $\Phi(-0.674489)>0.250000238$. Evaluating the density formulas and \eqref{eq:curvature} over that bracket, with $F(g_*)=1/4$ exactly, encloses the curvature in
\[
 [4.40037496282478,\ 4.40039330270280],
\]
where the displayed endpoints have been rounded outward. This proves the deliberately looser bound in the proposition. The replication code records the full intervals, precision, series remainder, and coefficient inputs. It uses no estimated parameters.

\begin{table}[htbp]\centering\small
\caption{Smooth full-support costs under the interior technology. The gamma and lognormal laws both have mean $0.021$ and variance $0.002^2/12$. Positive curvature occurs at one-quarter participation but is absent at the displayed median states.}\label{tab:interior}
\begin{tabular}{lrrrrr}
\toprule
Cost law & $a$ & $\mu$ & $m'$ & $m''$ & $\psi''$ \\
\midrule
lognormal & 0.250000 & 0.825060 & 1.173250 & 11.545310 & 4.400386 \\
lognormal & 0.500000 & 0.845806 & 1.280465 & 0.184871 & -0.080796 \\
gamma & 0.250000 & 0.825086 & 1.170225 & 11.683487 & 4.454331 \\
gamma & 0.500000 & 0.845860 & 1.280337 & 0.308177 & -0.031302 \\
\bottomrule
\end{tabular}

\end{table}

\subsection{Finite gaps and execution-parameter sensitivity}\label{app:smoothscale}
Table~\ref{tab:smoothscale} holds each binary target law $D=\mu_*\pm0.02$ fixed within its coarse/refined comparisons. The first two rows use the certified lognormal economy. The third uses the same cost law and a uniform schedule with $(\lambda,\kappa,\rho,\eta_0,\eta)=(1,1.5,1,4,4)$. Its carryover is $H=0.948181$, its integrated impact is $K=0.551819$, and its impact half-life is $0.693147$ pre-event horizons. At quarter participation, $\mu_*=0.477822$ and $\psi''(\mu_*)\approx0.103310$. Equal temporary charges and positive decay therefore admit a positive numerical gap as well. Concentrated activation costs and high carryover remain features of this illustration. The additional example is verified from the original cash objective at 60-digit precision, but has no directed-rounding curvature certificate.

\begin{table}[htbp]\centering\footnotesize
\caption{Economic scale of smooth information effects. Coarse cost includes the fixed prior variance. The final column divides the gap by that row's coarse institutional bill. Costs are normalized cash units, not basis points of assets.}\label{tab:smoothscale}
\begin{tabular}{lrrrr}
\toprule
Economy & $\epsilon$ & Coarse cost & Gap & Gap / cost (\%) \\
\midrule
Certified economy & 0.001 & 0.35090830 & 0.0000022010 & 0.000627 \\
Certified economy & 0.010 & 0.35090830 & 0.0002278461 & 0.064930 \\
Equal charges, positive decay & 0.010 & 0.20528799 & 0.0000060913 & 0.002967 \\
\bottomrule
\end{tabular}

\end{table}

\begin{figure}[htbp]\centering
\includegraphics[width=\textwidth]{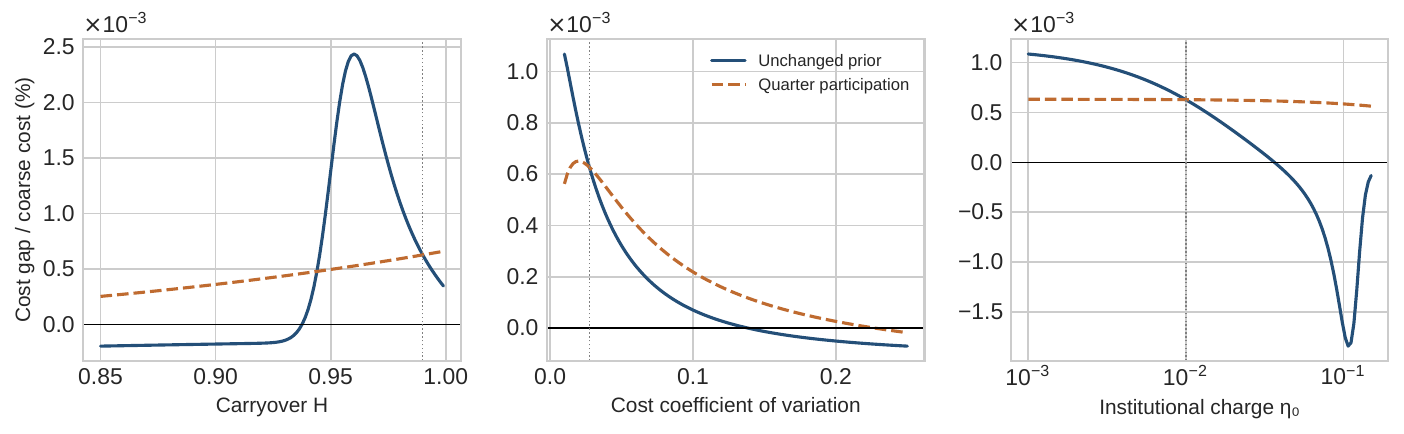}
\caption{Smooth-interior sensitivity. Each panel changes only its indicated primitive from the certified economy. Left: zero decay with $\kappa=H$ and $K=H/2$. Middle: lognormal dispersion with mean cost fixed at $0.021$. Right: institutional temporary charge with supplier charge fixed at $4$. The vertical scale is percent of each comparison's own coarse cost. The curves are numerical illustrations.}\label{fig:smoothsensitivity}
\end{figure}

Figure~\ref{fig:smoothsensitivity} varies one primitive at a time with $\epsilon=0.001$ and $\delta=0.02$. Solid curves retain the original target center. Dashed curves recenter at quarter participation in each economy, changing the prior across economies while preserving it within every information comparison. Along the unchanged-prior paths, the numerical zero crossings are $H\approx0.93753$, cost coefficient of variation $0.13847$, and institutional temporary charge $\eta_0\approx0.03650$. The respective root brackets are $[0.90,0.95]$, $[0.10,0.20]$, and $[0.02,0.05]$. These are one-dimensional calculations under the stated restrictions, not global parameter boundaries or calibration estimates.

\subsection{Components of the finite information cost}\label{app:cashcomponents}
Table~\ref{tab:cashcomponents} evaluates \eqref{eq:cashdecomposition} for the same lognormal economy and unchanged scalar prior $D=\mu_*\pm.02$. The inventory response outweighs the direct forecasting and institutional readjustment savings. Its coarse bill is $.350908300279$. This differs from the portfolio bill in Table~\ref{tab:directionalmain} by the additional target-shock variance, although the radial cost gaps coincide. Values are high-precision numerical evaluations of the original cash objective.
\begin{table}[htbp]\centering\footnotesize
\caption{Exact accounting components of a finite refinement. All columns use normalized scalar cash units. The first and third terms are savings, and their sum with inventory incidence equals the total change. Intermediate accounting steps are not separate equilibria.}\label{tab:cashcomponents}
\begin{tabular}{rrrrr}
\toprule
$\epsilon$ & Forecasting & Inventory incidence & Readjustment & Total \\
\midrule
0.001 & -0.000000500049 & +0.000002873060 & -0.000000172044 & +0.000002200966 \\
0.010 & -0.000050004902 & +0.000294687460 & -0.000016836429 & +0.000227846129 \\
\bottomrule
\end{tabular}

\end{table}

\FloatBarrier
\Needspace{8\baselineskip}
\subsection{A globally beneficial economy}\label{app:favorable}
\begin{proposition}[Exponential costs without pre-event impact]\label{prop:exponential}
Let $\lambda=\eta_0=\eta=1$, $K=H=0$, and $F(g)=1-\exp(-g/\theta)$ on $g\geq0$, where $\theta>0$. Then $\psi$ is globally strictly concave. Every feasible public refinement weakly lowers institutional cost, strictly whenever the posterior means change with positive probability.
\end{proposition}
\begin{proof}
For $\mu>0$, set $x=g/\theta$, $F=1-e^{-x}$, $B=1+(2x-1)e^{-x}$, and $u=B/(3+2B)$. Formula~\eqref{eq:primitivecurvature} gives
\[
 \psi''=-\frac43-\frac{2u}{3}-\frac{u^2}{3}
 -\frac{6x(1+F)(3-2x)e^{-x}}{(3+2B)^3}.
\]
For $0<x\leq3/2$, all terms are nonpositive and the first is strictly negative. For $x>3/2$, $B\geq1$ and $1+F\leq2$. The only positive term is at most
\[
 \frac{12}{125}x(2x-3)e^{-x}\leq\frac{108e^{-3}}{125}<0.044,
\]
because the function $x(2x-3)e^{-x}$ attains its maximum on $[3/2,\infty)$ at $x=3$. Thus $\psi''<0$ there as well. At zero, $m=O(\mu^3)$, so $\psi$ has matching first derivatives and $\psi''(0)=-4/3$. Reflection proves global strict concavity. Conditional strict Jensen then gives the information ranking.
\end{proof}

\subsection{Gaussian precision under an unchanged prior}\label{app:gaussian}
For $D\sim N(\bar\mu,\sigma_D^2)$, $\sigma_D>0$, observing $D$ plus independent Gaussian noise of variance $\sigma_D^2(1-r)/r$ gives posterior mean distribution $N(\bar\mu,r\sigma_D^2)$, $0<r<1$. Lower precision can be implemented by adding independent noise to a higher-precision signal. The endpoints $r=0,1$ are no information and revelation, respectively. For $Z\sim N(0,1)$,
\begin{equation}
 \mathcal C(r)=\lambda(\bar\mu^2+\sigma_D^2)
   +\E[\psi(\bar\mu+\sigma_D\sqrt r\,Z)].\label{eq:gaussianpath}
\end{equation}
Under the uniform-cost specifications, $\psi$ has quadratic growth, bounded piecewise curvature, and finitely many slope jumps. The Gaussian density $p_r$ obeys $\partial_r p_r=(\sigma_D^2/2)\partial_{xx}p_r$. Differentiation and distributional integration by parts, with vanishing Gaussian boundary terms, give for $r>0$
\begin{equation}
 \mathcal C'(r)=\frac{\sigma_D^2}{2}
 \left[\int_{\R}p_r(x)\psi''_{\rm reg}(x)\dd x
       +\sum_\tau p_r(\tau)J_\tau\right].\label{eq:gaussianderivative}
\end{equation}
The result is the usual Gaussian smoothing identity applied to the cost function. Numerical integration splits at all four participation boundaries, with independent checks in both posterior-mean and standard-normal coordinates.

\subsection{Sensitivity to estimated posterior means}\label{app:forecastbound}
For fixed primitives write $L_m=s/c$ and $t_1=s/(\eta+c)$. The monotone equilibrium equation implies $|m(u)-m(v)|\leq L_m|u-v|$, including nondifferentiable cost laws, and $|m(u)|\leq t_1|u|$. To see the first assertion, write $\mu=(\eta q+cm)/s$ and note that $q$ and $m$ are increasing together, also across zero. The second follows from $m/\mu=as/(\eta+ca)$ and $a\leq1$. Consequently,
\begin{align}
 |\psi(u)-\psi(v)|&\leq C_\psi(|u|+|v|)|u-v|,\label{eq:psiestimationbound}\\
 C_\psi&=\frac{d^2}{2A}+\frac{|e|}{2}(L_m+t_1)+kt_1L_m.\nonumber
\end{align}
Indeed, use $u m(u)-v m(v)=\{(u+v)[m(u)-m(v)]+[m(u)+m(v)](u-v)\}/2$ for the middle term of $\psi$ and factor both squares. Cauchy--Schwarz gives, for square-integrable $U,V$,
\[
 |\E\psi(U)-\E\psi(V)|\leq C_\psi(\|U\|_2+\|V\|_2)\|U-V\|_2.
\]
Apply this inequality to both the coarse and refined estimates and add the bounds. The target second moment cancels in a same-prior cost difference. This controls sensitivity to a given mean-error bound, without estimating that error or the structural coefficients from ETF bars.

\section{Notation}\label{app:notation}
\begingroup\setstretch{1.0}\small
\begin{center}\begin{tabular}{p{.16\textwidth}p{.76\textwidth}}
\toprule Symbol & Meaning \\
\midrule
$D,Y$ & Terminal target in fixed benchmark units and public signal observed before pre-event trading.\\
$\mu,v$ & Conditional target mean and variance given the public signal.\\
$F,f$ & Distribution and realization of private fixed entry cost per normalized trader.\\
$a,q,m$ & Entered population mass, inventory per entrant, and aggregate inventory $m=aq$.\\
$z,J,n$ & Institutional pre-event quantity, completion order $J=D-z$, and net auction demand $n=D-z-m$.\\
$\phi,I,\rho,\kappa$ & Common execution schedule, transient impact, decay rate, and impact loading.\\
$K,H$ & Schedule-averaged price impact and pre-auction carryover per unit of pre-event trading.\\
$\lambda,\eta_0,\eta$ & Auction inverse depth and institutional and normalized minor temporary-charge coefficients.\\
$A,b,d$ & Coefficients of the institutional cash objective in \eqref{eq:abd}.\\
$s,c$ & Effective signal loading and crowding coefficient after solving the institutional first-order condition.\\
$e,k$ & Linear inventory-incidence coefficient and quadratic institutional-adjustment coefficient.\\
$f_0,f_1,\ell_0,\ell_1$ & Cost endpoints and one-sided endpoint densities.\\
$\mu_0,\mu_1,L_0,L_1$ & Participation thresholds and interior one-sided inventory slopes.\\
$\psi,J_\tau$ & Posterior-mean cost function and its derivative jump at a boundary.\\
$\eta_N,c_N,\bar m_N$ & Finite-game coefficients and conditional mean aggregate inventory.\\
$N$ & Number of potential inventory allocations in the finite game.\\
$\xi_0,\gamma_0$ & Threshold-shift and adverse-information coefficients in the joint finite limit.\\
$\bar\mu,\sigma_D,r$ & Gaussian prior mean, standard deviation, and fraction of variance explained.\\
$\delta,\epsilon$ & Binary target half-spread and posterior-mean displacement.\\
$\chi,j$ & Outside completion charge and institutional auction quantity when an outside facility is available.\\
$\beta,A_0,R$ & Daily leverage objective, assumed initial fund assets, and benchmark simple return over the specified reset period.\\
\bottomrule
\end{tabular}\end{center}
\endgroup

\section{Cash accounts, units, and round trips}\label{app:cash}
\subsection{Derivation of the two objectives}
Let $\Phi_t=\int_0^t\phi_s\dd s$. The institution has inventory $z\Phi_t$ before the event, an entrant has $q\Phi_t$, and impact is $(z+m)\kappa\int_0^te^{-\rho(t-s)}\phi_s\dd s$. Integration of the pre-event cash purchases gives $Kz(z+m)$ for the institution and $Kq(z+m)$ for an entrant. Adding each temporary charge and the terminal purchase or sale gives
\begin{align*}
 C_0^{\rm realized}&=Kz(z+m)+\eta_0z^2/2+(D-z)p_A,\\
 \pi^{\rm realized}(q)&=q p_A-Kq(z+m)-\eta q^2/2.
\end{align*}
Taking expectations and using $\E[(D-z)^2\mid Y]=(\mu-z)^2+v$ produces \eqref{eq:major} and \eqref{eq:minor}. In particular, replacing the institution's completion payment by a diagonal residual penalty would remove the terms involving $m$ and would change the game.

The unscaled institution and mass-one anticipators use compatible aggregate units. The expected gross anticipator profit is $a\eta q^2/2$, and total fixed entry expenditure is $\int_0^{\eta q^2/2}f\dd F(f)$. If these expenditures are interpreted as resource costs, net trading profit subtracts that integral. If they are fees paid to another service provider, the recipient must also be included before making a welfare statement.

\subsection{Counterparty accounting}
Define the outside supplier's marked revenue as its pre-event receipts, temporary charges, and auction receipts, less inventory valued at the unaffected benchmark. Conditional expected revenue is
\begin{equation}
 \begin{split}
 \Pi_D={}&K(z+m)^2+\eta_0z^2/2+a\eta q^2/2\\
 &+H(z+m)(\mu-z-m)+\lambda\{(\mu-z-m)^2+v\}.
 \end{split}\label{eq:dealer}
\end{equation}
Direct subtraction of the cash accounts gives
\begin{equation}
 C_0-a\pi(q)=\Pi_D.\label{eq:cashidentity}
\end{equation}
This identity is checked numerically for every baseline equilibrium in the reported grid. It records transfers. A social objective would also require dealer costs, capital requirements, and risk preferences, which are not modeled here. Adding an independently chosen temporary-displacement penalty would therefore define a new welfare criterion.

\subsection{Admissibility and zero-target deviations}
When $D=0$, $m=0$, and the institution trades $z$ before undoing that trade at the event, its cash cost is $Az^2/2\geq0$. A sole finite minor with actual quantity $u/N$ has normalized profit $-(\eta/2+b/N)u^2\leq0$ before its nonnegative fixed cost. Thus the retained technology does not reward these mechanical round trips. The statement applies to the prescribed schedule and terminal convention. It does not establish absence of manipulation for arbitrary multivenue or multi-asset trading paths.

For each public state, square completion gives the institution a finite lower bound because $A>0$, and the minor a finite upper bound because $\eta>0$. Theorem~\ref{thm:eq} gives $|q|\leq s|\mu|/\eta$, $|m|\leq s|\mu|/(\eta+c)$, and
\[
 |z|\leq\frac{d+bs/\eta}{A}|\mu|.
\]
Since conditional expectation contracts $L^2$, all quadratic cash terms are integrable. The expected entry-cost integral is bounded by $a\eta q^2/2$, so it is integrable even when the unconditional entry-cost distribution has an infinite first moment. These bounds establish the admissibility of the proposed strategies without an ellipticity condition or a state-space compactness assumption.

\section{Endpoint expansions and cost-tail regularity}\label{app:information}
For a positive posterior mean, parameterize equilibrium by the entrant quantity $q$, with $g=\eta q^2/2$ and
\[
 \mu=\frac{q}{s}\{\eta+cF(g)\},\qquad m=qF(g).
\]
At onset, $q_0=\sqrt{2f_0/\eta}$ and $\mu_0=\eta q_0/s$. Write $x=g-f_0\downarrow0$. Expansion of the square root and the endpoint CDF gives
\[
 \mu-\mu_0=\left(\frac{1}{sq_0}+\frac{cq_0\ell_0}{s}\right)x+O(x^2),
 \qquad m=q_0\ell_0x+O(x^2).
\]
The coefficient of $x$ is positive, so inversion gives $m(\mu_0+h)=L_0h+O(h^2)$ with the constant in Theorem~\ref{thm:boundaries}. Below the boundary the unique equilibrium has $a=m=0$.

At saturation let $q_1=\sqrt{2f_1/\eta}$ and write $x=f_1-g\downarrow0$. Then $q=q_1-x/(\eta q_1)+O(x^2)$ and $F(g)=1-\ell_1x+O(x^2)$, so
\begin{align*}
 \mu_1-\mu&=\left(\frac{\eta+c}{s\eta q_1}+\frac{cq_1\ell_1}{s}\right)x+O(x^2),\\
 m(\mu_1)-m(\mu)&=\left(\frac{1}{\eta q_1}+q_1\ell_1\right)x+O(x^2).
\end{align*}
The ratio is $s(1+2f_1\ell_1)/[\eta+c(1+2f_1\ell_1)]=L_1$, proving the left expansion and its quadratic remainder without assuming global smoothness of $F$. Above the boundary $m=t_1\mu$ exactly. The symmetric-difference expansion of $\psi$ therefore has the remainder asserted in \eqref{eq:boundaryinfo}.

For completeness, in the binary experiment write $S\in\{-1,1\}$ for the reported sign and impose $\Pr(S=B\mid B)=(1+\epsilon/\delta)/2$. Then $\Pr(S=1)=1/2$, $\E[B\mid S]=S\epsilon/\delta$, and $\operatorname{Var}(D\mid S)=\delta^2-\epsilon^2$. Forgetting $S$ recovers the coarse experiment. This verifies feasibility under the unchanged target distribution, including the variance component of institutional expenditure.

\subsection{Proof of the power-law onset classification}\label{app:power}
Let $h=\mu-\mu_0>0$ and $x=g-f_0$. Continuity and uniqueness of equilibrium imply $x\downarrow0$ with $h\downarrow0$, and the parametrization above gives
\[
 h=\frac{x}{sq_0}(1+o(1))+\frac{cq_0 C}{s}x^\alpha(1+o(1)),
 \qquad m=q_0Cx^\alpha(1+o(1)).
\]
If $\alpha>1$, the first term dominates, yielding $x\sim sq_0h$ and the stated $M_\alpha$. If $\alpha<1$, the second term dominates and $m\sim sh/c$. When $\alpha=1$, both terms contribute and their sum yields $2sf_0C/(\eta+2cf_0C)$. In the lower posterior branch entry is zero, so the exact cost change is
\[
 -\frac{d^2}{2A}\epsilon^2+
 \frac12\{e(\mu_0+\epsilon)m(\mu_0+\epsilon)-km(\mu_0+\epsilon)^2\}.
\]
Substituting the three equivalents proves Proposition~\ref{prop:power}. In the case $1<\alpha<2$, the leading positive term is of order $\epsilon^\alpha$, while for $\alpha>2$ all inventory terms are $o(\epsilon^2)$. At $\alpha=2$ the displayed coefficient follows, with no sign conclusion at equality because the hypotheses specify only a little-$o$ remainder.

\section{An outside completion facility}\label{app:facility}
Conditional on terminal information, the institution chooses $j$ to minimize
\[
 j\{H(z+m)+\lambda(j-m)\}+\frac\chi2(D-z-j)^2.
\]
The curvature is $\chi+2\lambda>0$, and its unique minimizer is \eqref{eq:facility}. Put $B_\chi=\chi+2\lambda$ and
\[
 L_\chi=\chi\mu-(\chi+H)z+(\lambda-H)m.
\]
After substitution, conditional expected total institutional expenditure is
\begin{equation}
 C_\chi=Kz(z+m)+\frac{\eta_0}2z^2
 +\frac\chi2\{(\mu-z)^2+v\}
 -\frac{L_\chi^2+\chi^2v}{2B_\chi}.\label{eq:outsidecost}
\end{equation}
Define the institutional coefficients
\begin{align*}
 A_\chi&=2K+\eta_0+\chi-\frac{(\chi+H)^2}{B_\chi},&
 d_\chi&=\frac{\chi(2\lambda-H)}{B_\chi},\\
 b_\chi&=K+\frac{(\chi+H)(\lambda-H)}{B_\chi},&
 l_\chi&=\frac{\lambda\chi}{B_\chi},
\end{align*}
and the minor's two aggregate price loadings
\[
 b_{z,\chi}=K-H+\frac{\lambda(\chi+H)}{B_\chi},\qquad
 b_{m,\chi}=K-H+\frac{\lambda(\chi+\lambda+H)}{B_\chi}.
\]
The institution's first-order condition is $A_\chi z=d_\chi\mu-b_\chi m$. An entrant's expected profit is
\[
 q(l_\chi\mu-b_{z,\chi}z-b_{m,\chi}m)-\eta q^2/2.
\]
Thus, with
\[
 s_\chi=l_\chi-\frac{b_{z,\chi}d_\chi}{A_\chi},\qquad
 c_\chi=b_{m,\chi}-\frac{b_{z,\chi}b_\chi}{A_\chi},
\]
the proof of Theorem~\ref{thm:eq} applies whenever $A_\chi>0$, $s_\chi>0$, and $c_\chi\geq0$, replacing $(s,c,A,d,b)$ by the appropriate displayed coefficients. These inequalities are explicit primitive checks, not estimated stability constants. The numerical routine rejects a parameter set failing them, and all reported facility cases pass.

The optimized expression can be written directly as
\[
 C_\chi^*=l_\chi(\mu^2+v)
 -\frac{\chi(\lambda-H)}{B_\chi}\mu m
 -\frac{(\lambda-H)^2}{2B_\chi}m^2
 -\frac{(d_\chi\mu-b_\chi m)^2}{2A_\chi}.
\]
Its no-entry value sets $m=0$, giving the decomposition in Proposition~\ref{prop:facility}. To verify the incidence identity without assuming its sign, write $r=\lambda-H$ and $d=2\lambda-H$. Since $d_\chi=\chi d/B_\chi$,
\[
 e_\chi=\frac{\chi}{B_\chi A_\chi}(d b_\chi-rA_\chi).
\]
Substitution of the displayed coefficients cancels the $\chi$ terms and gives
\[
 d b_\chi-rA_\chi=HK+rH-r\eta_0
 =H(K+r)-r\eta_0=Hb-(\lambda-H)\eta_0=Ae.
\]
This proves the identity and its sign under $\chi,A_\chi>0$.

For completeness, $A_\chi\to A$, $d_\chi\to d$, $b_\chi\to b$, $l_\chi\to\lambda$, and both minor loadings tend to $b$ as $\chi\to\infty$. The unique roots of the continuous scalar entry equations converge by compactness and uniqueness, and the corresponding quantities and cash costs then converge. This establishes the baseline limit without claiming a capacity-constrained auction theorem.

\section{Additional information comparisons}\label{app:robustexhibits}
\subsection{Cost distributions and diffuse forecasts}
Moment matching and small CDF perturbations answer different robustness questions. Table~\ref{tab:matched} fixes mean cost at $0.03$ and variance at $0.02^2/12$, replacing the original uniform law by gamma and lognormal alternatives. Table~\ref{tab:smoothing} instead uses $U\exp(\sigma Z)$, with $U\sim\operatorname{Uniform}[0.02,0.04]$ and independent standard normal $Z$. This smooths both endpoints, but its moments change with $\sigma$. Each comparison retains the original target law and $\epsilon=0.01$. Neither exercise implies a distribution-independent sign.

\begin{table}[htbp]\centering\small
\caption{Moment-matched full-support costs under unchanged priors. Positive entries indicate harmful information.}\label{tab:matched}
\begin{tabular}{lrr}
\toprule
Cost law & Onset prior & Saturation prior \\
\midrule
uniform & +0.000304852 & +0.000443379 \\
gamma & +0.000078630 & -0.000060538 \\
lognormal & +0.000098836 & -0.000061918 \\
\bottomrule
\end{tabular}

\end{table}
\begin{table}[htbp]\centering\small
\caption{Multiplicative smoothing under unchanged priors. The cost law is $U\exp(\sigma Z)$, and entries are refined-minus-coarse cost.}\label{tab:smoothing}
\begin{tabular}{rrr}
\toprule
$\sigma$ & Entry onset & Saturation \\
\midrule
0.000000 & 0.000305 & 0.000443 \\
0.005000 & 0.000289 & 0.000230 \\
0.010000 & 0.000274 & 0.000102 \\
0.025000 & 0.000233 & -0.000005 \\
0.050000 & 0.000181 & -0.000042 \\
0.100000 & 0.000099 & -0.000061 \\
0.200000 & 0.000035 & -0.000070 \\
\bottomrule
\end{tabular}

\end{table}

\begin{figure}[htbp]\centering
\includegraphics[width=\textwidth]{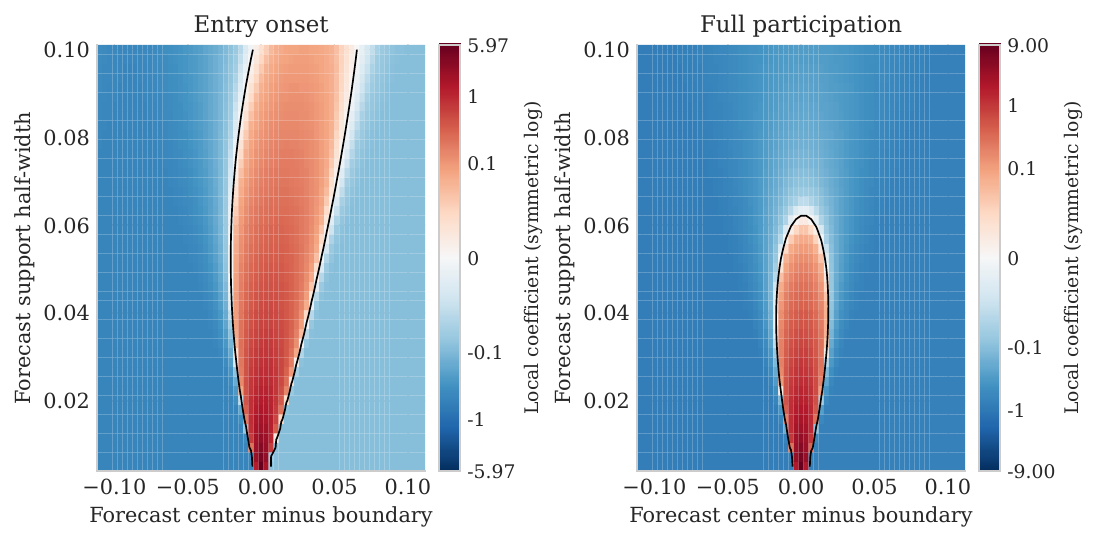}
\caption{Diffuse coarse forecasts. Red denotes a positive local coefficient and blue a negative one, with a black zero contour. Each colorbar uses symmetric-log normalization, linear on $[-0.05,0.05]$ and logarithmic outside, spanning the full data range without clipping. The panels have separate ranges. The vertical axis is the support half-width.}\label{fig:diffuse}
\end{figure}

Figure~\ref{fig:diffuse} uses $p_w(x)=15[1-((x-\xi)/w)^2]^2/(16w)$ on $|x-\xi|\leq w$, with coarse information revealing $X$ and $D=X+\delta B$ as in Theorem~\ref{thm:diffuse}. Each grid point changes the coarse-forecast distribution and defines its own target law, held fixed within its refinement. The displayed coefficient is evaluated by quadrature, splitting at all participation boundaries. Table~\ref{tab:diffusecenter} gives untransformed values at the boundary center, where the inverse-width jump contribution is most visible.

\begin{table}[htbp]\centering\small
\caption{The local coefficient $\lim_{\epsilon\downarrow0}\Delta\mathcal C/\epsilon^2$ at a coarse-forecast center equal to the participation boundary.}\label{tab:diffusecenter}
\begin{tabular}{rrr}
\toprule
Half-width & Onset & Full participation \\
\midrule
0.00500 & 5.96915 & 9.00277 \\
0.01000 & 2.83991 & 4.10743 \\
0.02000 & 1.27526 & 1.65978 \\
0.05000 & 0.33637 & 0.19127 \\
0.10000 & 0.02326 & -0.29810 \\
\bottomrule
\end{tabular}

\end{table}

\subsection{Smooth interior with finite strategic size}
The strict smooth-interior comparison also survives sufficiently large finite populations under an unchanged prior. Theorem~\ref{thm:exactfinite} gives convergence at each of the three posterior means in a fixed binary refinement. The finite cost gap therefore converges to its positive continuum value. Table~\ref{tab:finiteinterior} uses the lognormal construction of Proposition~\ref{prop:smoothinterior}, center $\mu_*=0.8250600786$, target half-spread $0.02$, and displacement $0.001$. Selected positive finite values do not establish positivity at every larger integer. Here $N$ counts normalized potential allocations, not observed trading firms.
\begin{table}[htbp]\centering\small
\caption{Smooth-interior information under one unchanged prior.}\label{tab:finiteinterior}
\begin{tabular}{lrr}
\toprule
Potential allocations & Participation & Cost change \\
\midrule
2 & 0.00005919 & $-0.00000036$ \\
3 & 0.00668475 & $+0.00000600$ \\
5 & 0.06735403 & $+0.00001214$ \\
10 & 0.15805744 & $+0.00000569$ \\
100 & 0.24132309 & $+0.00000242$ \\
$\infty$ & 0.25000000 & $+0.00000220$ \\
\bottomrule
\end{tabular}

\end{table}

\subsection{Joint finite-size scale}
Table~\ref{tab:jointscale} holds the prior at the continuum onset boundary and sets $\epsilon_N=1/N$. The coefficient $\gamma_0(1-\xi_0)$ is positive, but the exact cost sign at $N=100$ is negative. This illustrates the difference between the joint asymptotic scale and a finite-$N$ decision rule.
\begin{table}[htbp]\centering\small
\caption{Convergence on the joint scale, under the principal onset technology.}\label{tab:jointscale}
\begin{tabular}{rrr}
\toprule
$N$ & $N\Delta_N(1/N)$ & Limit \\
\midrule
100 & -0.001503 & 0.001564 \\
1000 & 0.001257 & 0.001564 \\
10000 & 0.001533 & 0.001564 \\
100000 & 0.001561 & 0.001564 \\
\bottomrule
\end{tabular}

\end{table}

\section{Numerical implementation and sensitivity}\label{app:numerical}
\begingroup\setstretch{1.05}
The uniform schedule has $K=\kappa(\rho-1+e^{-\rho})/\rho^2$ and $H=\kappa(1-e^{-\rho})/\rho$, with continuous limits at zero. The scalar participation equation is solved by a bracketed root method. Independent verification reconstructs the cash accounts and checks institutional and minor deviations alongside the scalar residual. The smoothing CDF is
\[
 F_\sigma(f)=\frac1{f_1-f_0}\int_{f_0}^{f_1}
 \Phi\!\left(\frac{\log(f/u)}\sigma\right)\dd u,\qquad f>0,
\]
evaluated with 256-node Gauss--Legendre quadrature and checked at 512 nodes. Diffuse-forecast integration splits at every participation boundary before adding the derivative-jump terms. Finite symmetric differences provide an independent check.

\begin{table}[htbp]\centering\footnotesize
\caption{Schedule comparisons with rate-cost coefficients $0.05$ multiplied by $\int\phi^2$ and displacement $0.01$. Priors are fixed at the original onset or recentered as labeled. Positive $e$ alone does not fix the sign at a finite displacement.}\label{tab:schedule}
\begin{tabular}{rlrrrrrr}
\toprule
$\rho$ & Schedule & $\int\phi^2$ & $K$ & $H$ & $e$ & Fixed prior & Recentered \\
\midrule
0.100000 & early & 1.333333 & 0.486940 & 0.935768 & 0.437538 & -0.000048 & 0.000318 \\
0.100000 & uniform & 1.000000 & 0.483742 & 0.951626 & 0.452291 & 0.000305 & 0.000305 \\
0.100000 & late & 1.333333 & 0.486940 & 0.967484 & 0.452611 & -0.000048 & 0.000352 \\
2.000000 & early & 1.333333 & 0.315165 & 0.296997 & 0.121505 & -0.000069 & -0.000012 \\
2.000000 & uniform & 1.000000 & 0.283834 & 0.432332 & 0.193809 & -0.000070 & 0.000026 \\
2.000000 & late & 1.333333 & 0.315165 & 0.567668 & 0.253261 & -0.000066 & 0.000093 \\
8.000000 & early & 1.333333 & 0.136390 & 0.031156 & -0.013243 & -0.000085 & -0.000091 \\
8.000000 & uniform & 1.000000 & 0.109380 & 0.124958 & 0.039260 & -0.000087 & -0.000070 \\
8.000000 & late & 1.333333 & 0.136390 & 0.218760 & 0.078162 & -0.000083 & -0.000043 \\
\bottomrule
\end{tabular}

\end{table}

\endgroup

\FloatBarrier
\section{Distributional robustness and dispersed forecasts}\label{sec:robustness}
\subsection{Continuous cost laws and robustness of a fixed experiment}
\begin{proposition}[Stability in the entry-cost distribution]\label{prop:stability}
Let $F,G$ be continuous cost distributions satisfying Assumption~\ref{ass:primitive}, with the execution coefficients unchanged, and let $\Delta_F,\Delta_G$ be the cost changes under the same pair of public experiments. With $d_{FG}=\|F-G\|_\infty$,
\begin{align}
 |a_F-a_G|&\leq d_{FG},&
 |m_F-m_G|&\leq\frac{s|\mu|}{\eta}d_{FG},\label{eq:cdfbound}\\
 |\psi_F(\mu)-\psi_G(\mu)|&\leq C_*\mu^2d_{FG},&
 C_*&=\frac{s}{\eta}\left(|e|+\frac{2ks}{\eta+c}\right),\label{eq:psibound}\\
 |\Delta_F-\Delta_G|&\leq2C_*\E[D^2]d_{FG}.&&\label{eq:gapbound}
\end{align}
Thus a fixed strictly harmful experiment remains harmful whenever the last bound is smaller than its original cost gap.
\end{proposition}
\begin{proof}
The root map $h_F(a)=a-F(\eta s^2\mu^2/[2(\eta+ca)^2])$ increases by at least the change in $a$. At the root for $G$, its absolute value is at most $d_{FG}$, proving the first bound. The function $as\mu/(\eta+ca)$ has derivative bounded in absolute value by $s|\mu|/\eta$, and $|m|\leq s|\mu|/(\eta+c)$. Factoring the difference of the inventory squares in \eqref{eq:psi} proves \eqref{eq:psibound}. Applying that bound to both experiments and using $\E[\E[D\mid\mathcal H]^2]\leq\E[D^2]$ proves \eqref{eq:gapbound}.
\end{proof}

A strictly positive finite information gap also survives sufficiently small changes in execution parameters when the experiment and its prior remain fixed. Recentering the prior at each new threshold gives a family of different experiments. The numerical results identify those cases explicitly. In either comparison, the permissible perturbation need not remain uniform as the information displacement tends to zero.

\subsection{How quickly entry begins}
A positive endpoint density is sufficient for the onset mechanism but is not necessary. Suppose
\begin{equation}
 F(f_0+x)=C x^\alpha(1+o(1)),\qquad C,\alpha,f_0>0.
 \label{eq:powercost}
\end{equation}
\begin{proposition}[Onset with a power-law cost tail]\label{prop:power}
Let $q_0=\sqrt{2f_0/\eta}$ and $\mu_0=\eta q_0/s$. As $h\downarrow0$,
\begin{equation}
 m(\mu_0+h)\sim
 \begin{cases}
 (s/c)h,&0<\alpha<1,\\
 \dfrac{2sf_0C}{\eta+2cf_0C}h,&\alpha=1,\\
 M_\alpha h^\alpha,&\alpha>1,
 \end{cases}
 \qquad M_\alpha=q_0C(sq_0)^\alpha.
 \label{eq:powerresponse}
\end{equation}
For $e>0$, a threshold-centered binary refinement is harmful for all sufficiently small displacements when $0<\alpha<2$, and beneficial when $\alpha>2$. At $\alpha=2$ its second-order coefficient is $(e\mu_0M_2-d^2/A)/2$, whose zero case requires higher-order information.
\end{proposition}
The proof in Appendix~\ref{app:power} compares the gain above the cost floor with the inventory feedback in the equilibrium equation. A sufficiently flat density can remove the local adverse effect, which explains why smoothing need not preserve it at every information scale.

\subsection{A diffuse distribution of coarse forecasts}
To study information across market states, take a random variable $X$ with continuously differentiable compactly supported density $p$, an independent symmetric sign $B$, and the fixed target $D=X+\delta B$. The coarse signal reveals $X$. A further noisy observation of $B$ with accuracy $(1+\epsilon/\delta)/2$ gives refined posterior means $X\pm\epsilon$. This construction keeps the target law unchanged and implements a feasible refinement for every $0<\epsilon<\delta$.

\begin{theorem}[Diffuse-forecast information effect]\label{thm:diffuse}
Suppose $\psi$ is continuous and piecewise $C^2$ on a neighborhood of the support of $p$, with finitely many derivative jumps $J_\tau$ and bounded piecewise second derivatives there. Then
\begin{equation}
 \lim_{\epsilon\downarrow0}\frac{\Delta\mathcal C}{\epsilon^2}
 =\frac12\left\{\int p(x)\psi''_{\rm reg}(x)\dd x
                         +\sum_\tau p(\tau)J_\tau\right\}.
 \label{eq:diffuse}
\end{equation}
In particular, let $p_w(x)=w^{-1}p_0((x-\tau)/w)$, where $p_0$ is a continuously differentiable density supported on $[-1,1]$ and $p_0(0)>0$. If $J_\tau>0$, the support includes no other boundary, and $|\psi''_{\rm reg}|\leq M$ there, then the limit is positive whenever $J_\tau p_0(0)/w>M$. If the support lies in a smooth region where $\psi''\leq-r<0$, the limit is at most $-r/2$.
\end{theorem}
\begin{proof}
The distributional second derivative is the locally finite signed measure $\nu=\psi''_{\rm reg}(x)\dd x+\sum_\tau J_\tau\delta_\tau$. Twice integrating the derivative gives
\[
 \psi(x+\epsilon)+\psi(x-\epsilon)-2\psi(x)
 =\int(\epsilon-|x-y|)_+\nu(\dd y).
\]
After multiplication by $p(x)/(2\epsilon^2)$ and integration, Fubini applies on the fixed compact neighborhood because $\nu$ has finite total variation there. The triangular kernels $\epsilon^{-2}(\epsilon-|\cdot|)_+$ integrate to one, and their convolution with the continuous $p$ converges uniformly to $p$. This proves \eqref{eq:diffuse}. The integral of the regular part lies between $-M$ and $M$, while the sole atomic contribution is $J_\tau p_0(0)/w$, giving the sufficient condition. The final assertion follows by integrating $\psi''\leq-r$ against a probability density.
\end{proof}

A diffuse coarse forecast places only order-$\epsilon$ probability within order-$\epsilon$ of a boundary, so its aggregate boundary effect is second order in displacement. An atom at a harmful boundary instead restores a first-order contribution proportional to its probability mass. The order thus depends on the distribution of coarse beliefs. The threshold-centered expansion cannot be read as a universal elasticity with respect to information precision.

\section{Smooth helpful supply and separated cost cohorts}\label{app:smoothcohorts}
\subsection{Proof of Proposition~\ref{prop:smoothcohorts}}
Fix $w\in(0,1)$ and an admissible execution environment with $e\leq0$. Choose $0<f_0<f_1<f_2<f_3$ and the continuous cost distribution
\[
 F_0(g)=w\,U_{[f_0,f_1]}(g)+(1-w)\,U_{[f_2,f_3]}(g),
\]
where $U_{[u,v]}$ is the uniform CDF on $[u,v]$. Define
\[
 q_w=\sqrt{2f_1/\eta},\qquad
 \tau=\frac{\eta+cw}{s}q_w,\qquad
 t_w=\frac{sw}{\eta+cw},
\]
so that $a(\tau)=w$ and $m(\tau)=t_w\tau$. Immediately above the cutoff $f_1$, participation is constant at $w$, because the second cohort starts strictly above $f_1$. Immediately below it, the cost density is $w/(f_1-f_0)$. Consequently,
\[
 B_w=w+\frac{2f_1w}{f_1-f_0},\qquad
 m'_-(\tau)=L_w:=\frac{sB_w}{\eta+cB_w},\qquad
 m'_+(\tau)=t_w.
\]
The function $x\mapsto sx/(\eta+cx)$ is strictly increasing, so $L_w>t_w>0$. Formula~\eqref{eq:jump} gives
\begin{equation}
 J_\tau=\tau(e-2kt_w)(t_w-L_w)>0.\label{eq:cohortjump}
\end{equation}
Both factors in parentheses are strictly negative. In particular, the inequality remains strict at $e=0$, since $k,t_w>0$.

Take a positive interval $J$ about $\tau$ so small that $|a_0(\mu)-w|<h/2$ throughout $J$. The baseline participation map is continuous. Its one-sided smoothness at $\tau$ and \eqref{eq:cohortjump} imply
\[
 \frac{\psi_0(\tau+\epsilon)+\psi_0(\tau-\epsilon)}2-\psi_0(\tau)
 =\frac{J_\tau}2\epsilon+O(\epsilon^2)>0
\]
for every sufficiently small $\epsilon>0$. Fix one such $\epsilon$ with $[\tau-\epsilon,\tau+\epsilon]\subset J$. This is a feasible comparison under the single binary prior $D=\tau\pm\delta$, for any fixed $\delta>\epsilon$: the coarse signal is uninformative and the refined binary signal has accuracy $(1+\epsilon/\delta)/2$.

Let $X$ have CDF $F_0$ and let $Z\sim N(0,1)$ be independent. For $\sigma>0$, replace costs by $X_\sigma=X\exp(\sigma Z)$ and denote their CDF by $F_\sigma$. The log-cost law is the convolution of the law of $\log X$ with a nondegenerate Gaussian. It therefore has a positive smooth density everywhere on $\R$, and $F_\sigma$ has a positive smooth density on $(0,\infty)$. Moreover,
\[
 F_\sigma(g)=\E[F_0(g\exp(-\sigma Z))],\qquad
 \|F_\sigma-F_0\|_\infty\longrightarrow0.
\]
For the second statement, the continuous CDF of $\log X$ is uniformly continuous on $\R$. Split the expectation into $|Z|\leq M$ and its complement, apply its modulus of continuity on the first event, and then let $\sigma\downarrow0$ followed by $M\uparrow\infty$. The argument starts from a continuous two-cohort law and does not require uniform approximation of an atomic CDF.

Proposition~\ref{prop:stability} now preserves the fixed positive information gap for all sufficiently small $\sigma$. It also gives $|a_\sigma-a_0|\leq\|F_\sigma-F_0\|_\infty<h/2$ on $J$. The equilibrium inverse map has derivative $(\eta+cB_F)/s>0$, so $\psi_\sigma$ is smooth on $J$. A strictly positive symmetric second difference forces $\psi_\sigma''$ to be positive at some point of $(\tau-\epsilon,\tau+\epsilon)$. Continuity gives an open neighborhood $I\subset J$ on which $\psi_\sigma''>0$ and $|a_\sigma-w|<h$. Full revelation of any nondegenerate binary target supported in this smaller interval is harmful by strict convexity. Finally, for every $\mu\ne0$ the positive density implies $a_\sigma(\mu)>0$, so
\[
 e\mu m_\sigma(\mu)-k m_\sigma(\mu)^2<0.
\]
This is the strict improvement due to inventory relative to no suppliers at the same signal, completing the proof.

\subsection{Cash verification and dispersion sensitivity}
For the mixture in Section~\ref{sec:smoothcohorts}, write
\[
 \sigma_f=\sqrt{\log(1+0.03^2)},\qquad
 \ell_j=\log(\bar f_j)-\sigma_f^2/2,\qquad
 (\bar f_1,\bar f_2)=(0.03,0.30).
\]
The CDF and density derivative used in the curvature calculation are
\begin{align*}
 F(g)&=\tfrac12\sum_{j=1}^2\Phi(x_j),&
 f_F(g)&=\tfrac12\sum_{j=1}^2\frac{\varphi(x_j)}{\sigma_f g},\\
 x_j&=\frac{\log g-\ell_j}{\sigma_f},&
 f'_F(g)&=-\tfrac12\sum_{j=1}^2
 \frac{\varphi(x_j)}{\sigma_f g^2}\left(1+\frac{x_j}{\sigma_f}\right).
\end{align*}
Substitution in \eqref{eq:primitivecurvature} and \eqref{eq:curvature} yields the main-text values. Double-precision calculations solve the participation equation by a bracketed root method with absolute tolerance $5\times10^{-15}$. An independent 70-digit calculation instead bisects the quantity equation
\[
 q\{\eta+cF(\eta q^2/2)\}=s\mu
\]
and evaluates terminal payments separately in each target state. For a signal branch with posterior mean $\mu$, the posterior probability of $D=1.033$ is $[1+(\mu-1.013)/0.02]/2$. Applying these probabilities directly to
\[
 \frac{\eta_0z^2}{2}+(D-z)(D-z-m)
\]
gives the coarse cost $0.2973387470735351$ and information gap $0.0000758931219877127$. The original-payment and reduced-form gaps differ by less than $10^{-60}$ in this check. The extra digits document numerical agreement, not estimated economic precision.

Table~\ref{tab:cohortdispersion} varies the two coefficients of variation together. The target prior, signal, execution coefficients, component means, and equal cohort weights remain unchanged. The adverse sign occurs at coefficients $0.02$ and $0.03$ on this grid and reverses at both narrower and wider costs. The replication output also includes every combination of lower-cohort weights $0.25,0.50,0.75$ and coefficients of variation $0.01,0.02,0.03,0.05,0.10,0.20$. The twelve cases with unequal weights are all beneficial at this same target location. No prior is moved to a new threshold. The comparison shows how the harmful region's relevance depends on the fixed target scale.

\begin{table}[htbp]\centering\small
\caption{Dispersion sensitivity under the unchanged prior $D=1.013\pm0.02$ and signal accuracy $0.75$. The two lognormal cohorts have equal weights and fixed means $0.03$ and $0.30$. Costs use $\lambda=\eta_0=\eta=1$, $K=H=0$.}\label{tab:cohortdispersion}
\begin{tabular}{rrrrr}
\toprule
Cohort CV & Entered mass & $\psi^{\prime\prime}(1.013)$ & $\Delta\mathcal C$ & Gap / coarse cost (\%) \\
\midrule
0.01 & 0.5000 & -1.4219 & -0.00007107 & -0.02393 \\
0.02 & 0.4998 & 0.0946 & +0.00002323 & +0.00782 \\
0.03 & 0.4957 & 1.7243 & +0.00007589 & +0.02552 \\
0.05 & 0.4802 & -0.2508 & -0.00001247 & -0.00418 \\
0.10 & 0.4474 & -1.1195 & -0.00005595 & -0.01859 \\
0.20 & 0.4089 & -1.3681 & -0.00006840 & -0.02251 \\
\bottomrule
\end{tabular}

\end{table}

\section{Finite populations: equilibrium selection and the critical scale}\label{app:finiteselection}

\subsection{Distributional bounds for uniqueness}
Proposition~\ref{prop:finiteall} applies to ordinary Bayesian Nash equilibria, including asymmetric strategies and independent private randomization. It does not assert uniqueness among correlated equilibria.

For $F=\mathrm{Uniform}[f_0,f_1]$,
\[
 \mathcal B_F=\frac{3f_1-f_0}{f_1-f_0}.
\]
Under the principal parameters $(\lambda,\kappa,\rho,\eta_0,\eta)=(1,1,0.1,0.05,0.05)$ and uniform costs on $[0.02,0.04]$, $\mathcal B_F=5$ and $b=0.532115983956$. Thus \eqref{eq:finiteallcondition} holds for every $N\geq32$. In particular, the opposite information rankings at $N=100$ and $N=200$ do not depend on equilibrium selection. For a lognormal cost law with log standard deviation $\sigma_f$,
\[
 \mathcal B_F=\Phi(\sigma_f/2)+\frac{2\varphi(\sigma_f/2)}{\sigma_f}.
\]
Indeed, with $x=(\log g-\mu_f)/\sigma_f$, the derivative of $\Phi(x)+2\varphi(x)/\sigma_f$ changes sign at $x=\sigma_f/2$. In the lognormal economy of Proposition~\ref{prop:smoothinterior}, $\mathcal B_F=29.5297328372$, and the uniqueness condition holds for every $N\geq4$. It covers the positive information gaps at $N=5,10,100$. Failure of this sufficient condition does not establish multiplicity.

\subsection{A strict asymmetric equilibrium at small population size}

The symmetry qualification matters outside the sufficient region. In the principal economy, take $N=2$ and a deterministic target $\mu=0.374843244353$. There is an equilibrium in which supplier 1 never enters and supplier 2 always enters. The entering quantities and gross payoffs are shown in Table~\ref{tab:asymmetricfinite}.
\begin{table}[htbp]\centering\small
\caption{A strict asymmetric equilibrium with two potential allocations.}\label{tab:asymmetricfinite}
\begin{tabular}{lrr}
\toprule
 & Supplier 1 & Supplier 2\\
\midrule
Entry probability & $0$ & $1$\\
Hypothetical entering quantity & $0.223323192599$ & $0.411316920627$\\
Gross entry payoff & $0.014516007519$ & $0.049241659452$\\
\bottomrule
\end{tabular}
\end{table}
The first gross payoff is below the cost support $[0.02,0.04]$, and the second exceeds it. Thus both participation decisions are strict for every cost type. These values follow from
\[
 q_2=\frac{s\mu}{\eta_2-b^2/(2A)},\qquad
 q_1=q_2\left(1-\frac{b}{2\eta_2}\right),\qquad
 \bar m=\frac{q_2}{2},\qquad z=\frac{d\mu-b\bar m}{A}.
\]
They give $z=0.254472886490$ and $\bar m=0.205658460314$. The symmetric equilibrium at the same state instead has $a=0.475645283829$ and $\bar m=0.151460510208$. Institutional expenditure is $0.100701454200$ in the asymmetric equilibrium and $0.093972089412$ in the symmetric equilibrium. Swapping the two suppliers gives another asymmetric equilibrium. These equilibria share one set of model primitives. They are not estimates of supplier behavior.

\subsection{The second layer at the finite entry boundary}

Assume the positive-density endpoint expansion
\[
 F(f_0+x)=\ell_0x+O(x^2),\qquad f_0>0,\quad\ell_0>0,
\]
and $e>0$. Let $\mu_0=\sqrt{2\eta f_0}/s$ and put
\[
 \xi_0=\frac{b\mu_0}{\eta},\qquad
 B_0=\frac{b^2\mu_0}{2\eta^2},\qquad
 L_0=\frac{2f_0\ell_0s}{\eta+2f_0\ell_0c}.
\]
Use the symmetric-equilibrium gap $\Delta_N$ from Corollary~\ref{cor:nonuniform}. The prior is held fixed within each comparison. For example, a binary prior with center $\mu_0$ and a fixed positive half-spread implements all sufficiently small $\epsilon$ through noisy binary signals.

\begin{corollary}[Second critical layer]\label{cor:secondlayer}
For every fixed $y\in\R$,
\begin{equation}
 \lim_{N\to\infty}N^2\Delta_N\!\left(\frac{\xi_0}{N}+\frac y{N^2}\right)
 =-\frac{d^2\xi_0^2}{2A}+\frac{e\mu_0L_0}{2}(y+B_0)_+.
 \label{eq:secondlayer}
\end{equation}
In particular,
\[
 \lim_{N\to\infty}N^2\Delta_N(\xi_0/N)
 =\frac{b^2\mu_0^2}{4\eta^2}
   \left(eL_0-\frac{2d^2}{A}\right)<0.
\]
\end{corollary}

\begin{proof}
The finite entry threshold satisfies
\[
 \mu_{0,N}=\mu_0\sqrt{1+\frac{2b}{\eta N}}
 =\mu_0+\frac{\xi_0}{N}-\frac{B_0}{N^2}+O(N^{-3}).
\]
For $\epsilon_N=\xi_0/N+y/N^2$, the center and the lower posterior mean remain below this threshold. The upper posterior mean exceeds it by
\[
 (\mu_0+\epsilon_N-\mu_{0,N})_+
 =\frac{(y+B_0)_+}{N^2}+o(N^{-2}).
\]
Expanding the participation equation at its lower endpoint gives, uniformly for large $N$,
\[
 \bar m_N(\mu_{0,N}+\delta)
 = L_N\delta+O(\delta^2),\qquad
 L_N=\frac{2f_0\ell_0s}{\eta_N+2f_0\ell_0c_N},\qquad \delta\downarrow0.
\]
Here $L_N\to L_0$, and mean inventory is zero below the threshold. Consequently,
\[
 \bar m_N(\mu_0+\epsilon_N)
 =\frac{L_0(y+B_0)_+}{N^2}+o(N^{-2}).
\]
In the information difference, the direct quadratic term is $-d^2\epsilon_N^2/(2A)$. The only inventory term is
\[
 \frac12\left\{e(\mu_0+\epsilon_N)\bar m_N(\mu_0+\epsilon_N)
 -k\bar m_N(\mu_0+\epsilon_N)^2\right\}.
\]
Multiplying by $N^2$ proves \eqref{eq:secondlayer}. To obtain the strict sign at $y=0$, note that $L_0<s/c$, $e\leq db/A$, and $bs/c\leq d$. These inequalities follow from $\eta>0$, $H\leq\lambda$, and the definitions of $s,c,d,e$. Since $e>0$, they imply $eL_0<d^2/A$.
\end{proof}

The limiting sign changes at
\[
 y_* =\frac{d^2\xi_0^2}{Ae\mu_0L_0}-B_0>0.
\]
The local expansion is uniform for bounded $y$. Continuity therefore gives a zero at a scale $\epsilon_N=\xi_0/N+y_N/N^2$ with $y_N\to y_*$. This is an existence statement. The endpoint expansion alone does not imply that the nearby zero is unique.

\section{Heterogeneous marginal inventory costs}\label{app:heterogeneouscosts}

Identical marginal costs are not needed for the scalar equilibrium reduction. Let supplier type $(\eta_i,f_i)$ have joint law $G$, independently of the target and the entire family of compared public experiments, with $\eta_i\geq\underline\eta>0$ and $f_i\geq0$. Correlation between the two cost components is allowed. Assume that the activation index $2\eta_i f_i$ has no atoms. It suffices that the conditional law of $f_i$ given $\eta_i$ is atomless. The institution's technology is unchanged, so the coefficients $A,b,d,s,c,e,k$ retain their definitions.

\begin{proposition}[Heterogeneous-cost equilibrium]\label{prop:heterogeneouscosts}
For every posterior mean $\mu$, define
\[
 S(w)=\int\frac{w}{\eta_i}\,
          \1_{\{2\eta_i f_i\leq w^2\}}\,G(\mathrm d\eta_i,\mathrm df_i).
\]
There is a unique solution of $w+cS(w)=s\mu$. The equilibrium is
\[
 m=S(w),\qquad z=\frac{d\mu-bm}{A},\qquad
 q_i=\frac{w}{\eta_i}\ \text{when active},
\]
and supplier $i$ enters exactly when $f_i\leq w^2/(2\eta_i)$. The optimized institutional mean-dependent cost remains
\[
 \psi_G(\mu)=-\frac{d^2\mu^2}{2A}+e\mu m-km^2.
\]
\end{proposition}

\begin{proof}
For fixed $(z,m)$, a supplier's gross profit is $(\lambda\mu-bz-bm)q_i-\eta_iq_i^2/2$. Its unique maximizer and participation threshold have the stated forms, with $w=\lambda\mu-bz-bm$. The institution's first-order condition yields $w=s\mu-cm$. For each type, the function $w\mapsto(w/\eta_i)\1_{\{2\eta_i f_i\leq w^2\}}$ is odd and nondecreasing. Integration preserves both properties. The atomless-index assumption and the bound $\eta_i\geq\underline\eta$ imply continuity of $S$ by dominated convergence on bounded intervals. Thus $w+cS(w)$ is continuous, strictly increasing, and tends to the two respective infinities as $w\to\pm\infty$. It has a unique inverse. Moreover, $|w|\leq s|\mu|$ and $|q_i|\leq s|\mu|/\underline\eta$, so the original finite-second-moment assumption ensures integrability. Substitution of the institution's response into its original cash account gives $\psi_G$.
\end{proof}

This extension separates activation costs from marginal inventory technology. It does not identify either distribution from prices. The conditional Jensen characterization of information value applies to $\psi_G$ without a differentiability assumption. Where inventory is twice differentiable, the curvature identity in Theorem~\ref{thm:geometry} also applies.

\section{Accounting along a continuous information path}\label{app:informationclock}
The curvature and boundary formulas admit a common representation when information arrives continuously. Let $\mathcal H_t$, $0\leq t\leq T$, be an increasing family of public information sets and let $M_t=\E[D\mid\mathcal H_t]$ have continuous paths. The index $t$ records information. The physical execution horizon, schedule, and cost law remain fixed. Each information set is evaluated in the same one-shot game, without allowing traders to revise entry decisions along a trading path.

\begin{proposition}[Posterior-martingale cost accounting]\label{prop:informationclock}
Let $D\in L^2$ and suppose the continuous cost function $\psi$ is piecewise $C^2$ on $\R$, with bounded ordinary second derivative, at most linear growth of its one-sided first derivatives, and finitely many derivative jumps
\[
 J_\tau=\psi'_+(\tau)-\psi'_-(\tau).
\]
For a continuous version of the posterior-mean martingale $M$, the change in expected institutional expenditure is
\begin{equation}
 \E\psi(M_T)-\E\psi(M_0)
 =\frac12\E\int_0^T\psi''_{\rm reg}(M_t)\dd\langle M\rangle_t
 +\frac12\sum_\tau J_\tau\E L_T^\tau(M),
 \label{eq:informationclock}
\end{equation}
where local time uses the occupation-density normalization
\[
 \int_0^T h(M_t)\dd\langle M\rangle_t
 =\int_\R h(x)L_T^x(M)\dd x.
\]
\end{proposition}
\begin{proof}
The distributional second derivative of $\psi$ is
\[
 \psi''_{\rm reg}(x)\dd x+\sum_\tau J_\tau\delta_\tau.
\]
The generalized It\^o--Tanaka formula therefore gives the pathwise difference as the two terms on the right of \eqref{eq:informationclock}, before taking expectations, plus $\int_0^T\psi'_-(M_t)\dd M_t$.

For completeness, localization removes any need for a fourth moment. Write $N_t=M_t-M_0$ and stop when $|N_t|+\langle M\rangle_t$ first exceeds $n$, or at $T$, whichever comes first. Up to that time, the linear derivative bound and $\E M_0^2<\infty$ make the stopped stochastic integral square integrable, hence of mean zero. The cost has at most quadratic growth. Doob's inequality gives $\E\sup_{t\leq T}|M_t|^2<\infty$, so its stopped terminal values converge in $L^1$. Boundedness of $\psi''_{\rm reg}$ and $\E\langle M\rangle_T<\infty$ give dominated convergence for the regular term. Tanaka's formula for $|M_t-\tau|$, whose stochastic integrand is bounded, gives $\E L_T^\tau(M)<\infty$. Monotone convergence of the stopped local times, followed by the finite signed sum over jumps, proves \eqref{eq:informationclock}. The target-second-moment term in \eqref{eq:representation} is unchanged.
\end{proof}

Regular curvature is weighted by the quadratic variation of posterior beliefs. A participation boundary contributes its slope jump multiplied by the expected local time spent there. Thus negative smooth curvature need not offset a positive boundary contribution. Conversely, nonpositive regular curvature and nonpositive jumps make every such information path favorable. Equation~\eqref{eq:informationclock} uses the same distributional curvature as Theorem~\ref{thm:diffuse}, but accounts for an entire feasible information path rather than one small symmetric refinement. If posterior means jump, the formula must also include the usual jump remainders.

\section{A signed outside completion facility}\label{sec:facility}
Allow the institution, after observing $D$, to submit any signed auction order $j\in\R$ and complete $D-z-j$ through an outside provider charging $\chi(D-z-j)^2/2$, where $\chi>0$. The provider's price anchor is the unaffected benchmark, independent of inherited auction displacement. Anticipators still liquidate in the auction at $H(z+m)+\lambda(j-m)$. Orders may offset across the two venues, even when the total residual is zero. Routing is unrestricted in sign, so a venue need not receive a same-direction share of the residual order. The unique terminal choice is
\begin{equation}
 j^*(D,z,m)=\frac{\chi(D-z)-H(z+m)+\lambda m}{\chi+2\lambda}.
 \label{eq:facility}
\end{equation}
Appendix~\ref{app:facility} derives coefficients $(A_\chi,d_\chi,b_\chi,s_\chi,c_\chi,l_\chi)$ for the resulting continuum game, whose equilibrium retains the scalar form whenever $A_\chi>0$, $s_\chi>0$, and $c_\chi\geq0$.

\begin{proposition}[Information incidence with outside completion]\label{prop:facility}
Let $B_\chi=\chi+2\lambda$ and suppose the stated facility coefficient restrictions hold. Optimized institutional cost relative to no anticipators is $e_\chi\mu m-k_\chi m^2$, where
\begin{equation}
 e_\chi=\frac{d_\chi b_\chi}{A_\chi}-\frac{\chi(\lambda-H)}{B_\chi}
       =\frac{\chi A e}{B_\chi A_\chi},\qquad
 k_\chi=\frac{b_\chi^2}{2A_\chi}+\frac{(\lambda-H)^2}{2B_\chi}>0.
 \label{eq:facilityincidence}
\end{equation}
Thus every admissible facility preserves the sign of initial inventory incidence. Fixed-roster information remains beneficial, and when $e>0$ the harmful-onset theorem holds at the facility's own entry threshold.
\end{proposition}
\begin{proof}
After optimizing $j$, the constant target term is $l_\chi\E[D^2]$. Completing the remaining square in $z$ gives \eqref{eq:facilityincidence}, with the algebra detailed in Appendix~\ref{app:facility}. At fixed mass, write $m=t\mu$ with $t=as_\chi/(\eta+c_\chi a)\geq0$. The coefficient of $\E[\mu^2]$ after extracting the constant target term is
\[
 -\frac{\chi(\lambda-H)}{B_\chi}t
 -\frac{(\lambda-H)^2}{2B_\chi}t^2
 -\frac{(d_\chi-b_\chi t)^2}{2A_\chi}<0.
\]
The inequality is strict for $t>0$ from its first term and for $t=0$ from $d_\chi>0$. The positive multiplier relating $e_\chi$ to $e$ then extends the onset proof.
\end{proof}

\FloatBarrier
\section{Joint markets and basket activation}\label{app:jointmodel}
Related benchmarks can share inventory suppliers. A trader who prepares Nasdaq-100 inventory may also trade the broad market or semiconductors. A common activation expense makes participation depend on the whole basket. Cross-impact gives a second source of interaction. Portfolio execution with cross-impact has been studied by \citet{MinEtAl2022}. Here the additional question is whether a shared entry decision remains tractable when each entrant chooses several quantities.

Under the idealized no-flow balance sheet, a benchmark family's aggregate exposure adjustment is $W_bR_b$, where $W_b=\sum_{i:b(i)=b}A_i\beta_i(\beta_i-1)$. Dividing by a fixed reference price expresses this quantity in the benchmark units used below. Long and inverse geared funds both contribute positive coefficients to $W_b$. Cross-family return dependence gives correlated target proxies. Shared inventory suppliers can also link their implementation. The game below describes one institution with a vector target. Independent fund managers need not coordinate their execution. Secondary-market volume does not identify the unobserved historical $A_i$.

\subsection{A vector quantity game}
Let $D\in\R^p$ be the institution's target in fixed benchmark units, with posterior mean $\mu$ and covariance $V$. Quantities $z,q,m$ are vectors. Entrants pay one private fixed cost to activate the basket, after which every coordinate can be traded. Let $\Lambda$ denote terminal inverse depth and let $K,H$ denote schedule-integrated impact and terminal carryover. Assume these matrices are symmetric, $K,H\succeq0$, $\Lambda-H\succ0$, and the temporary-charge matrices $E_0,E\succ0$. The common scalar decay specification gives $K=k\Gamma$ and $H=h\Gamma$ for a symmetric positive-semidefinite impact matrix $\Gamma$. No simultaneous diagonalization with $\Lambda,E_0,E$ is assumed.

The terminal price displacement is
\[
 p_A=H(z+m)+\Lambda(D-z-m).
\]
Pre-event payments are $z^\top K(z+m)+z^\top E_0z/2$ for the institution and $q^\top K(z+m)+q^\top Eq/2$ for an entrant. Define
\begin{align}
 B&=\Lambda+K-H, & A&=E_0+2B, & L&=2\Lambda-H,\nonumber\\
 C&=B-BA^{-1}B, & S&=\Lambda-BA^{-1}L.\label{eq:vectorcoeff}
\end{align}
The matrix $S$ need not be symmetric. Direct expansion of the institution's cash account gives
\begin{equation}
 \begin{split}
 C_0(z,m\mid Y)={}&\operatorname{tr}(\Lambda V)+\mu^\top\Lambda\mu
 -\mu^\top(\Lambda-H)m\\
 &+\tfrac12z^\top Az-z^\top(L\mu-Bm).
 \end{split}\label{eq:vectorcash}
\end{equation}
An entrant's gross expected profit is
\[
 \pi(q)=q^\top\{\Lambda\mu-B(z+m)\}-\tfrac12q^\top Eq.
\]
As in the scalar game, the institution and entrants choose simultaneously after observing the common signal. The institution does not first commit to a reaction function. Its first-order condition is substituted only when solving the simultaneous conditions.

\begin{proposition}[Unique basket-activation equilibrium]\label{prop:vector}
Suppose $\E\|D\|^2<\infty$ and the nonnegative activation cost has a continuous distribution function $F$. There is a unique common entrant quantity and participation mass, with
\begin{align}
 q(a)&=(E+aC)^{-1}S\mu, & g(a)&=\tfrac12q(a)^\top Eq(a),\nonumber\\
 a&=F(g(a)), & m&=a q(a), & z&=A^{-1}(L\mu-Bm).\label{eq:vectorfixed}
\end{align}
The scalar equation for $a$ has exactly one root in $[0,1]$. Quantities have at most linear growth in $\|\mu\|$, and the quadratic cash accounts are integrable.
\end{proposition}
\begin{proof}
The assumptions imply $B\succ0$ and $A\succ2B$. Inversion reverses the positive-definite order, giving $BA^{-1}B\prec B/2$ and therefore $C\succ B/2\succ0$. Strict convexity of \eqref{eq:vectorcash} gives the displayed institutional quantity. Strict concavity of entrant profit gives
\[
 Eq=\Lambda\mu-B(z+m)=S\mu-Cm.
\]
All entering traders consequently choose the same quantity, and $m=aq$ gives \eqref{eq:vectorfixed}. At the optimum, gross profit is $q^\top Eq/2$, so entry occurs exactly when the fixed cost does not exceed $g(a)$, up to a null set.

Put $W=E^{-1/2}CE^{-1/2}\succ0$ and $x=E^{-1/2}S\mu$. Then
\[
 g(a)=\tfrac12x^\top(I+aW)^{-2}x,\qquad
 g'(a)=-x^\top W(I+aW)^{-3}x\leq0.
\]
Thus $a-F(g(a))$ is continuous and strictly increasing, with a nonpositive value at zero and a nonnegative value at one. Existence and uniqueness follow. Moreover,
\[
 \|q(a)\|\leq\frac{\|S\|}{\lambda_{\min}(E)}\|\mu\|,
 \qquad \|m\|\leq\|q(a)\|.
\]
The formula for $z$ gives the remaining linear bound. Conditional expectation contracts $L^2$, establishing integrability. Expected fixed expenditure is bounded by $a g(a)$ and is therefore integrable as well.
\end{proof}

\subsection{Information and the source of jointness}
Substituting the institutional optimum yields
\begin{align}
 \E[C_0^*]&=\E[D^\top\Lambda D]+\E[\Psi(M)],\label{eq:vectorinfo}\\
 \Psi(\mu)&=-\mu^\top(\Lambda-H)m(\mu)
 -\tfrac12\{L\mu-Bm(\mu)\}^\top A^{-1}\{L\mu-Bm(\mu)\}.
 \nonumber
\end{align}
For nested public experiments, $M_0=\E[M_1\mid\mathcal H_0]$. Concavity of $\Psi$ on a convex region containing the posterior means is sufficient for every refinement within that region to benefit the institution. At a smooth interior state, a symmetric displacement $\pm\epsilon u$ changes cost by
\[
 \tfrac12\epsilon^2u^\top\nabla^2\Psi(\mu)u+o(\epsilon^2).
\]
Information can therefore have different local values in different portfolio directions. The scalar fixed-roster and centered-target rankings are examined below. They do not hold automatically for general execution matrices.

Even with diagonal impact and charge matrices, the assets remain linked through the single equation for $a$. Gross entry profit is the sum of the coordinate profits. Independent scalar games instead require separately paid, separately chosen activation decisions. Conversely, correlated target innovations alone do not create cross-impact. Holding $\mu$ and the execution matrices fixed, $V$ enters the risk-neutral institutional bill through $\operatorname{tr}(\Lambda V)$ and does not change equilibrium quantities. This distinction matters when interpreting joint return forecasts.

\subsection{Favorable portfolio information rankings}\label{app:vectorranking}
The scalar fixed-roster and centered-target rankings require additional restrictions in the portfolio model. For the matrices $T_a,P_a$ in \eqref{eq:vectorP}, fixed participation gives $m=T_a\mu$ and $\Psi_a(\mu)=\mu^\top P_a\mu$. The following results distinguish a ranking of all refinements at one roster from a ranking against no information under optional activation.

\begin{proposition}[Information with a fixed portfolio roster]\label{prop:vectorranking}
Fix $a$ and the execution matrices. For nested public experiments with posterior means $M_0,M_1$, the change in expected institutional cost is
\begin{equation}
 \Delta\mathcal C
 =\E[(M_1-M_0)^\top P_a(M_1-M_0)].\label{eq:vectorgap}
\end{equation}
Every refinement weakly lowers cost for every square-integrable target law if and only if $P_a\preceq0$.

If $\Lambda,K,H,E_0,E$ are simultaneously orthogonally diagonalizable, then $P_a\prec0$ for every $a\in[0,1]$. In this class, every refinement that changes the posterior mean with positive probability strictly lowers cost at a fixed roster. Under optional basket activation, every signal with a nonzero posterior mean with positive probability strictly improves on no information for a centered vector target.
\end{proposition}
\begin{proof}
Subtract the two instances of \eqref{eq:vectorinfo}. The target-second-moment term cancels. Since $\E[M_1-M_0\mid\mathcal H_0]=0$, the cross term in the quadratic expansion has zero expectation, proving \eqref{eq:vectorgap}. Negative semidefiniteness is sufficient. If $u^\top P_a u>0$, a target equal to $u$ and $-u$ with equal probabilities and its complete revelation give a strict increase, proving necessity.

For the sufficient class, work in the common orthonormal basis and use lower-case letters for the diagonal entries. Write
\[
 b_j=\lambda_j+k_j-h_j,\quad A_j=\eta_{0j}+2b_j,\quad
 l_j=2\lambda_j-h_j,\quad
 c_j=b_j-b_j^2/A_j,\quad
 s_j=\frac{\lambda_j\eta_{0j}+b_jh_j}{A_j}>0.
\]
All $b_j,c_j,l_j$ are positive, and $t_j(a)=as_j/(\eta_j+ac_j)\geq0$. The $j$th diagonal entry of $P_a$ is
\begin{equation}
 - (\lambda_j-h_j)t_j(a)
 -\frac{\{l_j-b_jt_j(a)\}^2}{2A_j}<0.\label{eq:commutingcost}
\end{equation}
At $a=0$, the squared term is strictly negative. At $a>0$, the first term is strictly negative. This proves the fixed-roster assertion. With optional activation, the same expression holds at the common endogenous mass $a(\mu)$, so $\Psi(\mu)<0$ for $\mu\ne0$, whereas $\Psi(0)=0$. Comparing with the zero posterior mean proves the centered-target assertion.
\end{proof}

\begin{proposition}[Centered targets with optional basket activation]\label{prop:vectorcentered}
Fix execution matrices satisfying the assumptions of Appendix~\ref{app:jointmodel} and one continuous proper cost CDF $F$ on $[0,\infty)$ with $F(0)=0$. The following statements are equivalent:
\begin{enumerate}
\item For every square-integrable vector target with $\E D=0$ and every public signal, equilibrium institutional expenditure is no greater than under no information.
\item $P_a\preceq0$ for every $a\in[0,1]$.
\end{enumerate}
\end{proposition}
\begin{proof}
With no information, the posterior mean, quantities, and gross entry profit are zero. Thus $\Psi(0)=0$. If the second statement holds, then $\Psi(\mu)=\mu^\top P_{a(\mu)}\mu\leq0$ pointwise. Equation~\eqref{eq:vectorinfo} proves the first statement.

Conversely, suppose $u^\top P_a u>0$ for some nonzero $u$ and an interior mass $a\in(0,1)$. Put $q_a=(E+aC)^{-1}Su$ and $g_a=q_a^\top Eq_a/2$. Necessarily $g_a>0$: if $g_a=0$, positive definiteness of $E$ gives $q_a=0$ and $T_a u=0$, whence
\[
 u^\top P_a u=-\tfrac12(Lu)^\top A^{-1}(Lu)<0,
\]
a contradiction. Here $L\succ0$ follows from $\Lambda-H\succ0$ and $H\succeq0$. Continuity and properness of $F$ give a finite $g_*>0$ with $F(g_*)=a$. Set $r=(g_*/g_a)^{1/2}$ and take $D=ru$ or $-ru$, each with probability one-half. Under full revelation, the quantity $\pm rq_a$ has gross profit $g_*$ and implements mass $a$ in both states. Uniqueness in Proposition~\ref{prop:vector} identifies this as the equilibrium. The change from no information is $r^2u^\top P_a u>0$.

Finally, $P_0=-L^\top A^{-1}L/2\prec0$, and a positive direction at $a=1$ persists at a nearby interior mass by continuity. Thus any failure of the second statement contradicts the first.
\end{proof}

The cost law stays fixed throughout the theorem and each comparison uses one fixed prior. The necessity argument selects a prior within the class of centered laws. A particular centered prior need not expose every adverse matrix direction. Nor does the result rank two already informative signals, even when the matrices commute.

\begin{proposition}[Gaussian information in isotropic portfolios]\label{prop:vectorGaussian}
In the isotropic specialization of Section~\ref{sec:portfolioinformation}, let $D$ be a centered Gaussian vector and let $\mathcal H_0\subseteq\mathcal H_1$ be jointly Gaussian public experiments. Write $M_j=\E[D\mid\mathcal H_j]$ and $\Sigma_j=\operatorname{Cov}(M_j)$. Optional-entry institutional expenditure weakly decreases from $\mathcal H_0$ to $\mathcal H_1$, and decreases strictly if $\Sigma_1\ne\Sigma_0$.
\end{proposition}
\begin{proof}
The posterior means are centered Gaussian, and the tower property gives
\[
 \Sigma_1-\Sigma_0=\E[(M_1-M_0)(M_1-M_0)^\top]\succeq0.
\]
Order the eigenvalues of each covariance matrix increasingly. The variational characterization of eigenvalues implies $\lambda_i(\Sigma_1)\geq\lambda_i(\Sigma_0)$ for every $i$. For independent standard normals $Z_1,\ldots,Z_p$, define
\[
 R_j=\left\{\sum_{i=1}^p\lambda_i(\Sigma_j)Z_i^2\right\}^{1/2}.
\]
Then $R_j$ has the distribution of $\|M_j\|$ and $R_1\geq R_0$ almost surely. If the covariances differ, the positive-semidefinite difference has positive trace, so at least one eigenvalue inequality is strict and $R_1>R_0$ almost surely. Proposition~\ref{prop:beneficial} gives strict decrease of $\psi$ on $[0,\infty)$. Consequently $\E\psi(R_1)\leq\E\psi(R_0)$, with the stated strictness. The linear quantity bounds in Proposition~\ref{prop:vector} give quadratic growth of the cost and justify both expectations. Equation~\eqref{eq:vectorinfo} completes the comparison.
\end{proof}

The radial coupling compares the distributions of the two norms. It need not coincide with the coupling induced by the actual signals. No commutation of $\Sigma_0$ and $\Sigma_1$ is required. A Gaussian prior with nonlinear signals need not have Gaussian posterior means. The Gaussian-experiment restriction is therefore substantive. The proof also requires isotropic execution. Simultaneous diagonalization of the execution matrices alone is insufficient.

\paragraph{Execution-weighted forecast error.}\label{app:weightederror}
For a fixed roster, put $Q=\E[(M_1-M_0)(M_1-M_0)^\top]$ and
\[
 G_\Lambda=\E[(D-M_0)^\top\Lambda(D-M_0)]
 -\E[(D-M_1)^\top\Lambda(D-M_1)]
 =\operatorname{tr}(\Lambda Q)\geq0.
\]
If $\theta_-$ and $\theta_+$ are the extreme eigenvalues of $\Lambda^{-1/2}P_a\Lambda^{-1/2}$, then
\begin{equation}
 \theta_-G_\Lambda\leq\Delta\mathcal C
 =\operatorname{tr}(P_aQ)\leq\theta_+G_\Lambda.\label{eq:weightedbound}
\end{equation}
The two Loewner inequalities $\theta_-\Lambda\preceq P_a\preceq\theta_+\Lambda$, multiplied by $Q\succeq0$ and traced, prove the bound. Under an invertible change of quantity units $\widetilde D=RD$, the matrices become $\widetilde\Lambda=R^{-\top}\Lambda R^{-1}$, $\widetilde P_a=R^{-\top}P_aR^{-1}$, and $\widetilde Q=RQR^\top$. Both traces and the generalized eigenvalues are unchanged. The comparison therefore measures execution incidence per unit of forecast improvement without depending on asset denomination. Feasibility remains determined by the prior. The bound does not assert that every $Q\succeq0$ is attainable. It applies to a fixed roster, not a state-dependent activation mass.

\subsection{A portfolio counterexample with original cash payments}\label{app:vectorcounterexample}
The unrestricted matrix class admits a different mechanism. Consider
\begin{gather}
 \Lambda=\begin{pmatrix}26.1&15\\15&9.1\end{pmatrix},\qquad
 H=\begin{pmatrix}25&15\\15&9\end{pmatrix},\qquad K=H/2,\nonumber\\
 E_0=\operatorname{diag}(.1,10),\qquad E=\operatorname{diag}(2,10),\qquad
 a=\tfrac12,\qquad u=(.4,-1)^\top.\label{eq:matrixexample}
\end{gather}
Here $H=(5,3)^\top(5,3)\succeq0$ and $\Lambda-H=\operatorname{diag}(1.1,.1)\succ0$. A uniform schedule with zero decay and instantaneous impact matrix $H$ gives $K=H/2$. Both charge matrices are positive definite. Thus the example satisfies the common-schedule interpretation as well as the displayed matrix assumptions.

Let $D$ equal $u$ or $-u$ with probability one-half. No information gives zero quantities and expected expenditure $u^\top\Lambda u=1.276$. Under revelation at $D=u$, the equilibrium quantities are
\[
 q\approx(-.428237463,-.099612436)^\top,\qquad
 m=q/2,\qquad z\approx(.014887660,-.082724781)^\top.
\]
All quantities change sign at $D=-u$, leaving the cash payments unchanged. Table~\ref{tab:matrixcash} evaluates pre-event impact, temporary charges, and terminal completion directly. With the rational primitives in \eqref{eq:matrixexample}, the exact cost increase is
\begin{equation}
 u^\top P_{1/2}u
 =\frac{9196314467527330257}{235903720523056824500}>0.\label{eq:matrixexactgap}
\end{equation}
This is a $3.0551\%$ increase relative to the no-information bill. The generalized eigenvalues in \eqref{eq:weightedbound} are approximately $-.27450906$ and $.03974858$. Here $G_\Lambda=1.276$ and $\Delta\mathcal C/G_\Lambda=.03055121$. The positive gap persists under sufficiently small admissible perturbations. These are normalized model costs, not estimated ETF execution losses.

\begin{table}[htbp]\centering\small
\caption{Original institutional cash payments in the portfolio counterexample. The centered target law and the participation mass $a=1/2$ are unchanged between experiments. The exact identity \eqref{eq:matrixexactgap} verifies the positive total difference.}\label{tab:matrixcash}
\begin{tabular}{lrr}
\toprule
Cash component & No information & Revelation \\
\midrule
Pre-event impact & 0.000000000000 & 0.121072164981 \\
Temporary charge & 0.000000000000 & 0.034228029449 \\
Terminal completion & 1.276000000000 & 1.159683143549 \\
Total expenditure & 1.276000000000 & 1.314983337978 \\
\bottomrule
\end{tabular}

\end{table}

The same target law yields a harmful comparison with optional entry. Set $g_*=q^\top Eq/2\approx.233000511835$ at the displayed half-roster equilibrium and choose the fixed cost law $F$ to be uniform on $[0,2g_*]$. This one law is used in both experiments. Under revelation, $F(g_*)=1/2$ at either state. Under no information, gross profit and participation are zero. Proposition~\ref{prop:vector} gives uniqueness, and the original cash difference remains \eqref{eq:matrixexactgap}.

The adverse fixed-roster comparison comes from inventory direction. The term $-\mu^\top(\Lambda-H)m$ in \eqref{eq:vectorinfo} need not be nonpositive. Cross-impact and unequal trading charges can rotate supply. Here entrants sell the first asset despite its positive target component. This mechanism operates at fixed participation. Noncommutativity permits it without making every economy adverse.

\subsection{A spectral margin and sensitivity to portfolio technology}\label{app:portfoliospectral}
The relevant measure of robustness is the cost effect per unit of execution-weighted forecast improvement. Define
\[
 Q_a=\Lambda^{-1/2}P_a\Lambda^{-1/2},\qquad
 \theta_+(a)=\lambda_{\max}(Q_a).
\]
The generalized eigenvalues in \eqref{eq:weightedbound} do not change when asset quantities are redenominated. A strict margin also gives a sufficient condition for an entire neighborhood of technologies to retain the centered-target ranking.

\begin{corollary}[A favorable neighborhood]\label{cor:spectralmargin}
Suppose $\sup_{a\in[0,1]}\theta_+(a)\leq-\delta<0$. Any admissible perturbed technology with matrices $\widetilde Q_a$ satisfying
\begin{equation}
 \sup_{a\in[0,1]}\|\widetilde Q_a-Q_a\|_2<\delta
 \label{eq:spectralmargin}
\end{equation}
has $\widetilde P_a\prec0$ for every $a$. It therefore strictly improves on no information for every centered target and every signal whose posterior mean is nonzero with positive probability. Every simultaneously orthogonally diagonalizable admissible technology has an open neighborhood with this property.
\end{corollary}
\begin{proof}
For unit $x$, $x^\top\widetilde Q_a x\leq x^\top Q_a x+\|\widetilde Q_a-Q_a\|_2<0$, uniformly in $a$. Congruence gives $\widetilde P_a\prec0$, and the pointwise cost representation proves the ranking. At a simultaneously diagonalizable technology, \eqref{eq:commutingcost} is strictly negative for every coordinate and mass. Continuity on the compact interval $[0,1]$ gives a uniform negative margin. Matrix inversion and the positive-definite square root vary continuously on the admissible domain. This dependence is uniform over the same compact interval. Sufficiently small admissible perturbations of the primitive matrices therefore satisfy \eqref{eq:spectralmargin}.
\end{proof}

\begin{figure}[htbp]\centering
\includegraphics[width=\textwidth]{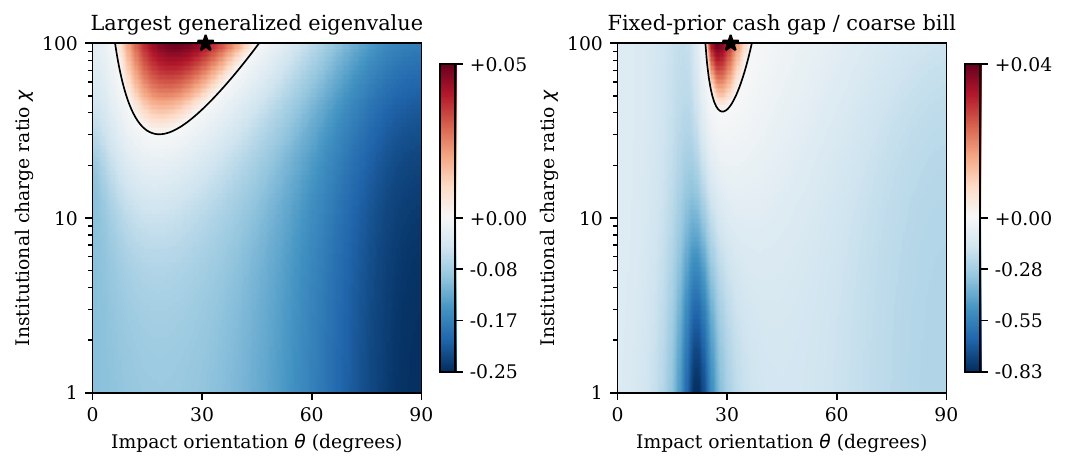}
\caption{Sensitivity to impact orientation and institutional charge heterogeneity at the fixed roster $a=1/2$. Red indicates a positive value. Black contours indicate zero. The star marks the original counterexample. The right panel keeps the centered two-point prior unchanged. The left measures the worst cost direction per unit of execution-weighted forecast gain. Technologies vary across scenarios, not within an information comparison.}\label{fig:portfoliospectral}
\end{figure}

Figure~\ref{fig:portfoliospectral} varies both impact orientation and institutional charge heterogeneity around \eqref{eq:matrixexample}. Its declared family is
\begin{gather}
 v_\theta=(\cos\theta,\sin\theta)^\top,\qquad
 H_\theta=34v_\theta v_\theta^\top,\qquad K_\theta=H_\theta/2,\nonumber\\
 \Lambda_\theta=H_\theta+\operatorname{diag}(1.1,.1),\qquad
 E_0(\chi)=\operatorname{diag}(.1,.1\chi),\qquad
 E=\operatorname{diag}(2,10).\label{eq:spectralfamily}
\end{gather}
Every scenario is admissible and has a uniform zero-decay schedule. The grid uses 181 equally spaced angles from $0$ to $\pi/2$ and 81 geometrically spaced charge ratios from $1$ to $100$. It holds the prior $D=\pm(.4,-1)^\top$ with equal probabilities and the roster $a=1/2$ fixed. Within each scenario, no information and revelation use exactly the same technology. The original counterexample is $\theta=\arctan(3/5)$ and $\chi=100$.

The left panel plots $\theta_+(1/2)$. Positive values identify an adverse target direction. The right panel instead evaluates the one fixed prior throughout. Those signs need not agree: at $15$ degrees and $\chi=100$, some directions are adverse but the displayed prior still benefits from revelation. At the original parameter values, the largest generalized eigenvalue is $0.03974858$ and the fixed prior's cash gap is $0.03898334$, or $0.03055121$ times its coarse bill. Commuting technologies at either angular endpoint are favorable, as guaranteed by the proposition.

\begin{table}[htbp]\centering\small
\caption{Selected portfolio technologies from \eqref{eq:spectralfamily}. The prior and roster are unchanged. $\theta_+$ is the largest generalized eigenvalue of $(P_{1/2},\Lambda)$. Costs are normalized model quantities.}\label{tab:portfoliospectral}
\begin{tabular}{rrrrrr}
\toprule
$\theta$ (degrees) & $\chi$ & $\theta_+$ & Coarse bill & Cash gap & Gap / bill\\
\midrule
0.00 & 100 & -0.024390 & 5.716000 & -0.581684 & -0.101764 \\
15.00 & 100 & +0.036892 & 0.829157 & -0.151556 & -0.182784 \\
30.96 & 1 & -0.099047 & 1.276000 & -0.317718 & -0.248995 \\
30.96 & 10 & -0.069888 & 1.276000 & -0.134632 & -0.105511 \\
30.96 & 100 & +0.039749 & 1.276000 & +0.038983 & +0.030551 \\
60.00 & 100 & -0.063255 & 15.358055 & -1.324013 & -0.086210 \\
90.00 & 100 & -0.169857 & 34.276000 & -5.960548 & -0.173899 \\
\bottomrule
\end{tabular}

\end{table}

The replication program reconstructs original institutional cash payments and both agents' first-order conditions independently of the reduced cost expression at every grid point. The largest absolute discrepancies are below $1.5\times10^{-14}$. A separate change-of-denomination check preserves the generalized eigenvalues to $10^{-14}$. These numerical checks cover the declared finite grid. The analytical margin establishes the neighborhood result. The grid does not certify a continuum of parameter values or estimate the prevalence of adverse information in actual markets.

\subsection{Proof and magnitudes for directional information}\label{app:directional}
Section~\ref{sec:portfolioinformation} states the isotropic model and Corollary~\ref{cor:directional}. Basket profit depends only on the posterior-mean radius. The scalar equilibrium equation therefore applies at $r=\|\mu\|$, and \eqref{eq:vectorinfo} reduces to $\Psi(\mu)=\psi(r)$, giving the Hessian in \eqref{eq:radialhessian}.

\begin{proof}[Proof of Corollary~\ref{cor:directional}]
Bayes' rule gives $\E[B_j\mid T_j]=(\epsilon/\delta)T_j$ and $\E[B_{3-j}\mid T_j]=0$, hence the stated posterior means, each with probability one-half. Conditional on either signal, the posterior variance trace falls from $2\delta^2$ to $2\delta^2-\epsilon^2$. The $\Lambda$-weighted reduction is also equal, at $\lambda\epsilon^2$. Equation~\eqref{eq:vectorinfo} cancels the unchanged target-second-moment term. Isotropy gives the two exact cost differences, and Taylor expansion yields the radial expression. Strict decrease of $\psi$ on the positive half-line, from Proposition~\ref{prop:beneficial}, proves the finite tangential inequality. At smooth positive states the scalar proof also gives $\psi'(r)<0$, so the tangential eigenvalue in \eqref{eq:radialhessian} is negative.
\end{proof}

The signals need not be Blackwell comparable to each other. Each refines the same coarse information. The result shows why the amount of forecast-error reduction alone does not determine execution value across portfolio directions. A rotation to $n=(1,1)^\top/\sqrt2$ and $v=(1,-1)^\top/\sqrt2$ gives a common-basket direction and a spread direction without changing the isotropic economy.

Use the lognormal cost law and execution coefficients of Proposition~\ref{prop:smoothinterior}, with $r\approx.825060078573$ and $\delta=.02$. The radial and tangential Hessian eigenvalues are approximately $4.40038637$ and $-.41909807$. The common coarse bill is $.351308300279$, including both target shocks. Table~\ref{tab:directionalcash} reports direct high-precision evaluations of the original cash objective. The existing interval certificate proves the local radial sign, while these finite magnitudes are numerical evaluations.

\begin{table}[htbp]\centering\small
\caption{Two signals under one unchanged target law. Coarse MSE is $0.0008$. Both signals attain the reported fine MSE. Cost changes use normalized model units. Percent changes divide by the same coarse institutional bill, including both shocks.}\label{tab:directionalcash}\label{tab:directionalmain}
\begin{tabular}{rrrrrr}
\toprule
$\epsilon$ & Fine MSE & Radial cost & Tangential cost & Radial (\%) & Tangential (\%) \\
\midrule
0.001 & 0.000799 & +0.000002200966 & -0.000000209548 & +0.000627 & -0.000060 \\
0.010 & 0.000700 & +0.000227846129 & -0.000020946066 & +0.064856 & -0.005962 \\
\bottomrule
\end{tabular}

\end{table}
\begin{figure}[htbp]\centering
\includegraphics[width=\textwidth]{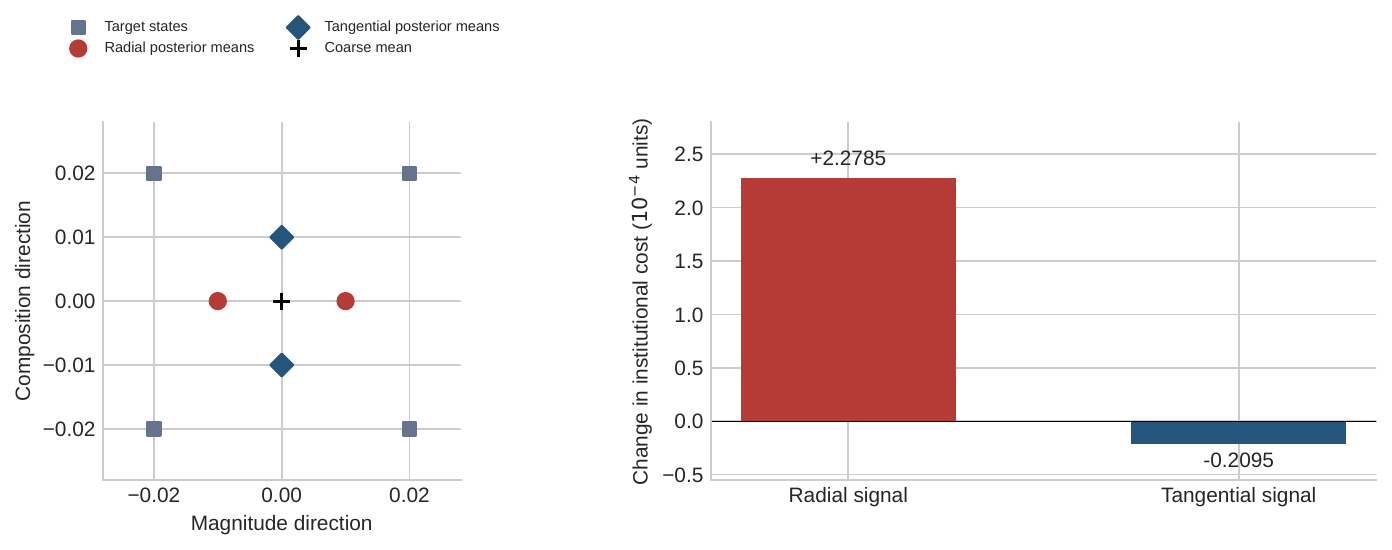}
\caption{Direction of information and execution value, with $\epsilon=.01$. Left: the four target states and the two alternative pairs of posterior means, shown relative to the coarse mean. Right: radial and tangential changes in the original institutional bill. Both signals reduce MSE by $0.0001$. The economy is constructed from the smooth-interior example, without fitting market data.}\label{fig:directionalinformation}
\end{figure}

\FloatBarrier
\section{Conditional execution risk and information}\label{app:riskextension}
With risk-neutral suppliers, conditional target variance affects the institution's bill but not their positions. Risk aversion introduces a supply response to that uncertainty. Retain the continuum of suppliers, the constant unaffected price, and Assumption~\ref{ass:primitive}. Keep the institution risk neutral, and let each supplier evaluate its cash profit by its conditional mean minus $\gamma/2$ times its conditional variance, where $\gamma\geq0$ is common across suppliers. This preference introduces exposure to target uncertainty without changing the timing or execution technology.

\subsection{Equilibrium with conditional target risk}
For an entrant with fixed cost $f$, cash profit is
\[
 \pi=q\{\lambda D-b(z+m)\}-\frac\eta2q^2-f.
\]
Conditional on the public signal, its variance is $\lambda^2q^2v$. The supplier therefore maximizes
\begin{equation}
 q\{\lambda\mu-b(z+m)\}-\frac{r_v}{2}q^2-f,
 \qquad r_v=\eta+\gamma\lambda^2v.\label{eq:riskobjective}
\end{equation}
When $D$ is conditionally Gaussian, this objective is also the exact certainty equivalent for exponential utility with absolute risk aversion $\gamma>0$. For a general conditional target law, it is the stated mean--variance preference.

\begin{proposition}[Equilibrium with execution risk]\label{prop:riskequilibrium}
Under Assumption~\ref{ass:primitive} and preference~\eqref{eq:riskobjective}, every posterior state $(\mu,v)$ has a unique participation equilibrium. With $s,c$ as in~\eqref{eq:sc}, it satisfies
\begin{align}
 a&=F\!\left(\frac{r_vs^2\mu^2}{2(r_v+ca)^2}\right),\label{eq:riskentry}\\
 q&=\frac{s\mu}{r_v+ca},\qquad m=aq,\qquad
 z=\frac{d\mu-bm}{A}.\label{eq:riskquantities}
\end{align}
Writing $M=\E[D\mid Y]$ and $V=\operatorname{Var}(D\mid Y)$, expected institutional expenditure is
\begin{align}
 \E C_0&=\lambda\E D^2+\E\psi(M,V),\label{eq:riskrepresentation}\\
 \psi(\mu,v)&=-\frac{d^2}{2A}\mu^2+e\mu m(\mu,v)-km(\mu,v)^2.\label{eq:riskpsi}
\end{align}
Equilibrium positions have finite second moments, and expected expenditure is finite.
\end{proposition}
\begin{proof}
The institution's first-order condition remains $Az+bm=d\mu$. The supplier's condition is $r_vq=\lambda\mu-b(z+m)$, so substitution gives $r_vq+cm=s\mu$. Its maximized gross certainty equivalent is $r_vq^2/2$. The entry rule therefore gives~\eqref{eq:riskentry}. The difference between $a$ and the right-hand side of that equation is continuous and strictly increasing on $[0,1]$, with opposite weak signs at its endpoints. Its unique zero verifies both agents' globally optimal best responses. Completing the square in the institution's cash bill gives
\[
 \min_z C_0(z,m;\mu,v)
 =\lambda(\mu^2+v)-(\lambda-H)\mu m-\frac{(d\mu-bm)^2}{2A}.
\]
Expansion proves~\eqref{eq:riskpsi}. The conditional second-moment identity proves~\eqref{eq:riskrepresentation}. Finally, $r_v\geq\eta$ gives $|m|\leq|q|\leq s|\mu|/\eta$, and $\E M^2\leq\E D^2$ completes the moment argument.
\end{proof}

\subsection{The supply response to uncertainty}
Higher uncertainty increases the effective quadratic charge $r_v$. Its effect on the number of suppliers is ambiguous, but its effect on their aggregate position has a definite sign.

\begin{proposition}[Risk and aggregate supply]\label{prop:risksupply}
Fix $\mu>0$. Aggregate inventory $m$ is nonincreasing in $r_v$, and decreases strictly whenever the initial inventory is positive. If $F$ is differentiable at the equilibrium cutoff, write
\[
 g=\frac{r_vq^2}{2},\qquad
 R_v=r_v+c\{a+2gf_F(g)\}>0.
\]
Then
\begin{align}
 \frac{\partial m}{\partial r_v}
 &=-\frac{q\{a+gf_F(g)\}}{R_v}\leq0,\label{eq:riskmderivative}\\
 \frac{\partial a}{\partial r_v}
 &=\frac{f_F(g)q^2(ca-r_v)}{2R_v}.\label{eq:riskaderivative}
\end{align}
Consequently, wherever these derivatives exist,
\begin{equation}
 \frac{\partial\psi}{\partial v}
 =(e\mu-2km)\gamma\lambda^2\frac{\partial m}{\partial r_v}.
 \label{eq:riskincidence}
\end{equation}
For negative posterior means, the corresponding monotonicity holds for $|m|$.
\end{proposition}
\begin{proof}
Consider $r_2>r_1$ and suppose $m_2\geq m_1$. The identities $r_iq_i+cm_i=s\mu$ imply $r_2q_2\leq r_1q_1$, hence $q_2<q_1$ and $r_2q_2^2<r_1q_1^2$. The entry rule gives $a_2\leq a_1$, so $m_2\leq m_1$, strictly if $m_1>0$. This proves the global statement without differentiability of $F$. At a differentiable cutoff, implicit differentiation of $r_vq+cqF(r_vq^2/2)=s\mu$ gives
\[
 \frac{\partial q}{\partial r_v}
 =-\frac{q\{1+cgf_F(g)/r_v\}}{R_v},\qquad
 \frac{\partial g}{\partial r_v}
 =\frac{q^2(ca-r_v)}{2R_v}.
\]
Differentiating $a=F(g)$ and $m=qF(g)$ yields~\eqref{eq:riskmderivative}--\eqref{eq:riskaderivative}. Differentiating~\eqref{eq:riskpsi} yields~\eqref{eq:riskincidence}. Odd symmetry of the positions proves the negative-mean statement.
\end{proof}

Equation~\eqref{eq:riskincidence} uses the same marginal cost coefficient as the baseline model: reducing uncertainty brings additional supply, whose value depends on $e\mu-2km$. It concerns the mean-dependent part of the bill. At a fixed posterior mean, the derivative of the full conditional bill is $\lambda+\partial\psi/\partial v$. In an ex ante information comparison, however, the prior is fixed and $\E[M^2+V]=\E D^2$. Mean and variance changes must therefore arise from a feasible common experiment. They cannot be selected independently.

\subsection{A feasible Gaussian precision comparison}
The Gaussian precision result in Proposition~\ref{prop:beneficial} uses risk neutrality. To see why the restriction matters, take
\[
 \lambda=1,\quad K=0.495,\quad H=0.99,\quad
 \eta_0=0.01,\quad\eta=0.05,
 \]
 \[
 F=\operatorname{Uniform}[0.02,0.04],\quad D\sim N(0,1).
\]
These coefficients satisfy Assumption~\ref{ass:primitive}. They can be implemented by a uniform schedule with zero decay and $\kappa=0.99$, which gives $K=\kappa/2$ and $H=\kappa$.

Let independent Gaussian increments $U_1,U_2,U_3$ have variances $1/2,1/4,1/4$, and set $D=U_1+U_2+U_3$. Reveal these increments successively. The nested information sets $\sigma(U_1)\subset\sigma(U_1,U_2)\subset\sigma(U_1,U_2,U_3)$ have posterior means $U_1$, $U_1+U_2$, and $D$. If $r$ denotes explained target variance, these experiments give $M_r\sim N(0,r)$ and $V_r=1-r$ for $r\in\{0.50,0.75,1.00\}$, under the same $N(0,1)$ target law.

\begin{table}[tbp]
\centering
\caption{Institutional expenditure under nested Gaussian information}\label{tab:riskprecision}
\begin{tabular}{lrrr}
\toprule
Supplier risk parameter & $r=0.50$ & $r=0.75$ & $r=1.00$ \\
\midrule
$\gamma=0$   & 0.981539466 & 0.973686347 & 0.965729227 \\
$\gamma=0.1$ & 0.967959873 & 0.964323826 & 0.965729227 \\
$\gamma=0.5$ & 0.914575193 & 0.922044258 & 0.965729227 \\
\bottomrule
\end{tabular}

\begin{minipage}{.95\textwidth}
\small\vspace{.5em}
\emph{Notes.} Entries are expected institutional execution costs from~\eqref{eq:riskrepresentation}, in model cash units. The no-information cost is one. The columns are nested experiments under a fixed standard normal target. Here $r$ is the variance of the posterior mean and $\gamma$ is the supplier's risk parameter. Quadrature is split at the entry and full-participation thresholds. Independent calculations using the reduced cost and the original cash bill agree within $4\times10^{-16}$. These are constructed model values, not estimated ETF costs.
\end{minipage}
\end{table}

For $\gamma=0.5$, refining the signal from $r=0.50$ to revelation raises expenditure by $0.0511540345$. For $\gamma=0.1$, the cost first falls and then rises. At revelation, conditional risk disappears and every row has the same value. All displayed costs remain below the no-information benchmark. The alternative cost representation in the proof of Proposition~\ref{prop:riskequilibrium} gives $\psi(\mu,v)\leq0$, with strict inequality for $\mu\ne0$. Information therefore never raises cost relative to no information for a centered target. Cost need not decrease at every increase in Gaussian precision.

This extension isolates execution risk from uncertainty about the unaffected price. Restoring a fundamental-price martingale generally adds inventory risk and covariance terms. The variance $\lambda^2q^2v$ then requires modification. The result also concerns the continuum model, not the finite-player game. Empirical interval scores can assess forecast calibration, but they neither identify $\gamma$ nor establish a feasible reduction in posterior variance state by state. A signal that leaves posterior means unchanged preserves average conditional variance. These restrictions determine how the uncertainty forecasts can be interpreted economically.

\FloatBarrier
\section{Further numerical comparisons}\label{app:numericalcomparisons}
\subsection{Normalization and the two mechanisms}
Choose a reference target quantity $D_*$ and normalize cash by $\lambda D_*^2$, inventory by $D_*$, and price-impact and temporary-cost coefficients by $\lambda$. The decay parameter is the physical decay rate multiplied by the pre-event horizon. All results below are constructed scenarios in these dimensionless units. The onset scenario uses
\[
 \lambda=\kappa=1,\quad\rho=0.1,\quad\eta_0=\eta=0.05,
 \quad F=\operatorname{Uniform}[0.02,0.04],\quad\phi_t=1.
\]
It has $H=0.951626$, $K=0.483742$, $e=0.452291$, and $\mu_0=0.089562$. Its impact half-life is $\log(2)/0.1=6.93$ pre-event horizons. The scenario therefore has deliberately persistent impact. The saturation scenario instead uses the zero-pre-event-impact parameters in \eqref{eq:saturationexample}, with threshold $\sqrt2$.

\begin{figure}[htbp]\centering
\includegraphics[width=.94\textwidth]{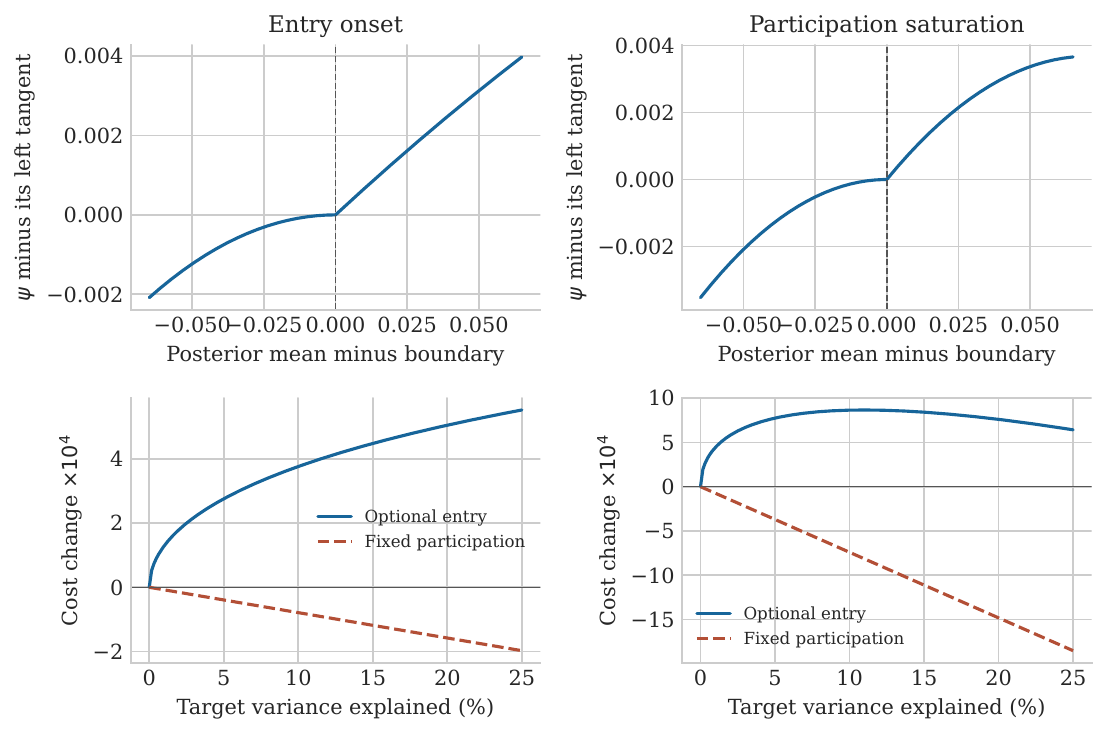}
\caption{Two information mechanisms. The upper panels subtract the left tangent from the posterior-mean cost function at each boundary. This preserves every information comparison and reveals the slope change. The lower panels compare optional with fixed participation. Each column holds its own binary target law fixed: $D=\mu_0\pm0.04$ at onset and $D=\sqrt2\pm0.1$ at saturation. The common horizontal scale is the percentage of target variance explained, $100(\epsilon/\delta)^2$, where $\delta$ is each fixed prior's half-spread. Cost units are normalized and differ across the two scenarios.}\label{fig:geometry}
\end{figure}

At $\epsilon=0.01$, the onset cost increase is $0.00030485$ and the saturation increase is $0.00044338$. At saturation, inventory is beneficial for every target mean. Cost nevertheless rises because the refinement changes participation asymmetrically. In both comparisons, the institution reoptimizes its pretrade. The constant-second-moment representation retains the posterior-variance term. The upper panels show an affine-detrended conditional-mean cost curve. The lower panels compare public experiments, which is a different calculation.

Table~\ref{tab:magnitudes} distinguishes raw cost changes from their economic scale. The onset effect is about $5.38\%$ of the coarse institutional bill, compared with $0.084\%$ at full participation. The latter has the larger raw gap. These are percentages of constructed execution cost, not basis points of assets or estimates of investor losses.
\begin{table}[htbp]\centering\footnotesize
\caption{Information magnitudes at $\epsilon=0.01$. Coarse cost includes target variance under $D=\mu_0\pm0.04$ and $D=\sqrt2\pm0.1$, respectively. The percentage denominator is that scenario's own coarse cost.}\label{tab:magnitudes}
\begin{tabular}{lrrrr}
\toprule
Scenario & $\mathcal C(0)$ & $\Delta\mathcal C$ & Cost increase, \% & Variance explained, \% \\
\midrule
Onset & 0.005665 & 0.000305 & 5.381175 & 6.250000 \\
Saturation & 0.530000 & 0.000443 & 0.083656 & 1.000000 \\
\bottomrule
\end{tabular}

\end{table}

\subsection{One prior, increasing precision}
\begin{figure}[htbp]\centering
\includegraphics[width=.97\textwidth]{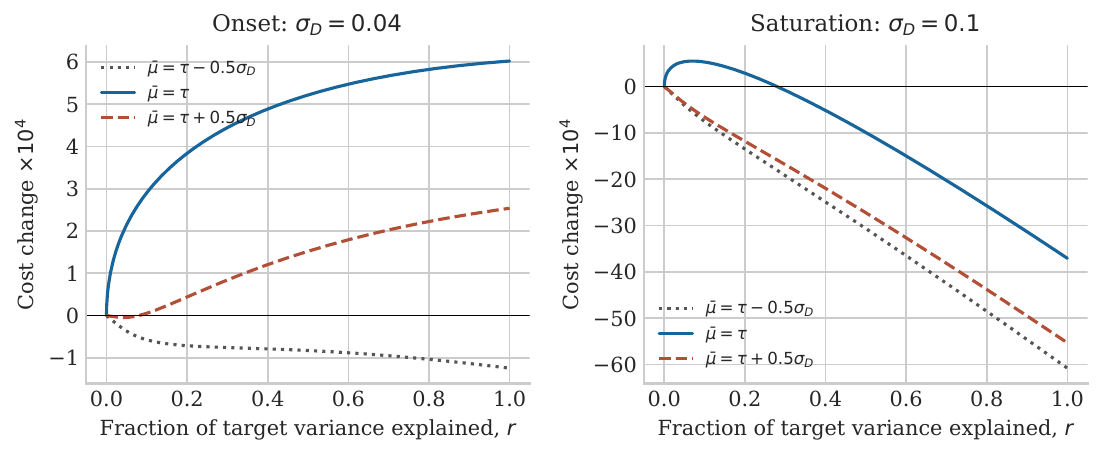}
\caption{Gaussian precision under unchanged priors. Each curve retains its indicated mean and the panel's target standard deviation. The boundary $\tau$ is $\mu_0$ on the left and $\mu_1$ on the right. Costs are relative to each curve's own no-information cost. Full revelation is $r=1$, and the cost scales differ across panels.}\label{fig:gaussian}
\end{figure}

Figure~\ref{fig:gaussian} follows a Gaussian target $D\sim N(\bar\mu,\sigma_D^2)$ as a public Gaussian signal explains an increasing fraction $r$ of its variance. Each curve holds both prior parameters fixed. Formula~\eqref{eq:gaussianpath} includes the same target second moment at every precision, and \eqref{eq:gaussianderivative} relates the path's slope to regular curvature and boundary contributions.

At the onset boundary, full revelation raises cost by $0.00060202$. At the full-participation boundary, small precision gains initially raise cost, but revelation lowers it by $0.00370761$. On the reported grid, the latter path peaks at $r=0.0694$. Its increase is $0.00055083$, or $0.10393\%$ of its coarse bill. This is a grid maximum, not a proved optimizer. With mean zero instead, full-revelation changes are $-0.00076215$ and $-0.00666667$, respectively, and Proposition~\ref{prop:beneficial} proves that both centered paths decrease throughout. Local harm consequently does not by itself determine the value of complete information.

\subsection{Strategic size and smooth heterogeneity}
\begin{figure}[htbp]\centering
\includegraphics[width=.94\textwidth]{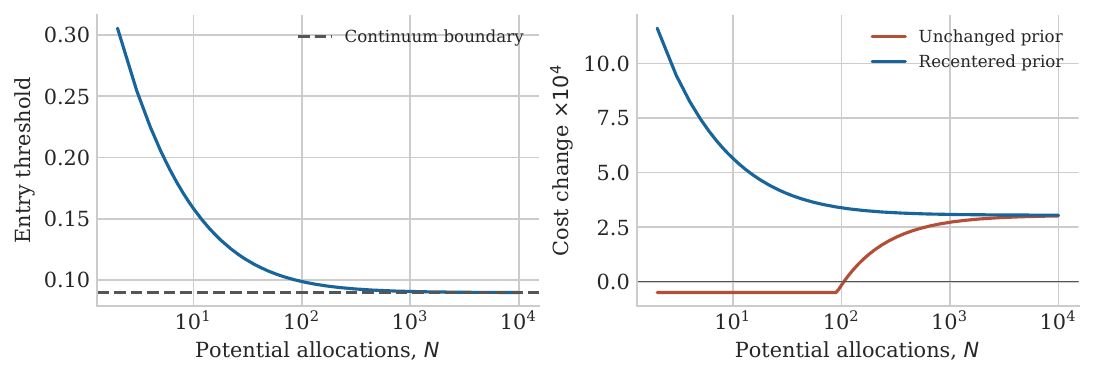}
\caption{Finite strategic size changes the information comparison. The left panel gives the exact entry boundary. The right fixes $\epsilon=0.01$ and binary target half-spread $0.04$, comparing a prior centered at the continuum threshold with priors centered at each finite threshold. Each individual coarse/refined pair has the same target law, while the recentered family changes that law across $N$.}\label{fig:finiteexact}
\end{figure}

Figure~\ref{fig:finiteexact} solves the exact symmetric finite game with the onset technology. With the prior held at the continuum center and $\epsilon=0.01$, the cost change is $-0.00001503$ at $N=100$, $+0.00014455$ at $N=200$, and $+0.00030485$ in the continuum. At $N=100$, the threshold is $0.098634$, exceeding the continuum threshold by about $0.009072$. Recentering the prior at that finite threshold gives $+0.00033884$ instead. This comparison changes the target law, unlike the experiment at the unchanged continuum center. The numerical $N$ counts potential inventory allocations and is not an estimate of the number of trading firms.

The full-support lognormal construction in Proposition~\ref{prop:smoothinterior} gives cost increases $0.0000022010$ and $0.0002278461$ at displacements $0.001$ and $0.01$ around its quarter-participation state. The target remains $D=\mu_*\pm0.02$, and its coarse institutional cost is $0.35090830$. The two increases are therefore $0.000627\%$ and $0.064930\%$ of that same execution bill. Existence of an adverse region need not imply a large finite effect. Appendix~\ref{app:smoothscale} reports the denominators, execution-parameter sensitivity, and one equal-charge example with positive decay. Table~\ref{tab:interior} reports both positive and negative curvature states. At the principal technology's unchanged onset prior, replacing uniform costs by moment-matched gamma or lognormal laws leaves positive gaps of $0.00007863$ and $0.00009884$. The analogous full-participation gaps become negative. Thus the two boundary examples have different distributional sensitivity. Appendix~\ref{app:robustexhibits} supplies these comparisons, multiplicative smoothing, and the diffuse-forecast display on an unclipped symmetric-log scale.

\subsection{Terminal technology and a consistent schedule comparison}
The outside-facility scenarios in Table~\ref{tab:facilityinfo} retain the onset parameters and use the unrestricted signed routing of Section~\ref{sec:facility}. The incidence coefficient stays positive throughout, as Proposition~\ref{prop:facility} requires. A sufficiently inexpensive facility moves the entry boundary far enough to make the fixed finite refinement beneficial at the original continuum center. The recentered comparisons remain positive in the reported cases, and the no-entrant column records the institution's direct forecasting benefit within each facility environment.

\begin{table}[htbp]\centering\footnotesize
\caption{Outside completion and information, with $\epsilon=0.01$. ``Fixed prior'' keeps the original onset center, ``Recentered'' uses $\mu_{0,\chi}$, and ``No entrants'' gives the matched zero-participation information change. All binary priors have half-spread $0.04$.}\label{tab:facilityinfo}
\begin{tabular}{rrrrrr}
\toprule
$\chi$ & $\mu_{0,\chi}$ & $e_\chi$ & Fixed prior & Recentered & No entrants \\
\midrule
2 & 0.134951 & 0.300167 & -0.000016 & 0.000168 & -0.000016 \\
5 & 0.107718 & 0.376057 & -0.000029 & 0.000230 & -0.000029 \\
10 & 0.098640 & 0.410666 & -0.000011 & 0.000262 & -0.000037 \\
100 & 0.090470 & 0.447752 & 0.000267 & 0.000300 & -0.000048 \\
10000 & 0.089571 & 0.452245 & 0.000304 & 0.000305 & -0.000049 \\
\bottomrule
\end{tabular}

\end{table}

Schedule sensitivity uses early, uniform, and late profiles $2(1-t)$, $1$, and $2t$. We keep the underlying rate-cost coefficients fixed and recompute both $K,H$ and the temporary charges. The early and late schedules therefore multiply each charge by $\int\phi^2=4/3$. Appendix~\ref{app:numerical} reports the full comparison at decay rates $0.1$, $2$, and $8$. At slow decay, early and late schedules preserve positive initial incidence but make the unchanged-prior information comparison beneficial by moving the threshold. Faster decay reduces carryover, and the early schedule at $\rho=8$ has $e<0$. Some finite recentered displacements are beneficial even with $e>0$, consistently with the onset theorem's sufficiently-small-displacement qualification.

\FloatBarrier
\section{Accounting and identification checks}\label{app:identification}
\subsection{The error in substituting intraday variation for a daily return square}
Taylor expansion of $\log(1+\beta R)-\beta\log(1+R)$ cancels the linear term and gives \eqref{eq:discrete}. For fixed $\beta$ and $|R|\leq r$ with $\max(1,|\beta|)r<1$, the third derivative is bounded, so the remainder is uniformly $O(|R|^3)$. Applying the same calculation at each five-minute reset instead gives $-\beta(\beta-1)\sum_k r_k^2/2$ plus the sum of the third-order remainders. The two approximations target different portfolios, even if their first-order benchmark exposures agree at the start.

In the numerical round trip, the benchmark rises eight percent in log-price units and returns to its initial value. The endpoint daily-reset multiplier is exactly one, while the repeated-reset multiplier is $\prod_k(1+\beta r_k)$. This is an accounting example selected independently of the empirical outcomes, with no claim about measured aggregate investor losses. The same-path calculation isolates reset accounting. The empirical target calculation uses the actual preceding-session reference.

\subsection{What a price-driven target forecast measures}
Let $R_{0,t}$ be the return from the previous reset reference to time $t$ and let $r_{t,T}$ be the remaining simple return. The exact composition is
\[
 R_{0,T}=R_{0,t}+(1+R_{0,t})r_{t,T}.
\]
Under the idealized mechanical rule, a forecast $\widehat r_{t,T}$ implies
\[
 \widehat d_t=\beta(\beta-1)\{R_{0,t}+(1+R_{0,t})\widehat r_{t,T}\},
\]
per assumed initial dollar, with forecast error $\beta(\beta-1)(1+R_{0,t})(r_{t,T}-\widehat r_{t,T})$. This transforms remaining-return uncertainty into mechanical target uncertainty under stated normalization. It excludes uncertainty about fund assets, flows, derivatives routing, and prior executions. Multiple leverage labels applied to the same benchmark return do not create independent observations.

\section{Fixed-prior information experiments from prices}\label{app:priceexperiments}
\subsection{Finite laws and nested signals}
The experiment uses the common forecasting dates and primary timestamp convention of Appendix~\ref{app:products}. Each period defines a joint finite law over dates and benchmark families. Dates receive equal weight, as do the four families within each date. The benchmark paths are retained jointly in the replication panel and in date-block resampling. Scalar cost comparisons average the family-specific decisions. Leverage labels remain linked to their common benchmark history, and portfolio cross-impact remains unestimated.

For each observation we compute the complete reset-period price return from its observed components,
\[
 R_{0,T}=R_{0,t}+(1+R_{0,t})r_{t,T},\qquad
 D_\alpha=\alpha\frac{a_\beta R_{0,T}}{0.01a_\beta}
          =\alpha\frac{R_{0,T}}{0.01}.
\]
The reference target is one percent of initial assets times the leverage coefficient. The hypothetical scale $\alpha$ sets target size relative to the stated cost law and remains fixed within every coarse--fine comparison. The observed return distribution retains its original mean, tails, and location relative to the activation threshold. At $\alpha=1$, the pooled target mean and standard deviation are $0.06448$ and $1.11436$ in 2023--2025, and $0.05255$ and $1.09821$ in 2026. The respective sample skewnesses are $0.47058$ and $-0.32022$.

For the observed-noon experiment, $Y_0$ records family identity and the sign of $R_{0,t}$. The fine signal $Y_1$ adds a magnitude category for $|10^4R_{0,t}|$, with boundaries $(25,50,100)$ basis points. A value at a boundary belongs to the category above it. Zero returns retain their own sign category. These rules give 36 occupied fine cells in the reference period and 32 in 2026. We calculate $M_j=\E[D_\alpha\mid Y_j]$ by exact finite-law averaging. At unit scale, $\E[(M_1-M_0)^2]$ is $0.26488$ and $0.29491$, respectively. The largest numerical residual in $\E[M_1\mid Y_0]=M_0$ is $2.3\times10^{-16}$.

The second channel supplies hypothetical information about the terminal return. Its coarse signal reveals the sign, and its fine signal adds absolute-magnitude categories with boundaries $(25,50,100,200)$ basis points. Dropping the magnitude category gives the coarse signal by deterministic garbling. There are 40 occupied fine cells in each period, with unit-scale mean-refinement variances $0.47235$ and $0.44718$. This channel applies a constructed public information technology to the target states. Historical traders did not possess that terminal information at noon. Both experiments calculate exact posteriors from the distribution under study, so neither provides an out-of-sample predictive test.

\subsection{Declared structural grid and cash verification}
The full Cartesian grid contains four impact environments, three temporary-cost pairs, four activation distributions, and seven target scales. The impact environments are zero pre-event impact and uniform schedules with $\kappa=1$ and $\rho\in\{0.1,2,8\}$. The temporary-cost pairs $(\eta_0,\eta)$ are $(0.05,0.05)$, $(1,1)$, and $(0.01,4)$. The activation laws are uniform on $[0.02,0.04]$ and lognormal with mean $0.03$ and coefficients of variation $0.03$, $0.30$, and $1$. The target scales are
\[
 \alpha\in\{1/16,1/8,1/4,1/2,1,2,4\}.
\]
Each continuum grid contains 336 cases, all of which are also computed at $N=100$ and $N=200$. The two periods and two channels give 4,032 calculations. The same grid is used throughout, including cases in which information benefits the institution.

For every case, the replication output records both bills and the posterior-mean refinement variance. It also gives the three terms of \eqref{eq:cashdecomposition}, positive and negative conditional contributions, participation, activation thresholds, and the fixed-roster comparison. The latter freezes the common roster at $\E[a(M_0)]$ and reoptimizes quantities at both signals. For finite populations it applies the same frozen participation probability to the exact symmetric finite equations. All fixed-roster gaps are negative.

Signed information contributions are defined at the coarse-state level. If $c$ is a coarse family--sign state, set
\[
 J_c=\E[\psi(M_1)\mid Y_0=c]-\psi(M_0(c)).
\]
The positive contribution is $\sum_c\Pr(c)\max\{J_c,0\}$ and the negative contribution is $\sum_c\Pr(c)\min\{J_c,0\}$. Their sum is the total gap. An information comparison averages over every fine-signal outcome within its coarse state. A positive difference at one realization alone cannot establish harm.

The finite uniqueness bound is satisfied in 2,464 of 2,688 finite cases. The remaining 224 cases are identified as symmetric-equilibrium calculations whose selection independence is unestablished. All finite cases in the main tables and figures satisfy the bound. The code solves the monotone participation equation by 52 bisection iterations. Across the complete grid, the maximum participation-equation residual is $1.6\times10^{-14}$. Independent Brent solutions agree within $2.5\times10^{-15}$ in the principal scenarios.

The independent cash check evaluates pre-event purchases and terminal payments under each policy at the fine-cell conditional target moments. Both policies are evaluated using $\E[D\mid Y_1]$, the target mean conditional on the fine cell. The largest discrepancy between original cash payments and the reduced expected cost is $7.6\times10^{-15}$. This check also retains the target second moment, which is identical under both signals. The finite institution's objective is affine in realized supplier inventory. Private entry is independent of target innovations, which allows expected inventory to enter the direct cash account.

\begin{table}[htbp]\centering\scriptsize
\caption{Three-term cash decomposition and signed conditional contributions at unit target scale. The direct, inventory, and readjustment columns sum to the information-cost change. Positive and negative columns sum to the same change after conditioning within coarse states.}\label{tab:pricecash}
\begin{tabular}{lllrrrrr}
\toprule
Period & Signal & Technology & Direct & Inventory & Readjust. & Positive & Negative \\
\midrule
2023--2025 & Noon & Persistent & -0.13064 & +0.19019 & -0.08592 & +0.00000 & -0.02636 \\
2023--2025 & Noon & Zero & -0.17658 & -0.02949 & -0.00276 & +0.00000 & -0.20884 \\
2023--2025 & Terminal & Persistent & -0.23297 & +0.33567 & -0.15030 & +0.00000 & -0.04759 \\
2023--2025 & Terminal & Zero & -0.31490 & -0.04660 & -0.00449 & +0.00000 & -0.36599 \\
2026 & Noon & Persistent & -0.14545 & +0.21037 & -0.09471 & +0.00000 & -0.02979 \\
2026 & Noon & Zero & -0.19661 & -0.03410 & -0.00326 & +0.00000 & -0.23397 \\
2026 & Terminal & Persistent & -0.22055 & +0.31758 & -0.14219 & +0.00000 & -0.04516 \\
2026 & Terminal & Zero & -0.29812 & -0.04456 & -0.00442 & +0.00000 & -0.34710 \\
\bottomrule
\end{tabular}

\end{table}

Adverse grid cases occur only with persistent impact and the low equal temporary charges. For the observed-noon signal, three activation laws give a positive gap at scale $1/8$ in both periods. The narrow lognormal law also gives a positive reference-period gap at scale $1/4$. The largest positive observed-noon gaps are $0.000242$ in 2023--2025 and $0.000479$ in 2026. In the terminal channel, the same three laws are positive at scale $1/8$ in both periods, with maxima $0.000412$ and $0.000477$. Broad lognormal dispersion eliminates the positive total in this grid. This dispersion comparison applies to the stated parameter cases. A distribution-free comparative static would require a separate result. For the uniform-cost noon refinement at scale $1/8$, harmful coarse states contribute $0.000284$ and beneficial states contribute $-0.000120$ in the reference period. The corresponding 2026 contributions are $0.000249$ and $-0.000024$. Harmful coarse states carry approximately $49.90\%$ and $75.00\%$ of the respective prior probability. Their economic contributions depend on the sizes of the conditional gaps as well as these probabilities.

The uniform-cost persistent-impact entry threshold is $0.08956$ in model target units. At scale $1/8$, it corresponds to a posterior forecast of about $71.65$ basis points for the complete reset-period return. At unit scale the corresponding value is $8.96$ basis points. Scaling therefore changes the return distribution's location relative to entry while preserving its shape. The raw gaps and their own coarse-bill denominators appear in the main table. The percentage denominator is the constructed execution bill, rather than fund assets.

\subsection{Sampling and reset-reference sensitivity}
The replication includes a five-session circular date-block sensitivity with 2,000 resamples and seed 20261125. Resampling carries all four benchmark paths together. Signal boundaries remain fixed, and every resample recomputes both posterior means under its resampled finite law. The resulting intervals describe sampling uncertainty in the price-informed distribution, conditional on the structural scenario. Uncertainty about the unobserved impact and activation parameters remains outside that calculation.

\begin{table}[htbp]\centering\footnotesize
\caption{Date-block sensitivity for the observed-noon channel and uniform activation costs. Percentile intervals are pointwise and descriptive of the empirical-prior calculation. They are not simultaneous inference over the structural grid.}\label{tab:pricebootstrap}
\begin{tabular}{llrrr}
\toprule
Period & Technology & Scale & Cost change & Date-block 95\% interval \\
\midrule
2023--2025 & Persistent & 0.125 & +0.000164 & $[+0.000013,+0.000237]$ \\
2023--2025 & Persistent & 1 & -0.026362 & $[-0.032329,-0.021944]$ \\
2023--2025 & Zero & 0.125 & -0.002759 & $[-0.003372,-0.002304]$ \\
2023--2025 & Zero & 1 & -0.208836 & $[-0.253434,-0.174698]$ \\
2026 & Persistent & 0.125 & +0.000225 & $[+0.000112,+0.000372]$ \\
2026 & Persistent & 1 & -0.029790 & $[-0.040953,-0.022556]$ \\
2026 & Zero & 0.125 & -0.003072 & $[-0.004245,-0.002344]$ \\
2026 & Zero & 1 & -0.233970 & $[-0.317887,-0.178505]$ \\
\bottomrule
\end{tabular}

\end{table}

The matched daily-reference panel supplies an additional measurement check. It contains 274 common reference-period sessions from November 19, 2024, and all 219 temporal-evaluation sessions. On this unchanged subset, we compare the original return, the price return using the matched daily previous close, and the corresponding cash-distribution-adjusted return. We transform the terminal reset return directly and add each cash distribution once, without reinvesting it through the remaining intraday return. The observed-noon signal uses the corresponding noon reference in each sensitivity.

At unit scale, the persistent-impact gap on the matched reference subset changes from $-0.02985$ to $-0.03142$ with the daily price reference and $-0.03146$ with cash distributions included. The 2026 values are $-0.02979$, $-0.02978$, and $-0.02974$. Under zero impact, the corresponding changes are $-0.23424$, $-0.24952$, and $-0.24976$ in the matched reference period, and $-0.23397$, $-0.23377$, and $-0.23328$ in 2026. Across both channels, both technologies, and all seven scales, each of the 56 scenario signs survives both reference adjustments. A small reference-price change can still affect a finite-law estimate by moving an observation between signal cells. Earlier dates remain a price-proxy sensitivity because the daily corporate-action panel does not cover them.

\paragraph{Terminal-share targets.}
Under the idealized one-period balance sheet, the normalized terminal-share adjustment is proportional to $R_{0,T}/(1+R_{0,T})$. We therefore replace every target observation by
\[
 D^{\mathrm{shares}}_\alpha
 =\frac{\alpha}{0.01}\frac{R_{0,T}}{1+R_{0,T}},
\]
before calculating either posterior mean. The benchmark return states, their probabilities, and both signal partitions remain unchanged. Applying the nonlinear conversion to $\E[R_{0,T}\mid Y]$ instead would change the target. Across the same 56 period--channel--technology--scale scenarios, this conversion preserves every information-gap sign. For the observed-noon signal with persistent impact, the reference-period gap at scale $1/8$ changes from $0.000164$ to $0.000158$, and the 2026 gap from $0.000225$ to $0.000220$. At unit scale the respective gaps change from $-0.026362$ to $-0.026483$ and from $-0.029790$ to $-0.029845$. Under zero impact, unit-scale gaps change from $-0.208836$ to $-0.209716$ and from $-0.233970$ to $-0.234340$. The comparison measures sensitivity to the quantity convention under fixed illustrative execution coefficients. Actual fund share orders remain uncalibrated.

\subsection{Forecast and target-uncertainty details}
The tables below give the full mean and interval summaries behind the compact main table. Range-scaled bands use the 241st order statistic of 252 prior absolute-return scores. The iid continuous-score benchmark is $241/253$, rather than a conditional-coverage guarantee for financial returns. The forecast models and activity increments are evaluated on identical dates, with training-only standardization and the same ridge penalty.

\begin{table}[htbp]\centering\footnotesize
\caption{Declared mean forecasts and predictive intervals. Mean-forecast entries are RMSE relative to zero remaining return. Widths and scores are in basis points.}\label{tab:historydetail}
\begin{tabular}{lrr}
\toprule
Mean forecast & 2023--2025 & 2026 \\
\midrule
Historical mean & 1.0022 & 1.0011 \\
Returns & 1.0062 & 1.0116 \\
Returns + range & 1.0159 & 1.0123 \\
Returns + range + count/size & 1.0168 & 1.0122 \\
Returns + range + volume & 1.0141 & 1.0121 \\
\bottomrule
\end{tabular}
\par\medskip
\begin{tabular}{llrrr}
\toprule
Period & Band & Coverage (\%) & Width (bp) & Score (bp) \\
\midrule
2023--2025 & Raw & 95.99 & 249.40 & 384.96 \\
2023--2025 & Range scaled & 95.64 & 230.50 & 342.38 \\
2023--2025 & RV scaled & 95.82 & 240.76 & 352.51 \\
2026 & Raw & 95.66 & 222.79 & 283.63 \\
2026 & Range scaled & 96.00 & 214.97 & 256.93 \\
2026 & RV scaled & 96.23 & 217.88 & 257.59 \\
\bottomrule
\end{tabular}

\end{table}

To translate uncertainty into target units, put $c_t=10^4a_\beta R_{0,t}$ and $k_t=a_\beta(1+R_{0,t})>0$. If $Y$ is the remaining return in basis points,
\begin{equation}
 D_t^{\mathrm{bp}}=c_t+k_tY,\qquad
 [l_t,u_t]\longmapsto[c_t+k_tl_t,c_t+k_tu_t].
 \label{eq:targetinterval}
\end{equation}
Coverage is unchanged, while width and interval score multiply by $k_t$. Each observation is transformed before averaging. Table~\ref{tab:targetwidth} gives the resulting mean half-widths. Recovering historical dollar orders would also require assets, flows, routing, and prior executions.

\begin{table}[htbp]\centering\footnotesize
\caption{Target-uncertainty half-widths, in percent of hypothetical initial assets, under the no-flow, no-fee fixed-reference exposure proxy.}\label{tab:targetwidth}
\begin{tabular}{lrrrr}
\toprule
Period & $\beta$ & Raw & Range & RV \\
\midrule
2023--2025 & $-3$ & 14.97 & 13.83 & 14.44 \\
2023--2025 & $+3$ & 7.49 & 6.91 & 7.22 \\
2026 & $-3$ & 13.38 & 12.90 & 13.07 \\
2026 & $+3$ & 6.69 & 6.45 & 6.54 \\
\bottomrule
\end{tabular}

\end{table}

\section{Reset-reference measurement audit}\label{app:measurement}
The reset-period proxy spans the preceding close. Corporate actions and the choice of closing reference can therefore affect both its level and its sign. We reconcile the five-minute marks against the unadjusted daily aggregates and corporate-action records used in Appendix~\ref{app:volume}. Their common interval is November 18, 2024 through November 13, 2026. The audit uses the immediately preceding exchange session, including early closes, under the primary start-label convention. Intraday observations within price-response windows remain unchanged.

Let $C_d$ be the unadjusted daily close, $c_d$ the final fresh five-minute mark, and
\[
 S_d=\prod_{u>d}\frac{1}{s_u},
 \qquad
 \delta_d=10^4\log\!\left(\frac{c_d}{S_d C_d}\right),
\]
where $s_u$ is a recorded split ratio. The product is over the corporate actions in the matched record. The four core proxies have $S_d=1$. For SOXX, $S_d=1/3$ before its recorded November 5, 2026 three-for-one split and $S_d=1$ thereafter. The five-minute SOXX prices therefore agree with a backward split adjustment. Applying the split once to the daily comparison removes the change in share units.

\begin{table}[htbp]\centering\footnotesize
\caption{Closing-reference reconciliation. The discrepancy $\delta_d$ compares the last regular-session five-minute mark with the daily close in the same split basis. Median and P99 absolute discrepancies are basis points. The 10-basis-point threshold flags sensitivity rather than declaring an erroneous observation. Ex-dates count recorded cash-distribution dates.}\label{tab:measurementaudit}
\begin{tabular}{lrrrrrr}
\toprule
Proxy & Closes & Median $|\delta|$ & P99 $|\delta|$ & $|\delta|>10$ & Ex-dates & Splits \\
\midrule
SPY & 499 & 0.72 & 5.50 & 3 & 8 & 0 \\
QQQ & 499 & 0.67 & 9.64 & 5 & 8 & 0 \\
IWM & 499 & 0.79 & 4.83 & 1 & 8 & 0 \\
DIA & 499 & 0.61 & 3.46 & 0 & 23 & 0 \\
SOXX & 499 & 1.60 & 14.74 & 26 & 8 & 1 \\
\bottomrule
\end{tabular}

\end{table}

The median absolute discrepancy is below one basis point for each core proxy and 1.60 basis points for SOXX. This agreement supports a split-only price adjustment in the matched interval. A second split correction would adjust the intraday returns twice. The largest core discrepancies are 96.16 basis points for SPY on April 9, 2025 and 74.32 basis points for IWM on June 26, 2026. SOXX differs by 75.34 basis points on its split date. A daily close and a last-bar mark are distinct endpoints. We therefore retain these observations and flag their discrepancies. The audit does not identify an official auction price or an institutional fill.

For a current intraday mark $p_d$, the daily-reference price proxy is
\[
 R_d^{\mathrm{price}}=\frac{p_d}{S_{d-1}C_{d-1}}-1.
\]
A separate total-return sensitivity adds the raw cash distribution $v_d$ in the current split basis exactly once:
\[
 R_d^{\mathrm{total}}
 =\frac{p_d+S_dv_d}{S_{d-1}C_{d-1}}-1.
\]
This formula handles a current-date split because $S_{d-1}=S_d/s_d$. For SOXX, scaling the raw cash amounts by $1/3$ agrees with the independently recorded split-adjusted distributions to less than $4\times10^{-7}$ dollars. The price agreement in Table~\ref{tab:measurementaudit} is inconsistent with a material cumulative dividend adjustment over the matched interval. Cash distributions are therefore added only in this labeled sensitivity, once and without multiplication by the subsequent noon-to-close price return. ETF total return also remains an imperfect proxy for its benchmark index.

There are 493 matched full-session noon observations per core proxy from November 19, 2024 onward. Substituting the daily reference changes the terminal reset-period return by a mean absolute 0.70--1.15 basis points across these proxies. On recorded ex-dates, adding the distribution changes that return by an average 12.58--28.48 basis points. The noon sign changes on 13 of 1,972 proxy-dates under the daily price reference and on 17 when distributions are also included.

We also re-sign the existing close-window outcomes without changing their price windows or eligibility. On common SPY/QQQ/IWM dates in the matched 2024--2025 interval, the original signed post-window mean is $-2.506$ basis points over 274 dates. It becomes $-2.291$ with the daily reference, with the same value after adding distributions. In 2026, the corresponding means over 218 dates are $-0.122$, $-0.136$, and $-0.087$ basis points. These paired measurement sensitivities preserve the finding that the later-period signed movement is small. They provide no additional identification of causal price pressure.

The replication output includes every matched closing mark, discrepancy flag, corrected reset proxy, re-signed event, and SHA-256 input hash. The panel lacks earlier daily and corporate-action records, leaving the full 2021--2024 adjustment history unresolved. We therefore describe the historical variable as a price-based target proxy. The matched-period audit cannot establish a total-return mandate or a consistent historical NAV series.

\section{Data and empirical methods}\label{app:products}
\subsection{Common sample and product scope}
All five-minute histories begin on November 9, 2021 and end on November 13, 2026 for this study. The start is the latest first observation among the 26 geared ETFs and five benchmark proxies in the sample. Earlier observations are excluded from normalization and estimation. Table~\ref{tab:products} reports the common study window. The dates do not denote product inception or the first raw observation. The main forecast exercise uses SPY, QQQ, IWM, and DIA. The joint-market extension adds SOXX, with SOXL and SOXS included in the daily volume and factor analysis.

The family assignments use issuer benchmark descriptions \citep{IssuerMapping2026,DirexionSemiconductor2026}. A proxy ETF's market return need not equal its benchmark index return. The semiconductor extension uses SOXX as an observable sector proxy. USD/SSG are not treated as exact SOXX exposures. Products in commodity, rates, international, and unrelated sector families do not enter these comparisons. Geared products sharing a benchmark are not additional independent realizations of that benchmark's return.

\begin{table}[htbp]\centering\setstretch{1}\footnotesize
\caption{Geared products and their common five-minute sample. Every row starts on November 9, 2021 and ends on November 13, 2026. The same start applies to the five benchmark proxies. The semiconductor family is used in the extensions.}\label{tab:products}
\begin{tabular}{llrr}
\toprule
Product & Benchmark proxy & $\beta$ & Sample start \\
\midrule
SDOW & DIA & $-3$ & 2021-11-09 \\
DXD & DIA & $-2$ & 2021-11-09 \\
DOG & DIA & $-1$ & 2021-11-09 \\
DDM & DIA & $+2$ & 2021-11-09 \\
UDOW & DIA & $+3$ & 2021-11-09 \\
SRTY & IWM & $-3$ & 2021-11-09 \\
TZA & IWM & $-3$ & 2021-11-09 \\
TWM & IWM & $-2$ & 2021-11-09 \\
RWM & IWM & $-1$ & 2021-11-09 \\
UWM & IWM & $+2$ & 2021-11-09 \\
TNA & IWM & $+3$ & 2021-11-09 \\
URTY & IWM & $+3$ & 2021-11-09 \\
SQQQ & QQQ & $-3$ & 2021-11-09 \\
QID & QQQ & $-2$ & 2021-11-09 \\
PSQ & QQQ & $-1$ & 2021-11-09 \\
QLD & QQQ & $+2$ & 2021-11-09 \\
TQQQ & QQQ & $+3$ & 2021-11-09 \\
SOXS & SOXX & $-3$ & 2021-11-09 \\
SOXL & SOXX & $+3$ & 2021-11-09 \\
SPXS & SPY & $-3$ & 2021-11-09 \\
SPXU & SPY & $-3$ & 2021-11-09 \\
SDS & SPY & $-2$ & 2021-11-09 \\
SH & SPY & $-1$ & 2021-11-09 \\
SSO & SPY & $+2$ & 2021-11-09 \\
SPXL & SPY & $+3$ & 2021-11-09 \\
UPRO & SPY & $+3$ & 2021-11-09 \\
\bottomrule
\end{tabular}

\end{table}

\subsection{Time, reference prices, and data provenance}\label{app:provenance}
All UTC timestamps are converted to America/New\_York. The exchange calendar supplies session dates and actual closing times, including ten early closes in each core series. The primary convention treats labels as interval starts. The last available bar is therefore the 11:55 bar at noon and the 15:25 bar at 15:30. The final regular reference is the 15:55 bar on a normal session and the 12:55 bar on an early close. The reset reference always comes from the immediately preceding exchange session, including a half-day. An ineligible full session is not skipped when selecting that reference. Normal-close days alone enter the main outcome samples.

The five-minute records contain prices, volume, weighted average price, and transaction count. The later historical record takes precedence at common symbol--timestamp keys. The original extraction settings and absolute volume units are undocumented. We therefore use the panel for price features, relative activity, and within-window weights. A valid bar has positive, internally ordered OHLC prices and positive count. Weighted-price windows additionally require positive volume and average price within the OHLC range, allowing 0.02 basis points for rounding. These rules are applied uniformly. The separately sourced daily data in Appendix~\ref{app:volume} identify share and dollar units directly.

The end-stamped sensitivity tests an alternative provisional alignment by shifting every relevant label forward by five minutes. This includes prior-session references, decision marks, and outcome windows. The sensitivity does not establish that both alignments are correct. Neither convention identifies an official auction print, NAV, or institutional fill. Prices may also reflect distributions, splits, and historical filtering that are undocumented in the records.

\subsection{Forecast definitions and chronological evaluation}\label{app:intervals}
Let $P_-$ be the previous session's final regular reference, $O$ today's first regular open, $P_t$ the completed noon mark, and $P_T$ the final regular reference. The response is $Y=10^4(P_T/P_t-1)$. The return block contains $10^4(P_t/P_--1)$ and $10^4(O/P_--1)$. The range block adds
\[
 v_t=10^4\sqrt{\sum_{b\leq t}(\log C_b-\log C_{b-1})^2},\qquad
 h_t=10^4\sqrt{\frac{\sum_{b\leq t}(\log H_b-\log L_b)^2}{4\log2}},
\]
where $C_0=O$, and the close's location in the morning high-low span, centered at one half. Both scales have a one-basis-point floor. The high-low statistic is used as a feature. Microstructure noise can prevent it from being an unbiased variance estimator \citep{MartensVanDijk2007}. Activity is the log ratio of cumulative count, volume, or volume/count to its previous 60-observation same-clock median. All 60 observations must exist. Volume and count-plus-size are alternative specifications because their logged levels have an exact accounting dependence.

For each family, the model uses the preceding 252 eligible observations. Training-only centering and scaling precede ridge regression, with an unpenalized intercept and penalty $0.1\|w\|^2$ added to mean squared error. We retain a historical-mean benchmark and zero remaining return. The training window, feature order, and penalty are fixed. They are not selected from the 2026 evaluation results. The reference evaluation is 2023--2025 and the temporal check is January 2--November 13, 2026. Exploratory 2022--2025 analyses informed the design, and some 2026 prices were examined before the specification was fixed. The evaluation is consequently retrospective. It is neither a wholly untouched holdout nor a preregistered test.

All forecasts use the same family-dates. Raw pooled RMSE ratios divide by the zero forecast's RMSE, while paired differences use mean squared loss in squared basis points.

\begin{table}[htbp]\centering\footnotesize
\caption{Annual raw RMSE ratios relative to zero remaining return, on common four-family dates.}\label{tab:annualmean}
\begin{tabular}{lrrrr}
\toprule
Forecast & 2023 & 2024 & 2025 & 2026 \\
\midrule
Historical mean & 1.0020 & 1.0012 & 1.0026 & 1.0011 \\
Returns & 1.0101 & 1.0049 & 1.0055 & 1.0116 \\
Returns + range & 1.0162 & 1.0099 & 1.0183 & 1.0123 \\
Returns + range + count/size & 1.0225 & 1.0137 & 1.0161 & 1.0122 \\
Returns + range + volume & 1.0176 & 1.0101 & 1.0145 & 1.0121 \\
\bottomrule
\end{tabular}

\end{table}
\begin{table}[htbp]\centering\scriptsize
\caption{Paired incremental mean-forecast loss. MSE differences and intervals have units of squared basis points. Each row compares the indicated forecast with its baseline on identical dates.}\label{tab:pairedmean}
\begin{tabular}{lllrr}
\toprule
Period & Forecast & Baseline & $\Delta$ MSE & 95\% interval \\
\midrule
2023--2025 & Returns & Zero & 56.16 & $[7.07,105.52]$ \\
2023--2025 & Returns + range & Returns & 87.78 & $[2.23,224.76]$ \\
2023--2025 & Returns + range + count/size & Returns + range & 8.22 & $[-39.15,43.39]$ \\
2023--2025 & Returns + range + volume & Returns + range & -16.89 & $[-66.12,17.05]$ \\
2026 & Returns & Zero & 57.26 & $[-15.06,136.90]$ \\
2026 & Returns + range & Returns & 3.34 & $[-68.62,67.15]$ \\
2026 & Returns + range + count/size & Returns + range & -0.37 & $[-49.12,44.68]$ \\
2026 & Returns + range + volume & Returns + range & -1.03 & $[-45.34,39.36]$ \\
\bottomrule
\end{tabular}

\end{table}

For scale evaluation, symmetric zero-centered bands use the 241st order statistic of 252 prior absolute scores. Raw scores are $|Y|$, range scores are $|Y|/h_t$, and RV scores are $|Y|/v_t$. Multiplication by the current scale gives the half-width. If these scores were iid and continuous, their rank benchmark would be $241/253=95.26\%$. Financial scores need not satisfy that assumption, so the rank calculation supplies no distribution-free conditional guarantee. With $\alpha=.05$, the interval score is
\[
 \operatorname{IS}_{\alpha}(l,u;y)=(u-l)+\frac{2}{\alpha}(l-y)\1_{\{y<l\}}
 +\frac{2}{\alpha}(y-u)\1_{\{y>u\}},
\]
following \citet{GneitingRaftery2007}. Lower values favor narrow intervals while penalizing misses. Low and high range states use each family's preceding 252-observation median, without a future threshold. State coverage and width weight family-date observations equally. For inference, we first average interval-score differences over the available state observations within each date. A state-specific difference can therefore differ from the difference between pooled state scores.

\begin{table}[htbp]\centering\footnotesize
\caption{Annual interval calibration. Low and high refer to the past-only range state. Scores are basis points.}\label{tab:annualscale}
\begin{tabular}{llrrrrr}
\toprule
Year & Band & Dates & All (\%) & Low (\%) & High (\%) & Score (bp) \\
\midrule
2023 & Raw & 221 & 98.53 & 98.48 & 98.74 & 330.43 \\
2023 & Range & 221 & 96.38 & 95.86 & 98.74 & 322.53 \\
2024 & Raw & 249 & 95.78 & 96.99 & 94.40 & 318.50 \\
2024 & Range & 249 & 95.28 & 93.61 & 97.20 & 321.23 \\
2025 & Raw & 247 & 93.93 & 98.38 & 89.49 & 500.75 \\
2025 & Range & 247 & 95.34 & 94.93 & 95.76 & 381.45 \\
2026 & Raw & 219 & 95.66 & 98.76 & 93.02 & 283.63 \\
2026 & Range & 219 & 96.00 & 95.29 & 96.62 & 256.93 \\
\bottomrule
\end{tabular}

\end{table}

Inference samples 10,000 circular blocks of five normal exchange sessions within each reported period, preserving the original calendar grid and missingness masks. The same date draws carry every family. Primary forecasts use complete four-family dates, and pooled close diagnostics use complete SPY/QQQ/IWM dates. Individual proxy and subgroup statistics retain their own eligible masks. The percentile intervals use the stated date-block approximation. They provide neither a finite-sample coverage guarantee nor familywise error control. The replication outputs give block lengths 1, 10, and 20 for the principal paired comparisons, all predictions and training endpoints, and the state and date counts. Table~\ref{tab:pairedintervals} adds the direct paired range--RV comparison using the same draws. The reference calendar contains 744 normal sessions, with 27 warm-up dates retained as missing, and the 2026 calendar contains 219. The seed is $20261115+n+100L$, where $n$ is calendar length and $L=5$. Section~\ref{app:calibrationdetails} examines concentration without changing the primary sample.

\begin{table}[htbp]\centering\footnotesize
\caption{Paired interval-score differences on common four-family dates. Negative values favor the first method. Units are basis points of prediction loss. Intervals are individual retrospective 95 percent date-block intervals.}\label{tab:pairedintervals}
\begin{tabular}{llrrr}
\toprule
Period & Comparison & Dates & Difference & 95\% interval \\
\midrule
2023--2025 & Range $-$ Raw & 717 & -42.59 & $[-101.07,-3.21]$ \\
2023--2025 & RV $-$ Raw & 717 & -32.45 & $[-88.34,5.54]$ \\
2023--2025 & Range $-$ RV & 717 & -10.13 & $[-15.22,-5.55]$ \\
2026 & Range $-$ Raw & 219 & -26.70 & $[-52.61,-3.28]$ \\
2026 & RV $-$ Raw & 219 & -26.04 & $[-53.95,-0.50]$ \\
2026 & Range $-$ RV & 219 & -0.66 & $[-7.08,6.71]$ \\
\bottomrule
\end{tabular}

\end{table}

\section{Close-related price movements}\label{app:close}
\subsection{Direction, windows, and paired clock comparison}
The event direction is the sign of $P_{15:30}/P_--1$, using the completed 15:30 mark under the primary convention. It is fixed before either price-response window. For both positive and negative leverage objectives, $\beta(\beta-1)>0$, so inverse products do not reverse the inferred direction of idealized underlying reset demand. Define $A$ as the volume-weighted average field over 15:50--16:00 and $P$ as its counterpart over 16:10--16:30. Every bar must have positive count, positive volume, valid prices, and an in-range average. These window calculations assume that average is a WAP and relative volume weights are meaningful.

The pre, post, and cumulative responses are respectively $10^4s\log(A/P_{15:30})$, $10^4s\log(P/A)$, and their sum. The earlier-clock placebo fixes its own direction at 13:30, uses a 13:50--14:00 anchor, and a 14:10--14:30 post window. The clock difference pairs eligible dates and preserves both directions in the bootstrap. This observational clock contrast compares different information sets. It has no interpretation as an exogenous treatment effect.

Nasdaq announced rollout of its Extended Trading Close on March 7, 2022, allowing eligible Nasdaq-listed interest to match at the closing-cross price through 16:05 \citep{NasdaqETC2022}. The main buffered window is separated from the immediate closing neighborhood. A 16:00 bar is not treated as an auction observation. The qualification is venue- and date-specific. The game itself ends at terminal completion and does not predict a post-16:00 decay path.

\begin{table}[htbp]\centering\scriptsize
\caption{Predetermined-direction close diagnostics. Pre means use the same dates as the corresponding post mean. $n_p$ counts dates eligible for both clocks. Intervals use date blocks.}\label{tab:closefamilies}
\begin{tabular}{llrrlrl}
\toprule
Period & Proxy & $n$ & Pre (bp) & Post (bp), 95\% interval & $n_p$ & Clock difference (bp) \\
\midrule
2022--2025 & SPY & 994 & -0.25 & $-1.31$ $[-2.32,-0.33]$ & 994 & $-0.86$ $[-2.38,0.64]$ \\
2022--2025 & QQQ & 994 & -0.48 & $-1.16$ $[-2.60,0.23]$ & 994 & $-1.12$ $[-3.17,0.97]$ \\
2022--2025 & IWM & 992 & -0.74 & $-1.18$ $[-2.26,-0.17]$ & 992 & $-0.60$ $[-2.56,1.32]$ \\
2022--2025 & DIA & 381 & 1.22 & $-0.85$ $[-2.19,0.50]$ & 381 & $-1.45$ $[-3.80,0.97]$ \\
2026 & SPY & 219 & -1.18 & $-0.51$ $[-2.32,1.11]$ & 219 & $0.59$ $[-1.63,2.68]$ \\
2026 & QQQ & 219 & -0.34 & $-0.15$ $[-2.91,2.49]$ & 219 & $1.64$ $[-1.79,5.00]$ \\
2026 & IWM & 218 & -1.30 & $0.29$ $[-1.60,2.28]$ & 218 & $2.55$ $[0.20,5.00]$ \\
2026 & DIA & 96 & 0.87 & $2.14$ $[-0.50,4.73]$ & 96 & $2.64$ $[-0.86,5.97]$ \\
\bottomrule
\end{tabular}

\end{table}
\begin{figure}[htbp]\centering
\includegraphics[width=\textwidth]{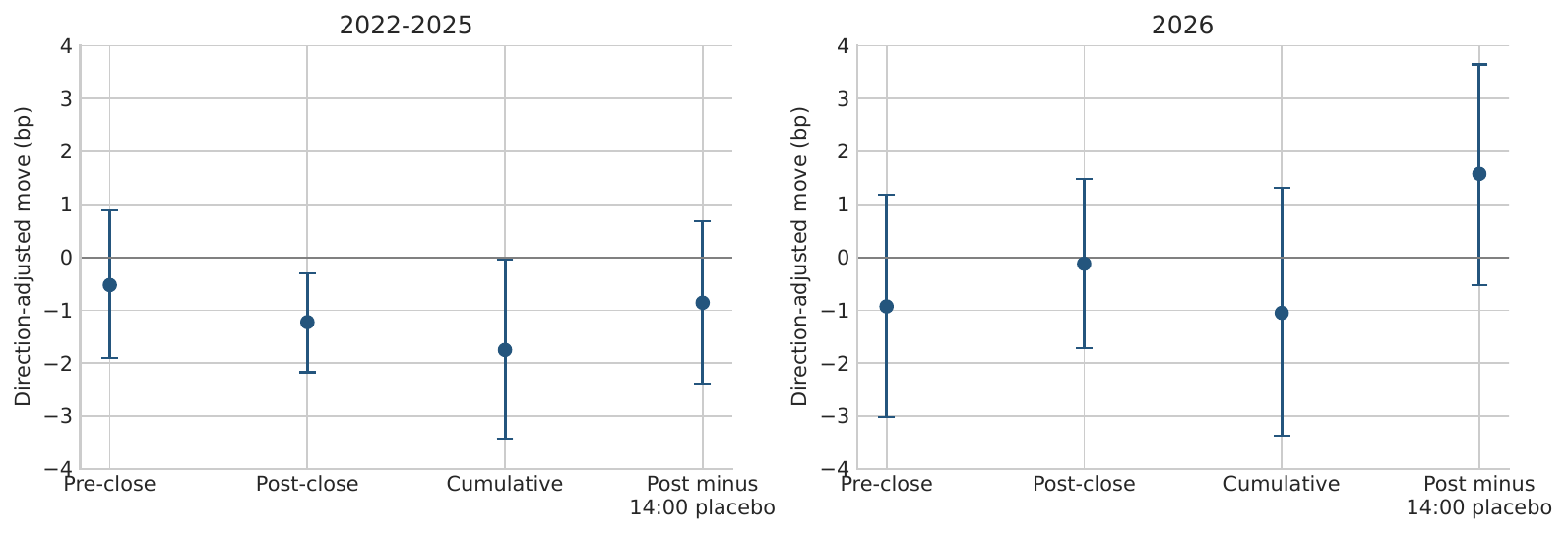}
\caption{Equal-family SPY/QQQ/IWM responses and individual 95 percent date-block intervals. Pre, post, and cumulative use common eligible dates. The clock contrast additionally requires the placebo window. No realized closing return is used to select the direction.}\label{fig:close}
\end{figure}

\subsection{Freshness, anchors, and robustness}
A shared noisy closing endpoint enters the pre return positively and the post return negatively. Their covariance can therefore be negative even without economic reversal. Predetermined direction, distinct window averages, buffered post windows, and cumulative responses reduce the most direct endpoint-selection problem. Microstructure noise can still remain.

For the high-low sensitivity, take the envelope $[L_A,H_A]$ of the 15:55 and 16:00 labeled bars. Given a fresh post reference $P$ and predetermined sign $s$, every anchor in the envelope implies a signed post move between
\[
 \min_{a\in\{L_A,H_A\}}10^4s\log(P/a)
 \quad\text{and}\quad
 \max_{a\in\{L_A,H_A\}}10^4s\log(P/a).
\]
A negative upper bound establishes a move opposite the earlier direction for every price in this recorded envelope. Interpreting it as an auction-price bound would require the official auction price to lie inside the envelope. The recorded bars do not establish that premise, and the bound does not require a positive preceding close move.

\begin{table}[htbp]\centering\footnotesize
\caption{Additional references and anchor sensitivities on common SPY/QQQ/IWM dates. Each horizon additionally requires the main post window, so counts can differ across rows. End-label and last-price sensitivities are paired with the primary eligibility rule.}\label{tab:closehorizons}
\begin{tabular}{llrl}
\toprule
Period & Post reference / sensitivity & Dates & Post move (bp), 95\% interval \\
\midrule
2022--2025 & 16:10 last price & 992 & $-0.96$ $[-1.86,-0.08]$ \\
2022--2025 & 16:10--16:30 WAP & 992 & $-1.23$ $[-2.17,-0.31]$ \\
2022--2025 & 16:40--17:00 WAP & 967 & $-0.89$ $[-2.09,0.27]$ \\
2022--2025 & Next-session 10:00 & 992 & $-0.84$ $[-5.37,3.69]$ \\
2022--2025 & End-label convention & 984 & $-0.96$ $[-1.85,-0.10]$ \\
2022--2025 & Last-price anchors & 992 & $-0.79$ $[-1.70,0.13]$ \\
2026 & 16:10 last price & 218 & $0.69$ $[-0.75,2.18]$ \\
2026 & 16:10--16:30 WAP & 218 & $-0.12$ $[-1.72,1.49]$ \\
2026 & 16:40--17:00 WAP & 217 & $-0.40$ $[-2.25,1.49]$ \\
2026 & Next-session 10:00 & 217 & $-2.43$ $[-10.93,5.55]$ \\
2026 & End-label convention & 216 & $0.13$ $[-1.37,1.67]$ \\
2026 & Last-price anchors & 218 & $0.46$ $[-1.17,2.00]$ \\
\bottomrule
\end{tabular}

\end{table}

Additional diagnostics retain year and sign splits, the large-move group defined by the prior 252-observation 80th percentile with at least 126 observations, leave-one-family-out and leave-one-year-out comparisons, and missingness by earlier direction, range, and activity. Half-days are excluded from the 16:00 exercise. Their small and holiday-selected sample does not provide an exogenous closing-time experiment. The next-session 10:00 reference additionally spans overnight news and potentially distributions, and remains secondary.

\subsection{Predetermined activity and path shape}
A descriptive regression relates the unsigned post-window log return in basis points to the earlier reset-period return $x$ in percentage points, log relative count $z$, $xz$, range scale, overnight return, recent 30-minute return, and family indicators. All regressors are available by 15:30. The last three return/scale controls are in percentage points. Three thousand five-session date-block draws give the interaction interval in Table~\ref{tab:activityinteraction}. Activity measured during the outcome window is excluded from the regressors.
\begin{table}[htbp]\centering\footnotesize
\caption{Predetermined return--activity interaction. Coefficients are basis points of post-window return per percentage point of earlier return times one log unit of relative count.}\label{tab:activityinteraction}
\begin{tabular}{lrrr}
\toprule
Period & Dates & Return $\times$ count & 95\% interval \\
\midrule
2022--2025 & 967 & -0.536 & $[-2.333,1.871]$ \\
2026 & 218 & 4.751 & $[-1.535,9.152]$ \\
\bottomrule
\end{tabular}

\end{table}

A further descriptive breakdown compares path efficiency $|\log(P_t/O)|/\sum_b|\Delta\log C_b|$, using denominator floor $10^{-8}$, above and below $0.25$, within fixed absolute-return bins of 0--50, 50--100, and more than 100 basis points and past-only range states. The replication outputs report all of these small cells without selecting those that favor a mechanism. Transaction count is not the number of strategic entrants, high-low range is not a bid-ask spread, and gross activity does not identify $F$, $N$, or causal inverse depth. Historical window prices likewise do not identify executable profits or the counterfactual impact of a changed institutional schedule.

\FloatBarrier
\section{Additional empirical evidence}\label{app:extended}
\subsection{Forecast stability across families and time}\label{app:forecaststability}
\begin{figure}[htbp]\centering
\includegraphics[width=\textwidth]{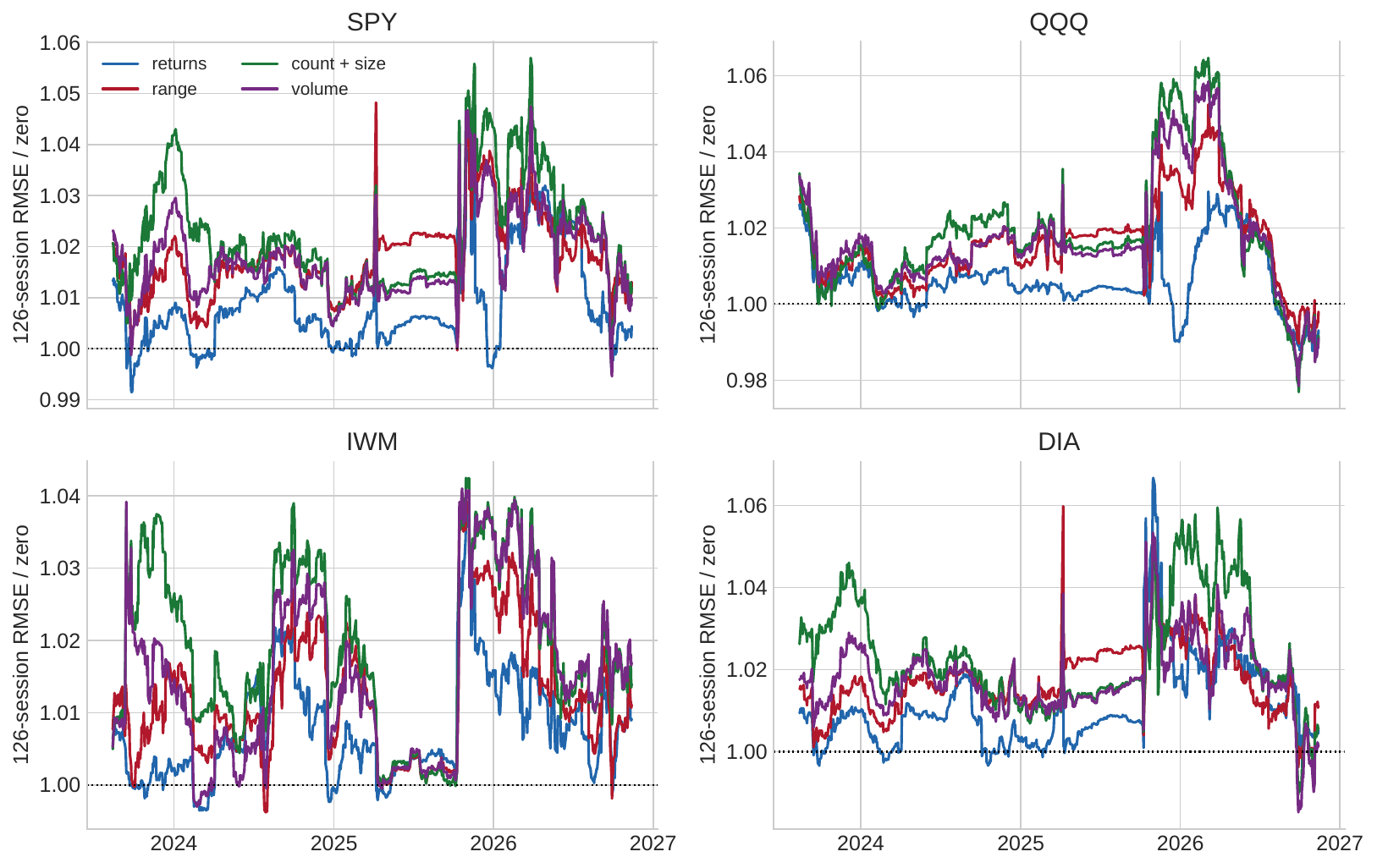}
\caption{Trailing 126-session RMSE ratios. The dotted line is the zero-forecast benchmark. These overlapping evaluation windows are highly dependent. Short periods below one do not establish a stable forecasting advantage.}\label{fig:rollingrmse}
\end{figure}

A pooled forecast score can hide differences across families and short periods of apparent predictability. Table~\ref{tab:familyrmse} gives each family's RMSE relative to the zero remaining-return forecast on the common sample. Figure~\ref{fig:rollingrmse} repeats the comparison in trailing 126-session windows. Every point uses forecasts already formed from earlier training observations. The moving evaluation window describes performance over time. It is not used to select a subsequent trading model.

\begin{table}[htbp]\centering\footnotesize
\caption{Family-level RMSE ratios on the common evaluation dates. A ratio below one favors the fitted forecast. All predictors and training windows are defined in Section~\ref{app:intervals}.}\label{tab:familyrmse}
\begin{tabular}{lllllll}
\toprule
period & symbol & count + size & mean & range & returns & volume \\
\midrule
2023--2025 & DIA & 1.0197 & 1.0017 & 1.0205 & 1.0079 & 1.0158 \\
2023--2025 & IWM & 1.0121 & 1.0027 & 1.0088 & 1.0057 & 1.0093 \\
2023--2025 & QQQ & 1.0197 & 1.0020 & 1.0183 & 1.0061 & 1.0175 \\
2023--2025 & SPY & 1.0176 & 1.0019 & 1.0199 & 1.0059 & 1.0150 \\
2026 & DIA & 1.0181 & 1.0006 & 1.0162 & 1.0152 & 1.0151 \\
2026 & IWM & 1.0152 & 0.9995 & 1.0119 & 1.0094 & 1.0171 \\
2026 & QQQ & 1.0022 & 1.0026 & 1.0090 & 1.0084 & 1.0029 \\
2026 & SPY & 1.0204 & 1.0017 & 1.0157 & 1.0184 & 1.0177 \\
\bottomrule
\end{tabular}

\end{table}

\subsection{Uncertainty calibration beyond a pooled score}\label{app:calibrationdetails}
Methods and calendar masks are defined in Section~\ref{app:intervals}. The range-minus-raw mean score difference is $-7.90$ basis points in 2023, $+2.72$ in 2024, and $-119.30$ in 2025. Removing 2025 leaves a mean of $-2.27$ basis points. The reference-period average therefore depends largely on 2025. The advantage is not uniform across years.

\begin{figure}[htbp]\centering
\includegraphics[width=\textwidth]{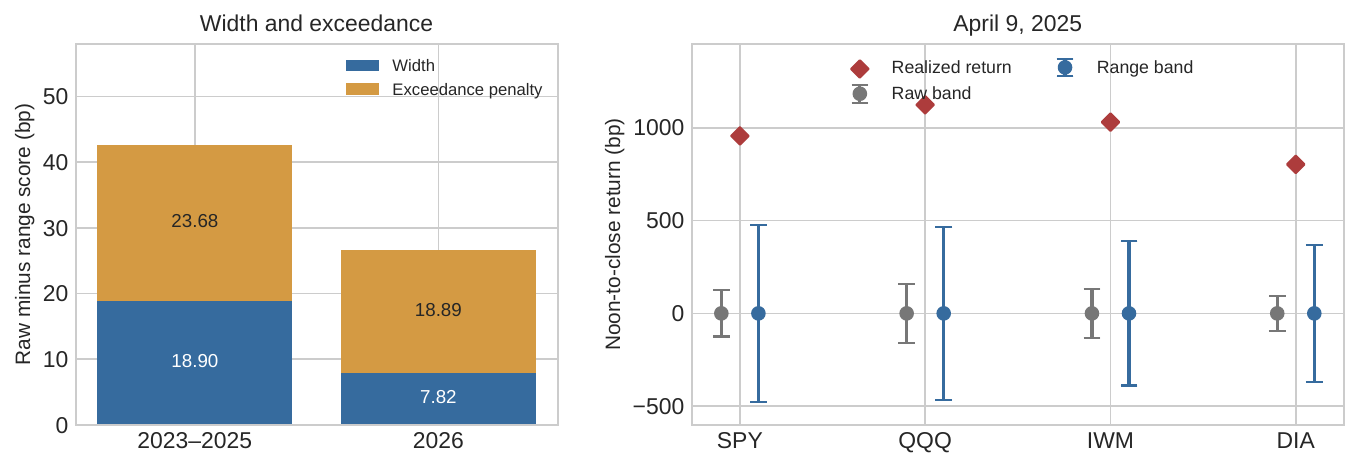}
\caption{Sources of interval-score improvement. Left: mean raw-minus-range width and exceedance-penalty differences, in basis points of prediction loss. Right: realized noon-to-close returns and symmetric half-widths on April 9, 2025. Every return exceeds the range band's upper endpoint. The date is selected retrospectively by its absolute contribution.}\label{fig:scoreconcentration}
\end{figure}

The interval score is full width plus forty times the exceedance beyond the nearest endpoint. In 2023--2025, the $42.59$-basis-point improvement divides into $18.90$ from narrower bands and $23.68$ from reduced exceedance penalties. The corresponding components in 2026 are $7.82$ and $18.89$. Figure~\ref{fig:scoreconcentration} displays this decomposition and the largest absolute-contribution date in the reference period. On April 9, 2025, the four remaining returns range from $803$ to $1{,}123$ basis points. The raw bands have mean full width $254.61$ and the range bands $849.60$ basis points. All four range bands still miss. The wider bands reduce exceedance losses even though they fail to contain the stress move.

Table~\ref{tab:dateinfluence} ranks dates by the absolute four-family mean score difference, allowing both favorable and adverse dates to enter. April 9, 2025 contributes $37.02\%$ of the reference period's net gain. June 11, 2026 contributes $30.38\%$ of the temporal period's gain. Omitting the five largest absolute contributions leaves negative point differences in both periods, but both intervals include zero. These omission results are conditional sensitivity summaries because dates were selected using realized outcomes. They have no interpretation as selection-adjusted tests or preferred trimmed estimates. Every valid date remains in the primary comparison.

\begin{table}[htbp]\centering\footnotesize
\caption{Concentration by date. Share is the selected dates' contribution divided by the full period's net gain. The last columns omit those dates while retaining their calendar positions as missing. The 95 percent intervals condition on the retrospectively selected omission set.}\label{tab:dateinfluence}
\begin{tabular}{lrrrr}
\toprule
Period & Dates omitted & Net-gain share (\%) & Mean (bp) & 95\% interval \\
\midrule
2023--2025 & 1 & 37.02 & -26.86 & $[-55.00,-1.25]$ \\
2023--2025 & 5 & 65.80 & -14.67 & $[-33.51,4.41]$ \\
2026 & 1 & 30.38 & -18.68 & $[-37.87,0.49]$ \\
2026 & 5 & 61.51 & -10.52 & $[-25.33,4.61]$ \\
\bottomrule
\end{tabular}

\end{table}

Table~\ref{tab:targetall} applies \eqref{eq:targetinterval} to all five geared objectives. Positive affine transformation preserves coverage observation by observation and scales squared prediction error by $k_t^2$. Averaging $k_t$ first would generally give a different result. The observed target component is included in the center, even though the remaining-return forecast is zero.

\begin{table}[htbp]\centering\footnotesize
\caption{Normalized target units across leverage objectives. Known component is the mean absolute already-observed target. Other entries are mean interval half-widths. All entries are percent of hypothetical initial assets.}\label{tab:targetall}
\begin{tabular}{lrrrrr}
\toprule
Period & $\beta$ & Known component & Raw & Range & RV \\
\midrule
2023--2025 & $-3$ & 7.76 & 14.97 & 13.83 & 14.44 \\
2023--2025 & $-2$ & 3.88 & 7.49 & 6.91 & 7.22 \\
2023--2025 & $-1$ & 1.29 & 2.50 & 2.30 & 2.41 \\
2023--2025 & $+2$ & 1.29 & 2.50 & 2.30 & 2.41 \\
2023--2025 & $+3$ & 3.88 & 7.49 & 6.91 & 7.22 \\
2026 & $-3$ & 8.81 & 13.38 & 12.90 & 13.07 \\
2026 & $-2$ & 4.41 & 6.69 & 6.45 & 6.54 \\
2026 & $-1$ & 1.47 & 2.23 & 2.15 & 2.18 \\
2026 & $+2$ & 1.47 & 2.23 & 2.15 & 2.18 \\
2026 & $+3$ & 4.41 & 6.69 & 6.45 & 6.54 \\
\bottomrule
\end{tabular}

\end{table}

\begin{figure}[htbp]\centering
\includegraphics[width=\textwidth]{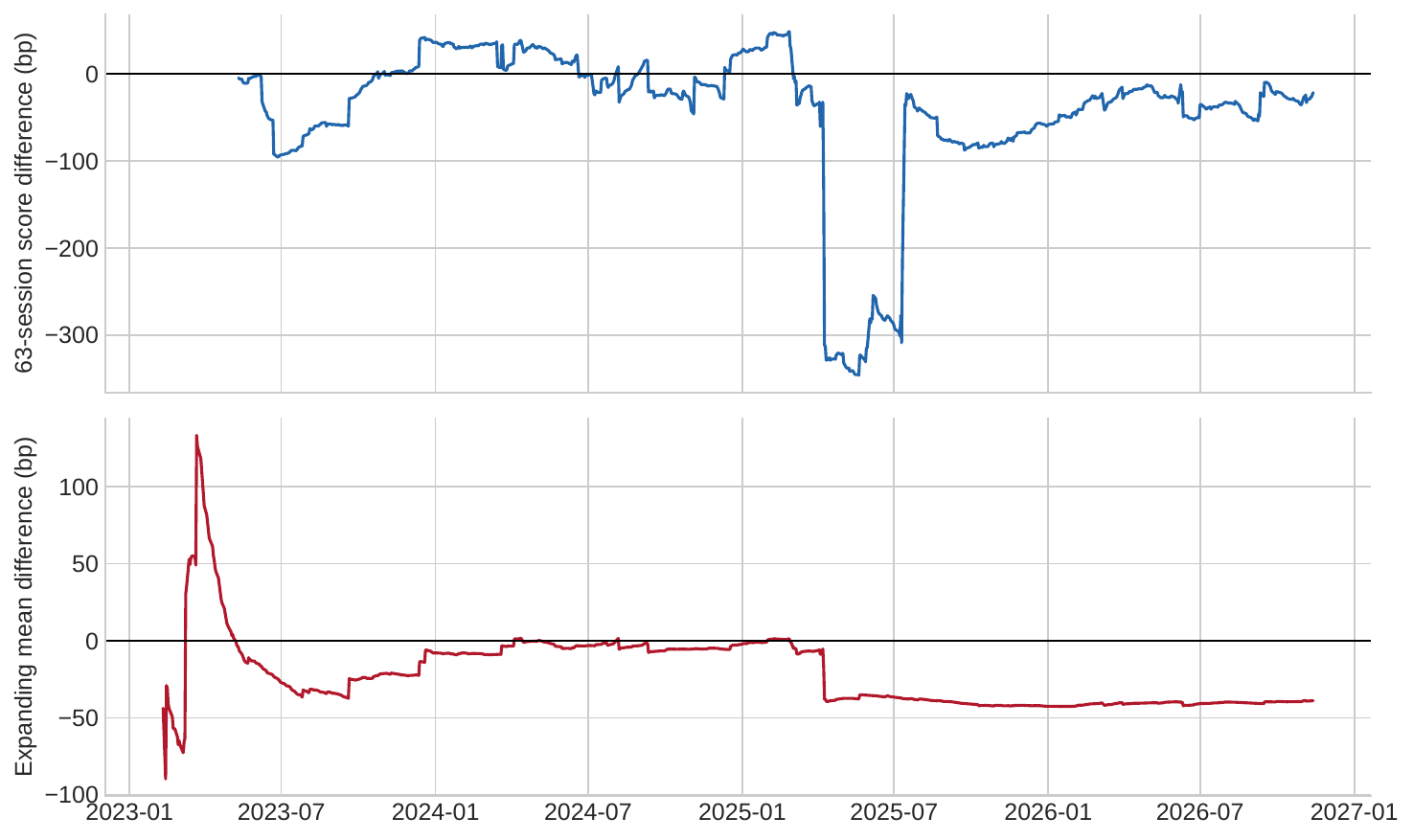}
\caption{Temporal stability of the interval-score difference. Families receive equal weight each day. Both panels report prediction loss, not trading returns. The expanding calculation starts at the first common evaluation date.}\label{fig:scorestability}
\end{figure}

The interval comparison keeps the center fixed and asks whether morning variation improves the scale estimate. Figure~\ref{fig:scorestability} reports the date-level range-minus-raw score difference through time. The upper panel uses a trailing quarter. The lower panel averages all evaluation dates available up to each point. A negative value favors range scaling. The local series can change sign even when the overall mean is negative. This distinguishes an average improvement from uniform performance across market conditions.

A finer state classification ranks today's morning range against the preceding 252 observations for the same family. Fixed rank intervals of width 0.2 define five states without using future outcomes. The replication results retain state-level coverage, counts, and widths. These summaries describe the observed states without assuming that their observations are independent.

\begin{figure}[htbp]\centering
\includegraphics[width=\textwidth]{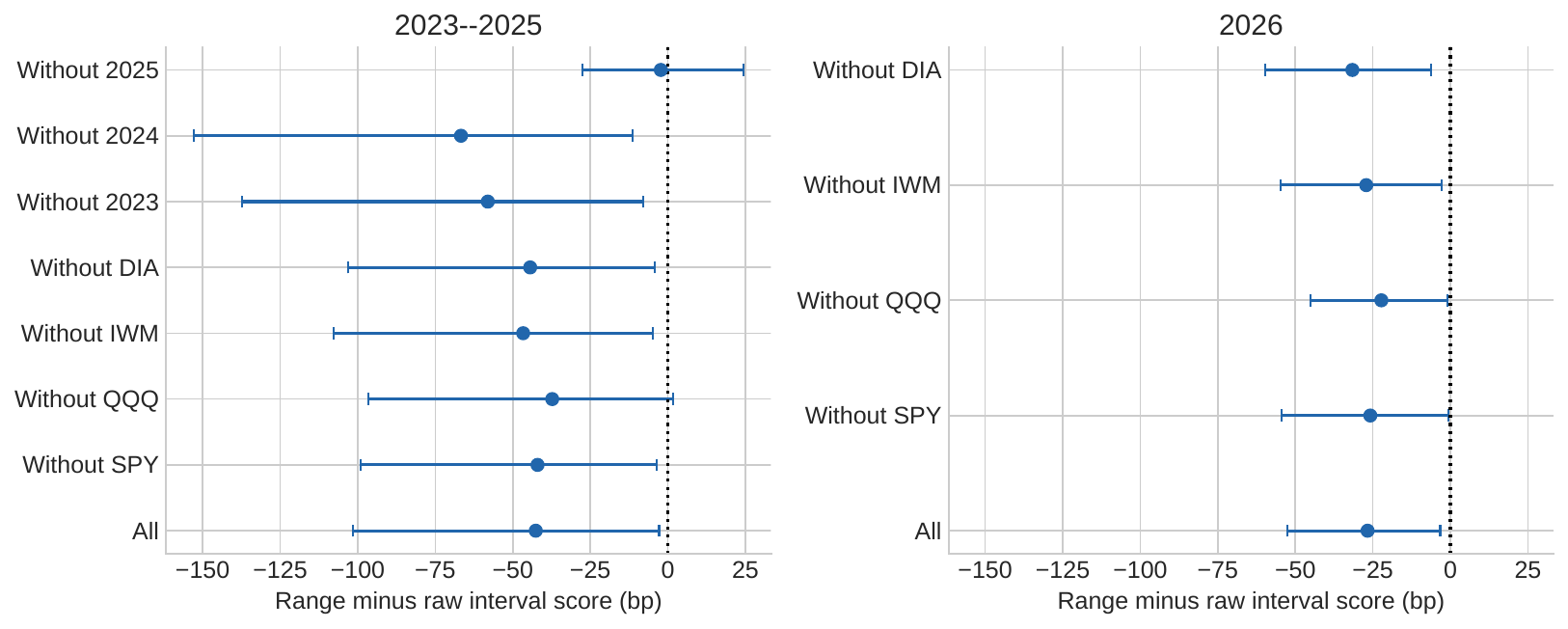}
\caption{Influence of individual families and years on the interval-score comparison. Whiskers are individual 95 percent circular five-session block intervals. The omitted family or year is removed before averaging the remaining families by date. These overlapping sensitivity comparisons are not separate confirmatory tests.}\label{fig:influence}
\end{figure}

Range-scaled intervals widen on mornings with large price movements, but large remaining returns can still exceed them. An overall coverage rate also leaves the dependence between misses unspecified. Figure~\ref{fig:influence} examines the concentration of score improvement by omitting each family and, in the reference sample, each year in turn. These comparisons leave the original forecast and primary sample definition unchanged.

\subsection{Intraday activity}\label{app:activityprofile}
\begin{figure}[htbp]\centering
\includegraphics[width=\textwidth]{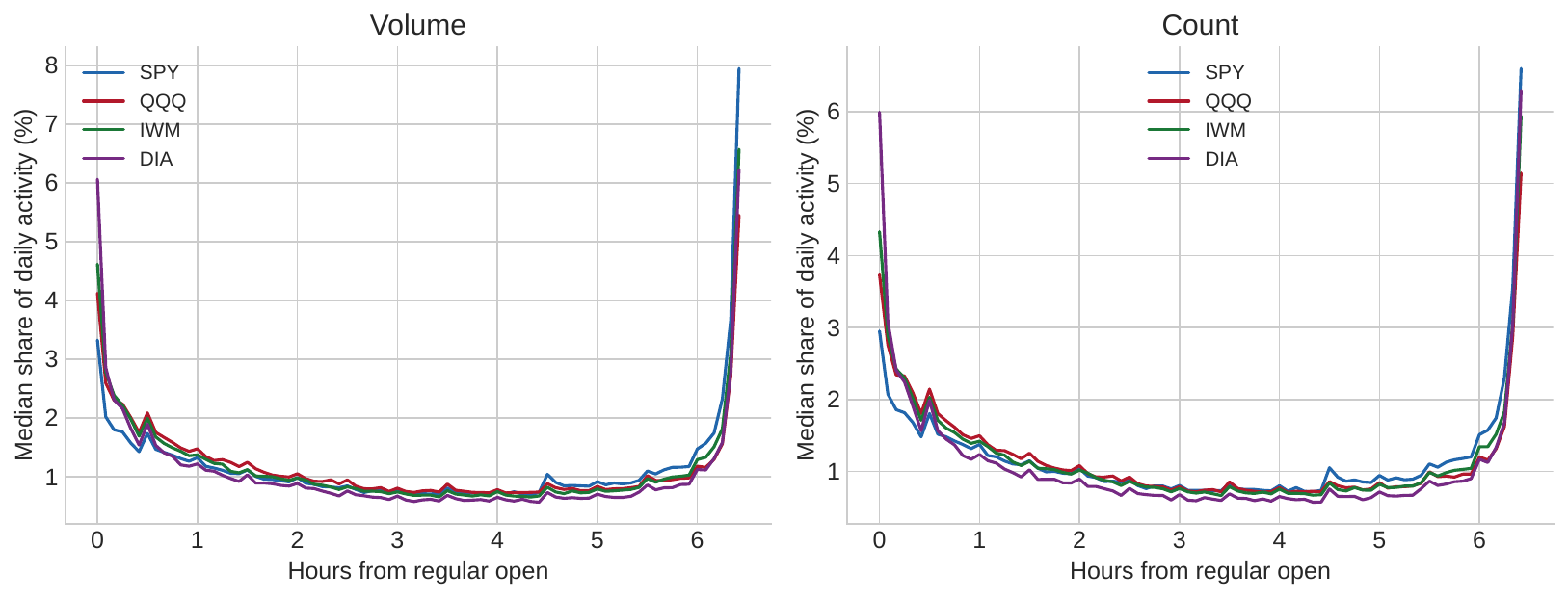}
\caption{Median within-day activity shares on normal sessions, 2022--2026. Each five-minute bar's activity is divided by its regular-session total before taking the median across dates. A median profile need not sum to one. Units cancel within each day.}\label{fig:activityprofile}
\end{figure}

Figure~\ref{fig:activityprofile} reports median intraday shares of regular-session activity. Each day's volume and count are divided by that day's corresponding total. The full-day denominator is confined to this descriptive clock profile. Trading decisions and noon predictors do not use it. Volume and count can have different shapes because average transaction size changes through the day. Neither observable estimates the model's participation mass.

\subsection{Close-related paths, annual variation, and anchor uncertainty}\label{app:paths}
\begin{figure}[htbp]\centering
\includegraphics[width=\textwidth]{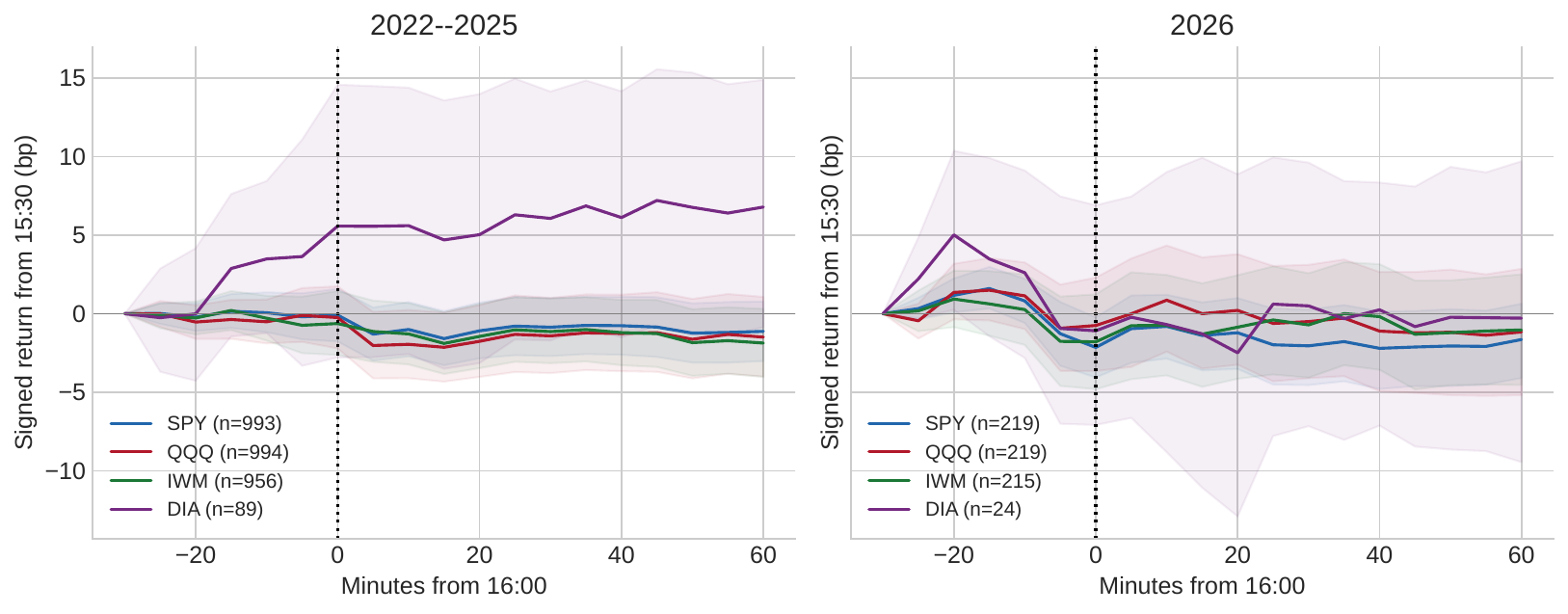}
\caption{Signed last-price paths from the completed 15:30 mark through 17:00. The sign is the earlier reset-period return. Shading gives individual pointwise 95 percent date-block intervals. Sample counts are fixed along each family path. The vertical line marks 16:00 and does not identify an official auction observation.}\label{fig:balancedpaths}
\end{figure}

The main close comparison uses two prespecified windows. Figure~\ref{fig:balancedpaths} adds the intervening path while keeping direction fixed at 15:30. Each family uses dates with valid, positive-count bars throughout the plotted interval. Requiring a balanced path makes the plotted dates comparable. It does not define an executable strategy sample. Its last-price observations also differ from the window-weighted references used in the principal tables.

\begin{figure}[htbp]\centering
\includegraphics[width=\textwidth]{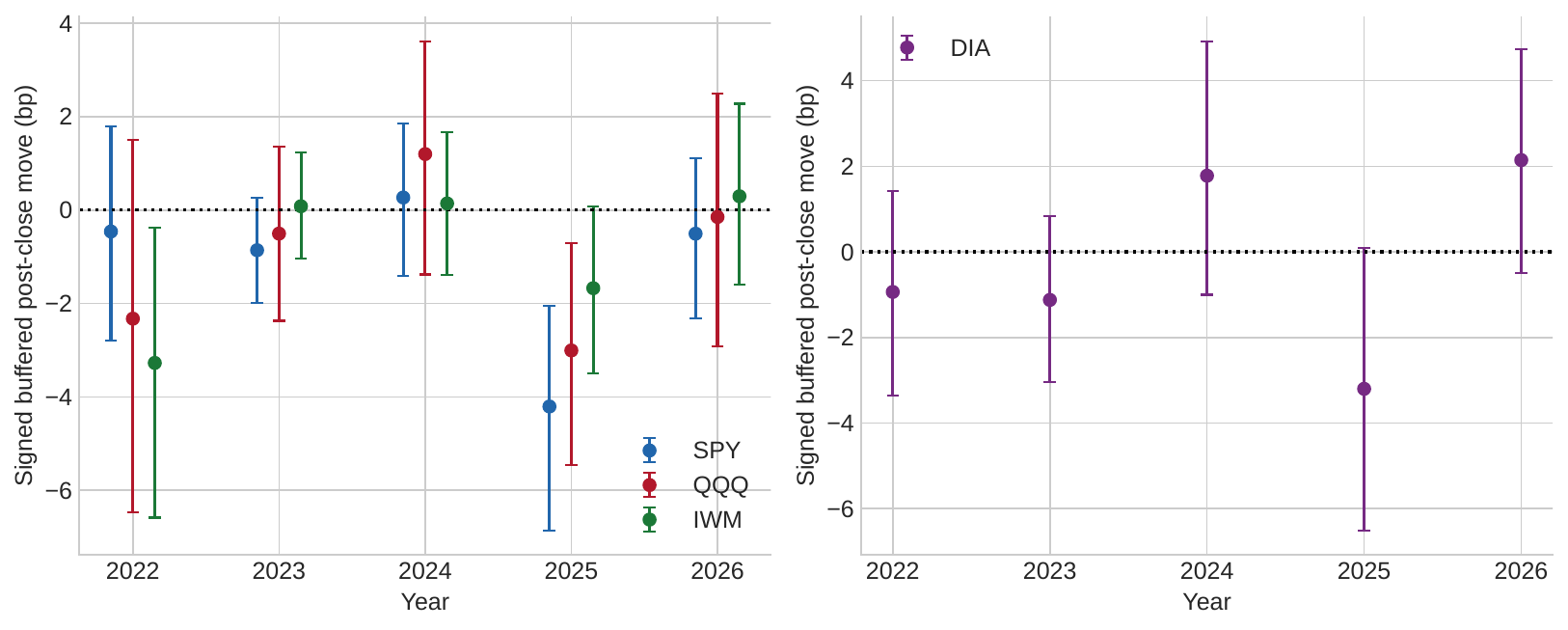}
\caption{Annual signed moves from the 15:50--16:00 reference to the 16:10--16:30 reference. Whiskers are individual 95 percent five-session block intervals. Horizontal offsets distinguish families within each year.}\label{fig:yearpost}
\end{figure}

Annual estimates in Figure~\ref{fig:yearpost} show variation that the full-period mean conceals. The intervals preserve within-family serial dependence through date blocks. They do not identify a causal effect of leveraged-fund demand. With many intervals displayed, an isolated exclusion of zero is weak evidence without an explicit multiple-testing design. DIA's estimates use their own eligible event sample.

\begin{figure}[htbp]\centering
\includegraphics[width=\textwidth]{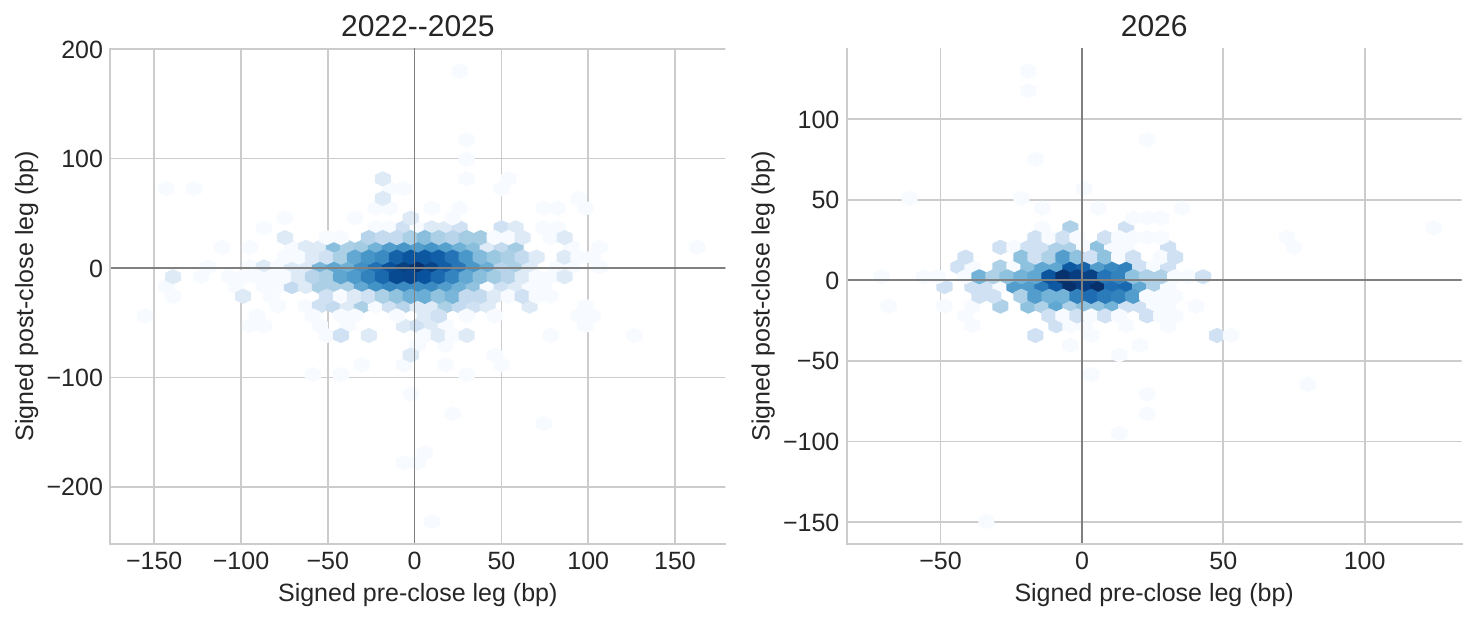}
\caption{Joint distribution of signed pre and post legs for SPY, QQQ, and IWM. Hexagon intensity is logarithmic in observation count. Both legs use the same 15:30 direction and share the close reference, so negative association alone cannot establish reversion.}\label{fig:jointlegs}
\end{figure}

The pre and post legs share the close reference with opposite signs, which affects their joint distribution. Noise in this reference can generate negative covariance even without an economic reversal. Figure~\ref{fig:jointlegs} reports the observed distribution but assigns no structural interpretation to its slope. Separating decision, reference, and post windows limits overlapping trades. The efficient close price remains unidentified by either that separation or the scatterplot.

\begin{figure}[htbp]\centering
\includegraphics[width=\textwidth]{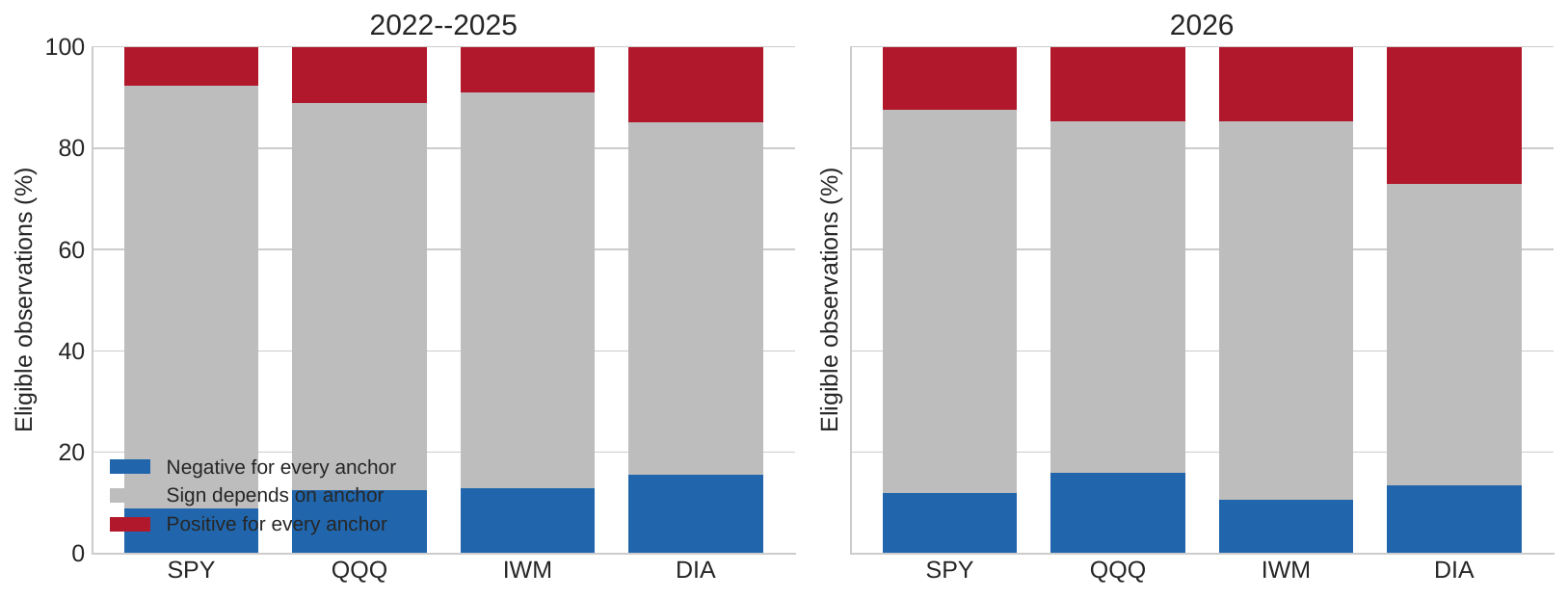}
\caption{Sensitivity of the signed post move to the close reference. Blue and red indicate a negative or positive move for every reference in the high-low envelope. Gray indicates that the sign changes within the envelope. Eligibility requires fresh envelope bars and the primary post window.}\label{fig:anchorclasses}
\end{figure}

Figure~\ref{fig:anchorclasses} asks whether the sign survives every reference in the high-low envelope of the two bars around 16:00. The middle category contains observations for which the inferred sign depends on the chosen reference. The envelope is a sensitivity set built from recorded bars. It is not a confidence interval for the official auction price and need not contain that price.

\FloatBarrier
\section{Econometric robustness}\label{app:econometrics}
\subsection{Serial dependence and multiple comparisons}
The loss comparisons retain the same dates and forecasts as Table~\ref{tab:history}. For each method pair, let $d_{it}$ be its interval-score difference for family $i$ and let $d_t=\frac14\sum_i d_{it}$. Equal weighting within a date prevents four correlated market observations from being treated as four independent experiments. The sign convention is always first method minus second method, so negative values favor the first.

We supplement the circular-block intervals with a Bartlett-kernel HAC covariance estimate \citep{NeweyWest1987}. For a vector of date-level moments $g_t$, define $u_t=g_t-\bar g$ and
\begin{equation}
 \widehat\Omega_L=\frac1n\sum_tu_tu_t^\top+
 \frac1n\sum_{k=1}^L\left(1-\frac{k}{L+1}\right)
 \sum_{t=k+1}^n(u_tu_{t-k}^\top+u_{t-k}u_t^\top).
 \label{eq:hac}
\end{equation}
The estimated covariance of $\bar g$ is $\widehat\Omega_L/n$. The main lag is ten evaluation sessions, with five and twenty as sensitivities. Dates are ordered within each reported evaluation period. These calculations allow contemporaneous dependence across families and short-range dependence across dates. The normal and chi-square reference distributions are asymptotic approximations that require suitable dependence and moment conditions. Heavy-tailed losses and a short temporal sample remain relevant to their accuracy.

Table~\ref{tab:hac} reports two-sided tests with Holm adjustments over six declared hypotheses: the three interval-method contrasts in each of the two periods. We apply the adjustment separately at each lag. Changing the lag assesses sensitivity of the same comparison and provides no independent confirmation. The reference-period range-versus-RV difference remains significant after adjustment. The range-versus-raw improvements are less decisive under HAC inference than under the percentile bootstrap. The reference-period HAC interval includes zero. The disagreement is consistent with the concentration of score gains in Appendix~\ref{app:calibrationdetails}. It limits the strength of the range-versus-raw comparison.

\begin{table}[htbp]\centering\footnotesize\setlength{\tabcolsep}{4pt}
\caption{Interval-score comparisons with ten-session HAC inference. Means and intervals are in basis points of prediction loss. Holm adjustment covers all six rows.}\label{tab:hac}
\begin{tabular}{llrrrrr}
\toprule
Period & Score contrast & Dates & Mean & HAC 95\% interval & $p$ & Holm $p$ \\
\midrule
2023--2025 & Range $-$ Raw & 717 & -42.59 & [-93.55, 8.38] & 0.101 & 0.304 \\
2023--2025 & Rv $-$ Raw & 717 & -32.45 & [-81.18, 16.27] & 0.192 & 0.383 \\
2023--2025 & Range $-$ Rv & 717 & -10.13 & [-14.94, -5.33] & $<0.001$ & $<0.001$ \\
2026 & Range $-$ Raw & 219 & -26.70 & [-50.44, -2.96] & 0.027 & 0.137 \\
2026 & Rv $-$ Raw & 219 & -26.04 & [-51.88, -0.21] & 0.048 & 0.193 \\
2026 & Range $-$ Rv & 219 & -0.66 & [-7.81, 6.49] & 0.857 & 0.857 \\
\bottomrule
\end{tabular}

\end{table}
\begin{figure}[htbp]\centering
\includegraphics[width=\textwidth]{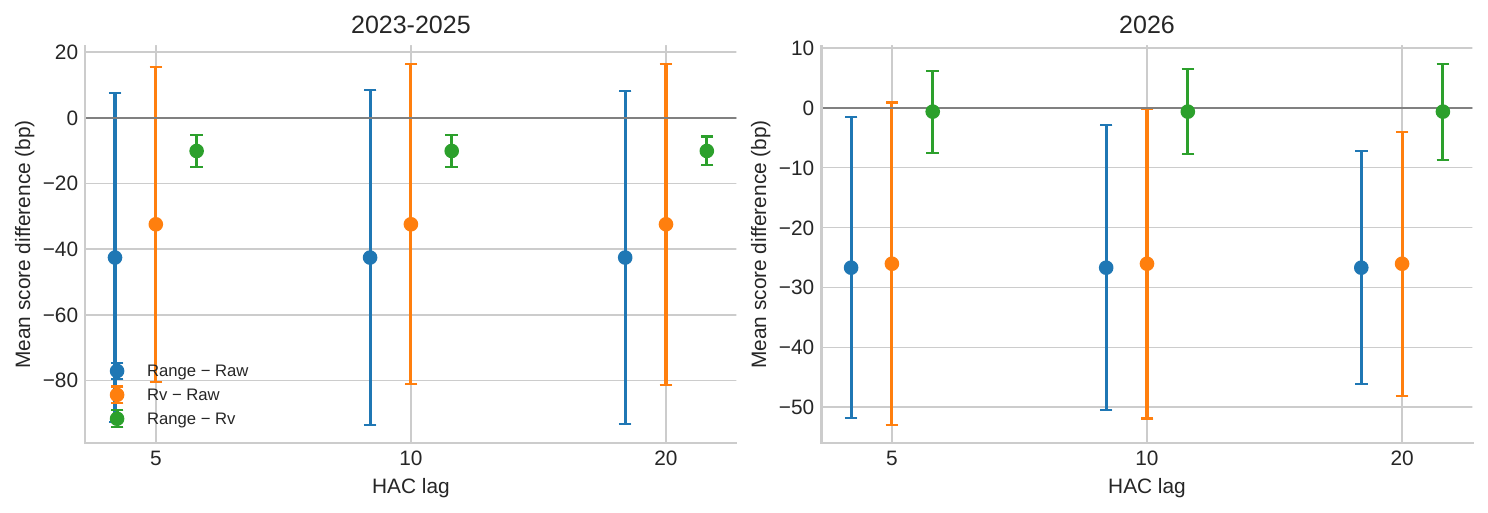}
\caption{Sensitivity to the HAC lag. Whiskers are individual asymptotic 95 percent intervals, before multiple-testing adjustment. The same forecasts and dates enter every lag choice.}\label{fig:hac}
\end{figure}

\subsection{Conditional loss and calibration moments}
Conditional forecast comparison asks whether observable conditions predict loss differences, extending the comparison of unconditional means \citep{GiacominiWhite2006}. Let $H_{it}$ indicate a morning range above its own preceding 252-observation median. For each method pair we test
\[
 \E\left[\frac14\sum_i d_{it}
 \begin{pmatrix}1\\H_{it}-1/2\end{pmatrix}\right]=0.
\]
The two-moment Wald statistic uses \eqref{eq:hac} with $L=10$. A rejection means that these moments distinguish the forecast procedures. Dominance in every state would be a stronger claim. The reference-period range-versus-RV contrast and temporal range-versus-raw contrast reject after the six-comparison adjustment in Table~\ref{tab:conditionalloss}.

\begin{table}[htbp]\centering\footnotesize
\caption{Conditional predictive-loss moments. The instruments are a constant and the observable morning range state. The reference distribution is $\chi^2_2$, and Holm adjustment covers six tests.}\label{tab:conditionalloss}
\begin{tabular}{llrrr}
\toprule
Period & Score contrast & $W_2$ & $p$ & Holm $p$ \\
\midrule
2023--2025 & Range $-$ Raw & 4.71 & 0.095 & 0.285 \\
2023--2025 & Rv $-$ Raw & 2.77 & 0.251 & 0.502 \\
2023--2025 & Range $-$ Rv & 19.23 & $<0.001$ & $<0.001$ \\
2026 & Range $-$ Raw & 9.43 & 0.009 & 0.045 \\
2026 & Rv $-$ Raw & 7.52 & 0.023 & 0.093 \\
2026 & Range $-$ Rv & 2.39 & 0.302 & 0.502 \\
\bottomrule
\end{tabular}

\end{table}

For calibration, write $I_{it}=\1\{|Y_{it}|>h_{it}\}$ for a miss by the relevant symmetric band. We test moments of $I_{it}-.05$ against three instruments: a constant, $H_{it}-1/2$, and the preceding evaluation date's average absolute noon-to-close return, expressed in percentage points and centered within the evaluation period. The last instrument is already known at today's noon decision. The first date of each period is excluded from this particular test, leaving 716 and 218 dates. All four family contributions are averaged before forming the HAC covariance. The joint null is assessed with three degrees of freedom, and Holm adjustment covers the three bands in each of the two periods.

The raw band rejects these calibration moments in both periods. Range scaling does not reject after the six-test adjustment, with adjusted $p$-values $0.083$ and $0.814$. The tests therefore find departures from the specified raw-band restrictions, with insufficient evidence to reject the range-band restrictions. Failure to reject does not establish conditional validity, independent errors, or an exact 95 percent guarantee.

\begin{table}[htbp]\centering\footnotesize
\caption{Conditional coverage moments. Coverage is calculated on the moment-test sample, which excludes the first date of each period. $W_3$ tests a constant, the morning range state, and lagged market movement.}\label{tab:conditionalcoverage}
\begin{tabular}{llrrrrr}
\toprule
Period & Band & Dates & Coverage (\%) & $W_3$ & $p$ & Holm $p$ \\
\midrule
2023--2025 & Raw & 716 & 95.98 & 26.08 & $<0.001$ & $<0.001$ \\
2023--2025 & Range & 716 & 95.64 & 9.12 & 0.028 & 0.083 \\
2023--2025 & Rv & 716 & 95.81 & 13.18 & 0.004 & 0.017 \\
2026 & Raw & 218 & 95.64 & 25.44 & $<0.001$ & $<0.001$ \\
2026 & Range & 218 & 95.99 & 1.40 & 0.706 & 0.814 \\
2026 & Rv & 218 & 96.22 & 2.90 & 0.407 & 0.814 \\
\bottomrule
\end{tabular}

\end{table}

\subsection{Training-window sensitivity}
The scale comparison is repeated with 126, 252, and 504 preceding eligible observations. For window length $n$, the empirical quantile is order $\lceil .95(n+1)\rceil$. Every method and window is evaluated on identical four-family dates. The longest window determines the first comparison date, February 15, 2024, leaving 465 reference dates and 219 temporal dates. Thus the differences in Table~\ref{tab:windowrobustness} do not reflect changing evaluation composition.

Range scaling has a lower point score than the raw band for every window. The reference-period inference remains sensitive to the window and tail observations. The windows are fixed sensitivity choices and are not ranked to select the most favorable result. The 252-observation specification remains the primary one.

\begin{table}[htbp]\centering\footnotesize\setlength{\tabcolsep}{4pt}
\caption{Alternative training windows on common evaluation dates. Coverage is in percent and score differences are range minus raw, in basis points. The HAC intervals are individual ten-session intervals.}\label{tab:windowrobustness}
\begin{tabular}{lrrrrrr}
\toprule
Period & Window & Dates & Raw cov. & Range cov. & $\Delta$ score & HAC 95\% interval \\
\midrule
2024--2025 & 126 & 465 & 94.78 & 95.86 & -61.12 & [-119.28, -2.96] \\
2024--2025 & 252 & 465 & 94.62 & 95.59 & -61.20 & [-135.70, 13.29] \\
2024--2025 & 504 & 465 & 94.84 & 96.18 & -64.79 & [-148.38, 18.81] \\
2026 & 126 & 219 & 94.86 & 95.66 & -25.99 & [-51.45, -0.54] \\
2026 & 252 & 219 & 95.66 & 96.00 & -26.70 & [-50.44, -2.96] \\
2026 & 504 & 219 & 96.35 & 96.00 & -21.42 & [-41.81, -1.03] \\
\bottomrule
\end{tabular}

\end{table}

\FloatBarrier
\section{Joint benchmark forecasts}\label{app:jointempirical}
\subsection{Own-market and joint conditional means}
We study two systems: QQQ/SPY/IWM and QQQ/SOXX. The first captures common broad-equity variation with a small-cap component. The second contrasts Nasdaq-100 and semiconductor exposure. Each equation's own-market predictor vector contains elapsed and overnight return, close-return RV, high-low range, close location, relative count, and relative transaction size. The joint specification adds the same completed-noon features from every other market in its system. Both use the preceding 252 common eligible dates, training-only standardization, an unpenalized intercept, and the same ridge penalty $0.1$. The joint specification expands the predictor set while excluding contemporaneous afternoon outcomes and future normalization from training.

The two systems are reported on common evaluation dates beginning February 13, 2023. Each own-versus-joint comparison uses identical target observations. The own and joint columns in Table~\ref{tab:jointforecast} are pooled raw RMSE divided by the zero-forecast RMSE. For inference, squared-error differences are divided by each equation's training-window target variance, then averaged across markets within the date. The resulting four period-by-system tests use ten-session HAC inference and a four-test Holm adjustment.

Joint mean forecasts do not improve point RMSE in either system or period. In the three-market reference comparison, the standardized loss increase remains significant after adjustment. These results provide no support for a return-trading rule based on the larger predictor set. Any different forecast procedure would need independent validation.

\subsection{Joint uncertainty}
The covariance comparison holds the mean at zero and scales each outcome by its own observed morning range. Let $v_t$ be this vector and let
\[
 M_t=\frac1{252}\sum_{s=t-252}^{t-1}v_sv_s^\top,\qquad
 D_t=\operatorname{diag}(M_t),\qquad
 \Sigma_t=.9M_t+.1D_t.
\]
The diagonal model uses $D_t$ and the joint model uses $\Sigma_t$. Shrinkage is fixed at ten percent. The zero-mean Gaussian negative log score is
\[
 \ell_t(\Sigma)=\tfrac12\{p\log(2\pi)+\log\det\Sigma+v_t^\top\Sigma^{-1}v_t\}.
\]
Restoring the original return units adds the same sum of log scales to both models, so the score difference is unchanged. The Gaussian score is used to evaluate the joint distribution without assuming Gaussian observed tails. Their evaluation uses the realized out-of-sample return vectors.

Both rolling joint forecasts improve on diagonal covariance. To separate the value of dependence from the value of updating it, we add a matched frozen-correlation forecast. Let $Q_0$ be the uncentered second moment of the standardized outcomes on the initial 252 common eligible dates, February 9, 2022 through February 10, 2023, and let $D_0=\operatorname{diag}(Q_0)$. Define
\begin{equation}
 R_0=D_0^{-1/2}Q_0D_0^{-1/2},\qquad
 \Sigma_t^{\rm frozen}=D_t^{1/2}(.9R_0+.1I)D_t^{1/2}.\label{eq:frozencovariance}
\end{equation}
The initial correlation estimate stays fixed throughout evaluation, including 2026. Both correlated models retain the same rolling marginal second moments, morning-range scaling, zero mean, ten-percent shrinkage, and evaluation dates. Only the correlation update differs. No evaluation outcome enters $R_0$. This additional benchmark was designed retrospectively. Its correlation window and shrinkage choice were not selected from alternatives.

Table~\ref{tab:jointforecast} retains the conditional-mean comparison. Table~\ref{tab:frozencorrelation} separates rolling and frozen dependence from independent marginals. Frozen correlation has the lower point score in both systems and periods. The rolling-minus-frozen differences for QQQ/SPY/IWM are inconclusive. For QQQ/SOXX, they are positive in both periods and remain significant after Holm adjustment over the four system-period comparisons. Each contrast has its own four-test family, with individual ten-session HAC intervals. Table~\ref{tab:directionalrisk} translates these covariance forecasts into two fixed portfolio directions.

Persistent correlation alone can produce a substantial improvement over independent marginals. If a zero-mean vector has covariance $D^{1/2}RD^{1/2}$, with $R$ a positive-definite correlation matrix, then
\begin{equation}
 \E\{\ell(D^{1/2}RD^{1/2},v)-\ell(D,v)\}
 =\tfrac12\log\det R.\label{eq:correlationscoreidentity}
\end{equation}
The two expected quadratic terms both equal the vector dimension, leaving the log determinant. This expectation identity uses the stated second moments and does not require Gaussian returns. The evidence supports accounting for dependence when describing joint uncertainty. These specifications provide no evidence that updating correlation adds predictive value.

Return covariance and structural cross-impact describe different objects. The execution matrices in Appendix~\ref{app:jointmodel} determine how orders affect prices. At a fixed posterior mean and execution environment, covariance enters the conditional expected institutional bill through $\operatorname{tr}(\Lambda V)$. Equilibrium quantities remain unchanged. The joint forecast exercise describes target uncertainty and does not estimate cross-impact or activation.

\begin{table}[htbp]\centering\footnotesize\setlength{\tabcolsep}{4pt}
\caption{Own-market and joint conditional means, with the original covariance comparison. RMSE is relative to the zero-forecast RMSE. The final columns report rolling joint minus diagonal negative log score, in natural-log units per return vector. Negative scores favor joint covariance.}\label{tab:jointforecast}
\begin{tabular}{llrrrrr}
\toprule
System & Period & Dates & Own RMSE & Joint RMSE & $\Delta$ NLL & HAC 95\% interval \\
\midrule
QQQ/SPY/IWM & 2023--2025 & 716 & 1.0162 & 1.0384 & -1.564 & [-1.795, -1.334] \\
QQQ/SPY/IWM & 2026 & 219 & 1.0111 & 1.0304 & -1.360 & [-1.579, -1.141] \\
QQQ/SOXX & 2023--2025 & 716 & 1.0160 & 1.0180 & -0.734 & [-0.819, -0.650] \\
QQQ/SOXX & 2026 & 219 & 1.0164 & 1.0215 & -0.630 & [-0.741, -0.519] \\
\bottomrule
\end{tabular}

\end{table}
\begin{table}[htbp]\centering\scriptsize\setlength{\tabcolsep}{3pt}
\caption{Frozen and rolling correlation on matched dates. Both retain rolling marginal variances. Positive rolling-minus-frozen values favor the frozen estimate. Intervals and Holm values concern that final contrast. The intervals are individual, and the Holm family contains its four system-period tests. Scores are natural-log units per return vector.}\label{tab:frozencorrelation}
\begin{tabular}{llrrrrrr}
\toprule
System & Period & Dates & Rolling $-$ diag. & Frozen $-$ diag. & Rolling $-$ frozen & HAC 95\% interval & Holm $p$ \\
\midrule
QQQ/SPY/IWM & 2023--2025 & 716 & -1.5643 & -1.5814 & +0.0170 & [-0.0318, +0.0659] & 0.9890 \\
QQQ/SPY/IWM & 2026 & 219 & -1.3604 & -1.4063 & +0.0459 & [-0.1231, +0.2149] & 0.9890 \\
QQQ/SOXX & 2023--2025 & 716 & -0.7342 & -0.7789 & +0.0447 & [+0.0326, +0.0569] & $<.001$ \\
QQQ/SOXX & 2026 & 219 & -0.6302 & -0.7052 & +0.0750 & [+0.0348, +0.1152] & $<.001$ \\
\bottomrule
\end{tabular}

\end{table}
\paragraph{Risks of a fixed sum and spread.}
For QQQ and SOXX, fix $w_+=(.5,.5)^\top$ and $w_-=(.5,-.5)^\top$. Let $y_t$ contain unscaled remaining returns in basis points and $S_t$ the diagonal of current completed-morning ranges. Return covariance is $\widehat V_t=S_t\Sigma_tS_t$. Table~\ref{tab:directionalrisk} compares observed RMS $\{\overline{(w^\top y_t)^2}\}^{1/2}$ with forecast RMS $\{\overline{w^\top\widehat V_tw}\}^{1/2}$. Forecast RMS takes the square root after averaging the variances. It therefore differs from the mean daily standard deviation. A diagonal forecast assigns equal variance to sum and spread. Dependence distinguishes the larger common-direction risk from the smaller spread risk.

\begin{table}[htbp]\centering\scriptsize\setlength{\tabcolsep}{3pt}
\caption{Fixed QQQ/SOXX directions on matched evaluation dates. Observed and forecast RMS values are basis points. The final columns are mean projected Gaussian quasi-score differences, with lower scores preferred. Weights are $(.5,.5)$ for the sum and $(.5,-.5)$ for the spread.}\label{tab:directionalrisk}
\begin{tabular}{llrrrrrrr}
\toprule
Period & Direction & Dates & Observed & Diagonal & Rolling & Frozen & Fr. $-$ diag. & Roll. $-$ fr. \\
\midrule
2023--2025 & Sum & 716 & 93.59 & 67.99 & 90.09 & 90.98 & -0.067284 & -0.000497 \\
2023--2025 & Spread & 716 & 26.31 & 67.99 & 33.59 & 31.11 & -0.480910 & +0.021314 \\
2026 & Sum & 219 & 86.19 & 73.07 & 91.62 & 93.62 & -0.052986 & -0.000801 \\
2026 & Spread & 219 & 36.12 & 73.07 & 47.78 & 43.74 & -0.257355 & +0.017973 \\
\bottomrule
\end{tabular}

\end{table}

The projected score is $\{\log(2\pi s^2_{w,t})+(w^\top y_t)^2/s^2_{w,t}\}/2$, where $s^2_{w,t}=w^\top\widehat V_tw$. Frozen correlation improves on diagonal forecasts in both directions and periods. Rolling correlation scores slightly better than frozen correlation for the sum and worse for the spread. The ranking therefore depends on the projection. These retrospective fixed-weight diagnostics introduce no additional significance tests, conditional-calibration claims, or trading rule. The return directions do not identify the model's harmful or beneficial signals.

\FloatBarrier
\section{Share volume and dollar turnover}\label{app:volume}
\subsection{Measurement on a common daily panel}
Daily unadjusted OHLC, share volume, VWAP, cash distributions, and splits are obtained from Massive \citep{Massive2026}. The balanced panel contains all 31 ETFs on 499 dates from November 18, 2024 through November 13, 2026. This is the common available window for the external daily series. The daily window is shorter than the five-minute study but is the same for every product in the turnover analysis.

Let $V_{it}$ be reported shares traded and $P^{\rm VWAP}_{it}$ the daily volume-weighted price. Dollar turnover is
\begin{equation}
 T_{it}=V_{it}P^{\rm VWAP}_{it}.\label{eq:dollarturnover}
\end{equation}
Share volumes retain each date's actual share basis. A split or reverse split changes those quantities mechanically. Adding them across unrelated ETFs would not measure economic exposure. Dollar turnover avoids that mechanical denomination effect. The data describe trades eligible for the provider's daily aggregates. They are secondary-market trading measures, not fund creations, redemptions, derivative notionals, or underlying-basket executions.

Table~\ref{tab:volumeproducts} reports mean shares, mean dollars, and the upper tail of daily dollar activity for every ETF. Table~\ref{tab:volumefamilies} sums geared-product dollar volume within each family, compares it with benchmark-ETF dollar volume, and reports the inverse products' share. Each family denominator is its proxy ETF's turnover, rather than total trading in all underlying constituents. The dollar sums measure recorded trading across distinct securities. They do not identify distinct investors or independent risk transfers.

\begin{table}[htbp]\centering\footnotesize\setlength{\tabcolsep}{4pt}
\caption{Daily share and dollar volume for all 31 ETFs. Shares and dollars are in millions. P95 is the empirical 95th percentile across the same 499 dates. Raw share quantities retain the contemporaneous share denomination.}\label{tab:volumeproducts}
\begin{tabular}{llrrrr}
\toprule
ETF & Family & $\beta$ & Mean shares (m) & Mean dollars (m) & P95 dollars (m) \\
\midrule
SDOW & DIA & $-3$ & 4.014 & 141.6 & 306.2 \\
DXD & DIA & $-2$ & 1.695 & 35.1 & 99.9 \\
DOG & DIA & $-1$ & 3.252 & 76.8 & 236.9 \\
DIA & DIA & $+1$ & 4.503 & 2,129.9 & 4,173.6 \\
DDM & DIA & $+2$ & 0.279 & 21.5 & 59.6 \\
UDOW & DIA & $+3$ & 2.760 & 208.9 & 389.5 \\
SRTY & IWM & $-3$ & 3.602 & 69.3 & 155.8 \\
TZA & IWM & $-3$ & 71.679 & 508.9 & 1,331.0 \\
TWM & IWM & $-2$ & 0.528 & 16.3 & 40.3 \\
RWM & IWM & $-1$ & 17.924 & 283.7 & 716.7 \\
IWM & IWM & $+1$ & 32.606 & 7,949.8 & 13,891.7 \\
UWM & IWM & $+2$ & 0.430 & 20.6 & 49.2 \\
TNA & IWM & $+3$ & 11.017 & 480.0 & 802.8 \\
URTY & IWM & $+3$ & 1.142 & 60.5 & 126.1 \\
SQQQ & QQQ & $-3$ & 75.767 & 2,579.6 & 5,257.4 \\
QID & QQQ & $-2$ & 17.810 & 346.0 & 740.3 \\
PSQ & QQQ & $-1$ & 8.777 & 265.3 & 580.7 \\
QQQ & QQQ & $+1$ & 47.236 & 28,313.7 & 49,526.2 \\
QLD & QQQ & $+2$ & 4.236 & 390.2 & 678.5 \\
TQQQ & QQQ & $+3$ & 75.143 & 5,101.4 & 8,339.8 \\
SOXS & SOXX & $-3$ & 207.578 & 1,846.8 & 4,623.3 \\
SOXX & SOXX & $+1$ & 7.481 & 2,582.9 & 6,809.7 \\
SOXL & SOXX & $+3$ & 100.901 & 5,248.8 & 13,626.8 \\
SPXS & SPY & $-3$ & 34.378 & 375.0 & 793.5 \\
SPXU & SPY & $-3$ & 14.273 & 357.2 & 652.3 \\
SDS & SPY & $-2$ & 9.926 & 253.6 & 523.7 \\
SH & SPY & $-1$ & 7.725 & 288.5 & 695.8 \\
SPY & SPY & $+1$ & 65.553 & 43,037.1 & 73,262.8 \\
SSO & SPY & $+2$ & 3.298 & 249.7 & 511.0 \\
SPXL & SPY & $+3$ & 3.268 & 626.6 & 1,280.5 \\
UPRO & SPY & $+3$ & 4.327 & 443.1 & 879.3 \\
\bottomrule
\end{tabular}

\end{table}
\begin{table}[htbp]\centering\footnotesize\setlength{\tabcolsep}{4pt}
\caption{Geared trading by benchmark family. The geared/benchmark ratios are calculated date by date before taking their median or P95. Inverse share is the average date-level fraction of geared dollars.}\label{tab:volumefamilies}
\begin{tabular}{lrrrrr}
\toprule
Family & Dates & Geared \$bn/day & Median ratio & P95 ratio & Inverse share (\%) \\
\midrule
DIA & 499 & 0.484 & 0.22 & 0.46 & 50.7 \\
IWM & 499 & 1.439 & 0.18 & 0.29 & 59.3 \\
QQQ & 499 & 8.683 & 0.30 & 0.51 & 35.8 \\
SOXX & 499 & 7.096 & 2.92 & 5.48 & 27.2 \\
SPY & 499 & 2.594 & 0.06 & 0.09 & 48.5 \\
\bottomrule
\end{tabular}

\end{table}

The Nasdaq family has mean geared turnover of $8.68$ billion dollars per day, compared with $7.10$ billion for SOXL and SOXS together. Relative to their benchmark ETF, the semiconductor products have the largest footprint, with a median ratio of $2.92$ and a 95th percentile of $5.48$. The Nasdaq median ratio is $0.30$. The Russell family has the largest average inverse share, about $59.3\%$. These comparisons describe differences in secondary-market scale. Turnover alone does not identify the committed demand $D$.

\begin{figure}[htbp]\centering
\includegraphics[width=\textwidth]{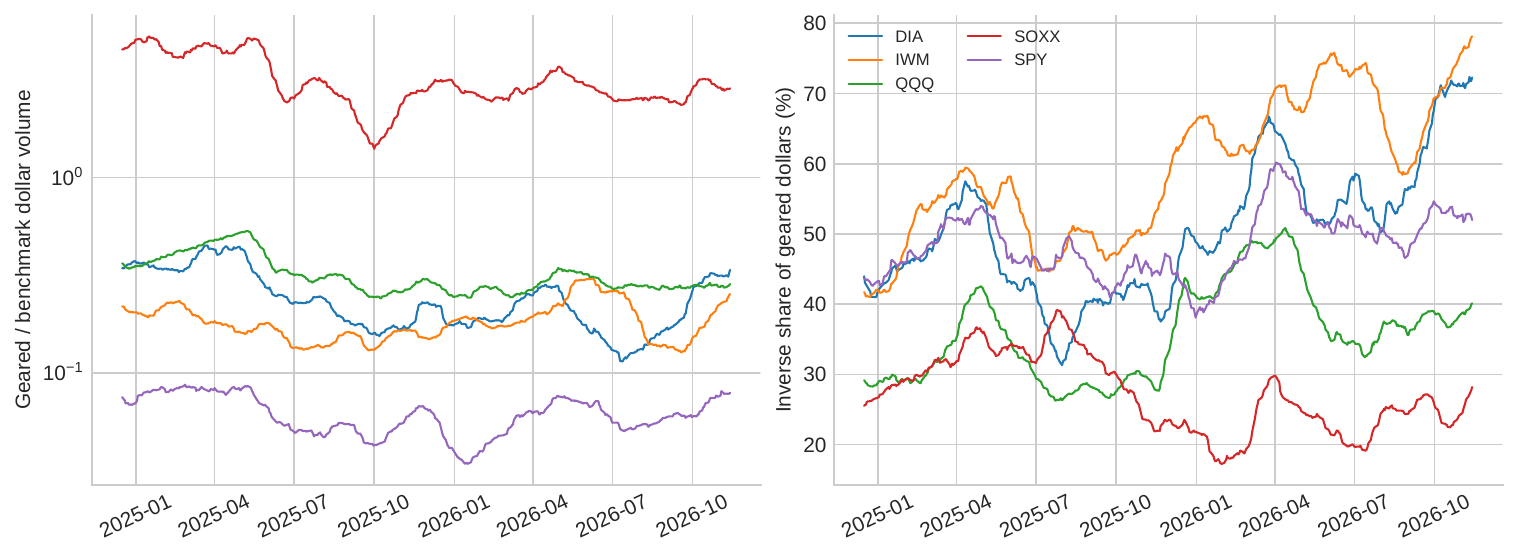}
\caption{Twenty-session averages of geared/benchmark dollar-turnover ratios and inverse shares. The left axis is logarithmic. All families use the same daily panel.}\label{fig:volumefamilies}
\end{figure}

\FloatBarrier
\subsection{Activity, market moves, and common shocks}
Define relative dollar activity using a strictly lagged denominator,
\[
 Z_{it}=\log\left(\frac{T_{it}}{20^{-1}\sum_{s=t-20}^{t-1}T_{is}}\right).
\]
We relate $Z_{it}$ to its benchmark's absolute daily total return, the negative part of that return, benchmark relative dollar activity, and the interaction of the negative return with an inverse-product indicator. Returns enter in percentage points. ETF fixed effects account for persistent product differences. The first twenty daily observations supply the denominator, leaving a balanced 479-date regression panel. Full-sample, broad-family, and semiconductor regressions use the same dates. A fourth specification removes both ETF and date means from the dependent variable and all regressors, thereby absorbing arbitrary common daily shocks.

The covariance calculation first sums each date's regression scores across products, then applies the ten-session Bartlett weights. This allows arbitrary within-date dependence across ETFs and serial dependence in the aggregated scores. Table~\ref{tab:volumeregression} reports Holm adjustments over the sixteen displayed slope tests. The regressions describe contemporaneous associations. Today's return and activity enter as realized contemporaneous quantities, rather than information assumed available before today's trading.

Benchmark activity is strongly associated with geared-product activity. After controlling for it, the associations with the magnitude and sign of benchmark returns are smaller and more sensitive to specification. These regressions do not identify how institutional reset orders respond to a benchmark move. The next decomposition separates changes in share activity from changes in price denomination.

\begin{table}[htbp]\centering\footnotesize\setlength{\tabcolsep}{4pt}
\caption{Daily geared-product activity regressions. The outcome is log dollar volume relative to the preceding twenty-day mean. ``ETF and date effects'' is the within-ETF, within-date specification. Standard errors use date-aggregated HAC scores.}\label{tab:volumeregression}
\begin{tabular}{llrrr}
\toprule
Sample & Term & Coefficient & HAC s.e. & Holm $p$ \\
\midrule
All geared ETFs & Absolute return & 0.004 & 0.009 & 1.000 \\
All geared ETFs & Negative return & 0.010 & 0.010 & 1.000 \\
All geared ETFs & Benchmark activity & 0.784 & 0.040 & $<0.001$ \\
All geared ETFs & Negative $\times$ inverse & 0.008 & 0.006 & 1.000 \\
Broad families & Absolute return & 0.004 & 0.011 & 1.000 \\
Broad families & Negative return & 0.016 & 0.015 & 1.000 \\
Broad families & Benchmark activity & 0.807 & 0.042 & $<0.001$ \\
Broad families & Negative $\times$ inverse & 0.002 & 0.009 & 1.000 \\
Semiconductors & Absolute return & 0.015 & 0.008 & 0.633 \\
Semiconductors & Negative return & 0.005 & 0.006 & 1.000 \\
Semiconductors & Benchmark activity & 0.577 & 0.053 & $<0.001$ \\
Semiconductors & Negative $\times$ inverse & 0.016 & 0.006 & 0.069 \\
ETF and date effects & Absolute return & 0.058 & 0.010 & $<0.001$ \\
ETF and date effects & Negative return & -0.017 & 0.009 & 0.633 \\
ETF and date effects & Benchmark activity & 0.265 & 0.030 & $<0.001$ \\
ETF and date effects & Negative $\times$ inverse & 0.007 & 0.006 & 1.000 \\
\bottomrule
\end{tabular}

\end{table}
\subsection{Separating share activity from price denomination}
Dollar turnover combines shares and prices. Let $c_{it}$ be the cumulative split ratio in new shares per initial share. Express daily volume and VWAP in a constant denomination as
\[
 V^*_{it}=V_{it}/c_{it},\qquad P^*_{it}=c_{it}P^{\rm VWAP}_{it},\qquad
 T_{it}=V^*_{it}P^*_{it}.
\]
Using the same strictly lagged twenty-session denominator, define
\begin{align}
 Z^V_{it}&=\log\!\left(\frac{V^*_{it}}{20^{-1}\sum_{j=1}^{20}V^*_{i,t-j}}\right),\nonumber\\
 Z^P_{it}&=Z_{it}-Z^V_{it}
 =\log\!\left(\frac{P^*_{it}}{
 \sum_{j=1}^{20}V^*_{i,t-j}P^*_{i,t-j}/\sum_{j=1}^{20}V^*_{i,t-j}}\right).
 \label{eq:volumedecomposition}
\end{align}
The second component compares current split-adjusted VWAP with a lagged share-weighted price reference. Its definition differs from the current return and carries no causal price interpretation.

Table~\ref{tab:volumedecomposition} uses exactly the same regressors, dates, and fixed-effect transformations for dollar activity, share activity, and the price component. OLS linearity makes every dollar coefficient the sum of the other two. There are 479 dates and 26 geared products in each specification. With ETF and date effects, the negative-return interaction for inverse products is $0.007181=-0.058678+0.065859$, up to rounding. Its small positive dollar-activity coefficient therefore combines a negative share-activity component with a positive price-related component. The decomposition describes the original coefficient through its two accounting components. No additional significance claim or causal interpretation is attached to it.

\begin{table}[htbp]\centering\footnotesize\setlength{\tabcolsep}{4pt}
\caption{Exact coefficient decomposition of relative dollar activity. Every row uses the same design for all three outcomes. Dollar activity equals share activity plus the price-related component. Return regressors remain in percentage points. No new coefficient-testing family is invoked.}\label{tab:volumedecomposition}
\begin{tabular}{llrrr}
\toprule
Effects & Regressor & Dollar activity & Share activity & Price component \\
\midrule
ETF effects & Absolute return & +0.004208 & +0.008612 & -0.004404 \\
ETF effects & Negative return & +0.009973 & +0.046288 & -0.036315 \\
ETF effects & Benchmark activity & +0.783672 & +0.771184 & +0.012488 \\
ETF effects & Negative return $\times$ inverse & +0.008452 & -0.057329 & +0.065780 \\
ETF and date effects & Absolute return & +0.057657 & +0.063225 & -0.005567 \\
ETF and date effects & Negative return & -0.017175 & +0.016623 & -0.033798 \\
ETF and date effects & Benchmark activity & +0.264642 & +0.261020 & +0.003622 \\
ETF and date effects & Negative return $\times$ inverse & +0.007181 & -0.058678 & +0.065859 \\
\bottomrule
\end{tabular}

\end{table}

\FloatBarrier
\section{Daily factor exposures and tracking}\label{app:factors}
\subsection{Factors, returns, and matching}
We use the daily US Fama--French market, size, value, profitability, and investment factors, together with momentum and the value-weighted electronic-equipment industry portfolio from the Kenneth French Data Library \citep{French2026}. These factors concern US listed equities and are appropriate broad risk descriptors for the equity benchmark families studied here. The industry series has code \texttt{Chips}, but its SIC definition includes electronic equipment beyond semiconductors. It supplies an additional exposure control rather than replicating the SOXX benchmark.

The factor files use the September 2026 CRSP database and end on September 30, 2026. Downloaded percentage returns are converted to decimal returns. For ETF $i$, let $P_{it}$ be the unadjusted close, $d_{it}$ the ex-date cash distribution per contemporaneous share, and $s_{it}$ the new-shares-per-old-share split ratio, equal to one on ordinary dates. The total return is
\begin{equation}
 r_{it}=\frac{s_{it}(P_{it}+d_{it})}{P_{i,t-1}}-1.\label{eq:dailyreturn}
\end{equation}
Cash amounts retain their ex-date basis. There are no simultaneous split and distribution events in this matched panel. The first available daily close supplies the return denominator. The common regression window is therefore November 19, 2024 through September 30, 2026, with 466 return dates for every ETF. Each factor and ETF return is matched by trading date, without forward filling. Market-price total returns are used throughout, rather than substituting NAV returns or assuming a leveraged ETF equals a fixed multiple of its proxy.

The daily regressions measure ex-post exposure. Their same-day factor returns are unavailable at noon and never enter the intraday forecast experiment. The factor vintage incorporates the French library's current historical construction. It cannot be interpreted as a point-in-time predictor dataset. All products use the same factor vintage and daily observation window.

\subsection{Exposure regressions}
For each ETF, estimate
\begin{equation}
 r_{it}-r_{ft}=\alpha_i+b_i^\top f_t+\varepsilon_{it},\label{eq:ffregression}
\end{equation}
where $f_t$ contains the five Fama--French factors and momentum. The augmented model adds the electronic-equipment return minus the same risk-free return. The intercept and loadings are estimated by OLS, with ten-session HAC standard errors. Table~\ref{tab:factorproducts} reports the six-factor intercept in basis points per day, its HAC statistic, market loading, and $R^2$ before and after the industry control. The industry loading comes from the augmented regression. All 31 products enter the same estimation dates.

For the benchmark ETFs, the six-factor $R^2$ is $99.2\%$ for SPY, $97.6\%$ for IWM, $94.7\%$ for QQQ, $86.4\%$ for DIA, and $72.9\%$ for SOXX. Adding electronic equipment raises SOXX's $R^2$ to $83.4\%$ and QQQ's to $96.3\%$. Broad-market factors explain much of the general equity exposure. Sector composition matters more for the semiconductor family. The increase in in-sample $R^2$ concerns exposure attribution and provides no evidence of out-of-sample forecast improvement.

No six-factor intercept rejects zero after Holm adjustment across the 31 intercept tests. These intercepts describe factor-model residual returns. An implementable strategy's net alpha would require a separate trading and cost calculation. We next examine departures from a financed benchmark position, which is a different outcome from the ETF's total return.

\begin{table}[htbp]\centering\footnotesize\setlength{\tabcolsep}{4pt}
\caption{Daily factor regressions on 466 common dates. FF6 denotes the five Fama--French factors plus momentum. The $R^2$ columns are percentages. Intercepts and market loadings come from FF6, and the final loading comes from FF6 plus electronic equipment.}\label{tab:factorproducts}
\begin{tabular}{lrrrrrrr}
\toprule
ETF & $\beta$ & $\alpha$ bp/day & $t_{HAC}$ & Market & FF6 $R^2$ & +Chips $R^2$ & Chips \\
\midrule
SDOW & $-3$ & 2.86 & 0.70 & -2.65 & 86.2 & 88.6 & 0.55 \\
DXD & $-2$ & 1.62 & 0.59 & -1.76 & 85.8 & 88.3 & 0.37 \\
DOG & $-1$ & 0.61 & 0.45 & -0.89 & 85.8 & 88.3 & 0.18 \\
DIA & $+1$ & -1.21 & -0.89 & 0.89 & 86.4 & 88.8 & -0.18 \\
DDM & $+2$ & -2.97 & -1.09 & 1.77 & 86.1 & 88.5 & -0.36 \\
UDOW & $+3$ & -4.42 & -1.08 & 2.64 & 86.3 & 88.7 & -0.54 \\
SRTY & $-3$ & 4.03 & 1.41 & -2.92 & 97.6 & 97.6 & -0.03 \\
TZA & $-3$ & 4.18 & 1.46 & -2.92 & 97.7 & 97.7 & -0.03 \\
TWM & $-2$ & 2.38 & 1.24 & -1.95 & 97.5 & 97.5 & -0.03 \\
RWM & $-1$ & 0.97 & 1.01 & -0.98 & 97.6 & 97.6 & -0.02 \\
IWM & $+1$ & -1.64 & -1.80 & 0.97 & 97.6 & 97.6 & 0.01 \\
UWM & $+2$ & -3.89 & -2.06 & 1.94 & 97.5 & 97.5 & 0.01 \\
TNA & $+3$ & -5.88 & -2.08 & 2.92 & 97.6 & 97.6 & 0.03 \\
URTY & $+3$ & -6.07 & -2.14 & 2.93 & 97.5 & 97.5 & 0.03 \\
SQQQ & $-3$ & -3.60 & -0.91 & -3.36 & 94.7 & 96.3 & -0.64 \\
QID & $-2$ & -2.49 & -0.94 & -2.25 & 94.7 & 96.3 & -0.43 \\
PSQ & $-1$ & -1.44 & -1.09 & -1.12 & 94.7 & 96.2 & -0.21 \\
QQQ & $+1$ & 0.81 & 0.62 & 1.13 & 94.7 & 96.3 & 0.22 \\
QLD & $+2$ & 1.14 & 0.43 & 2.24 & 94.7 & 96.3 & 0.43 \\
TQQQ & $+3$ & 1.70 & 0.43 & 3.36 & 94.8 & 96.3 & 0.64 \\
SOXS & $-3$ & -22.96 & -1.34 & -4.91 & 73.2 & 83.5 & -3.24 \\
SOXX & $+1$ & 7.09 & 1.24 & 1.64 & 72.9 & 83.4 & 1.10 \\
SOXL & $+3$ & 19.33 & 1.13 & 4.89 & 73.3 & 83.7 & 3.23 \\
SPXS & $-3$ & -0.59 & -0.79 & -2.94 & 99.4 & 99.4 & -0.03 \\
SPXU & $-3$ & -0.81 & -1.14 & -2.93 & 99.5 & 99.5 & -0.03 \\
SDS & $-2$ & -0.75 & -1.60 & -1.95 & 99.6 & 99.6 & -0.02 \\
SH & $-1$ & -0.58 & -2.42 & -0.98 & 99.5 & 99.5 & -0.01 \\
SPY & $+1$ & -0.06 & -0.19 & 1.01 & 99.2 & 99.2 & 0.02 \\
SSO & $+2$ & -0.52 & -1.09 & 1.96 & 99.6 & 99.6 & 0.02 \\
SPXL & $+3$ & -0.80 & -1.04 & 2.92 & 99.4 & 99.4 & 0.03 \\
UPRO & $+3$ & -0.72 & -0.99 & 2.92 & 99.5 & 99.5 & 0.04 \\
\bottomrule
\end{tabular}

\end{table}
\begin{figure}[htbp]\centering
\includegraphics[width=\textwidth]{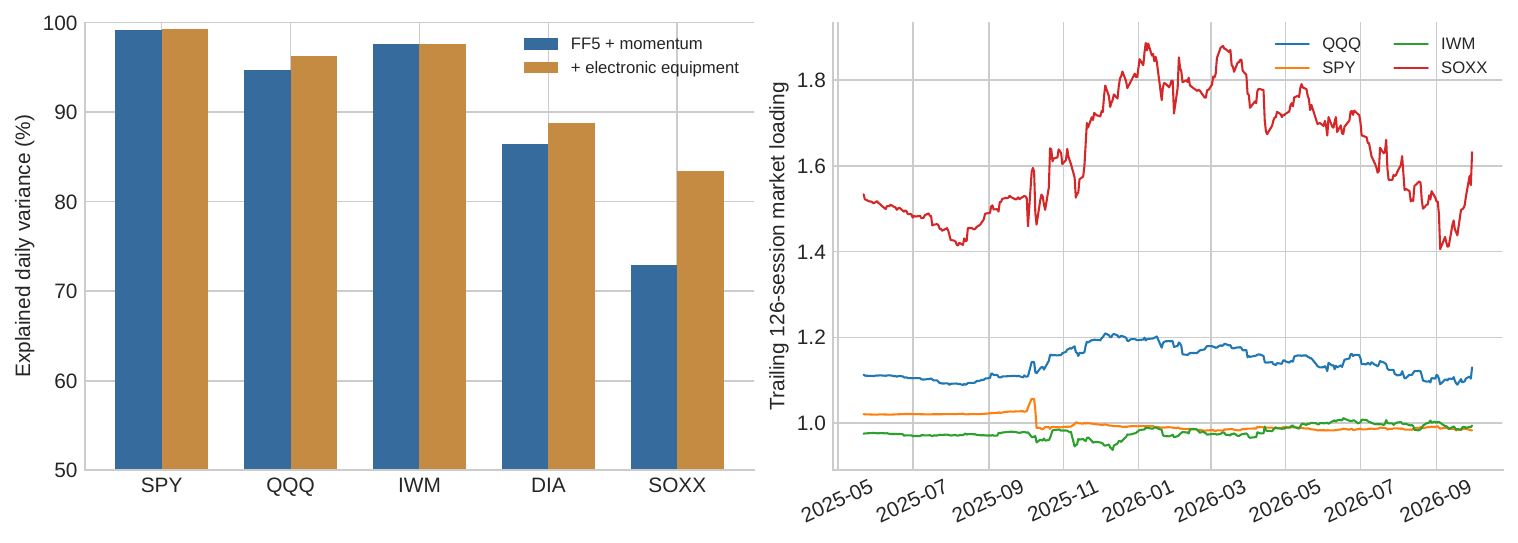}
\caption{Factor fit and time variation. Left: full-window $R^2$ with and without electronic equipment. Right: trailing 126-session six-factor market loadings for selected benchmark ETFs. The rolling estimates are descriptive exposures, not selected trading signals.}\label{fig:factorfit}
\end{figure}

\subsection{Tracking relative to a financed proxy position}
To distinguish systematic exposure from daily implementation differences, define
\begin{equation}
 e_{it}=r_{it}-\beta_i r_{b(i),t}-(1-\beta_i)r_{ft},\label{eq:tracking}
\end{equation}
where $b(i)$ is the associated benchmark ETF. This is the return difference from a hypothetical daily position with $\beta_i$ dollars in the proxy and $1-\beta_i$ dollars in the risk-free asset per initial dollar. This accounting benchmark assumes neither that the fund can finance at the risk-free rate nor that it holds the stated cash-ETF portfolio. Proxy fees, index-basis differences, market-price deviations, derivatives, and fund expenses can all enter the residual.

\begin{figure}[htbp]\centering
\includegraphics[height=.78\textheight,width=.95\textwidth,keepaspectratio]{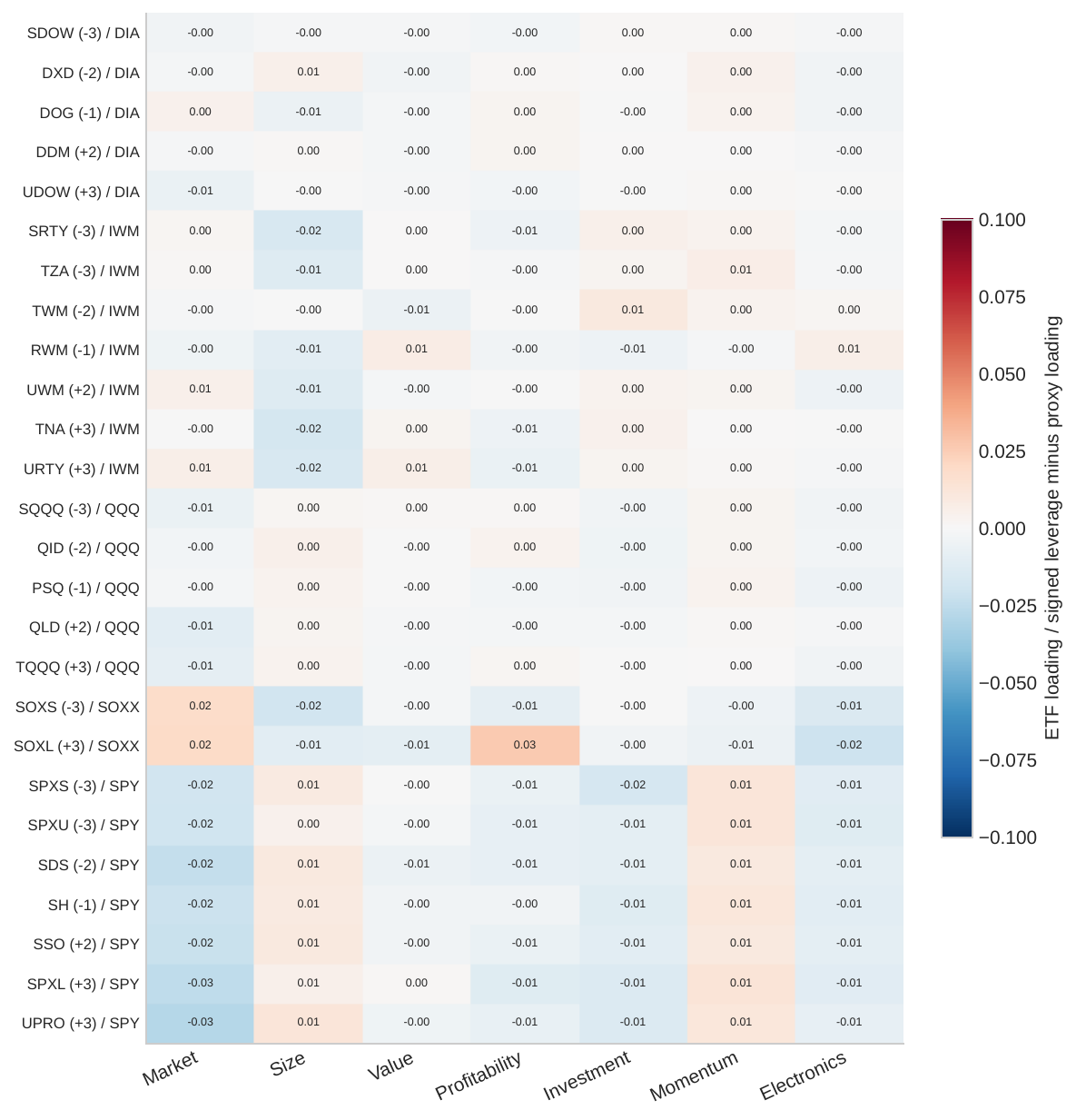}
\caption{Exposures beyond a financed proxy. Each cell is the fund loading divided by signed leverage minus the proxy loading, equivalently the tracking-difference loading divided by signed leverage. The model is FF6 plus electronic equipment. These are in-sample exposure differences, not selected investment signals.}\label{fig:factorrelative}
\end{figure}

The same-day excess-return regression makes the benchmark comparison exact. Write $\widehat\theta_i$ for the vector of intercept and loadings from the fund regression, $\widehat\theta_b$ for its proxy, and $\widehat\theta_e$ for the tracking difference. With the same design and dates,
\begin{equation}
 \widehat\theta_e=\widehat\theta_i-\beta_i\widehat\theta_b,\qquad
 \frac{\widehat b_e}{\beta_i}=\frac{\widehat b_i}{\beta_i}-\widehat b_b.
 \label{eq:factoridentity}
\end{equation}
This follows directly from linearity of OLS. Figure~\ref{fig:factorrelative} displays the normalized loading differences for the FF6-plus-electronics specification. The figure isolates exposures beyond the financed proxy by removing the component shared with that proxy. All products retain the same 466 dates.

Table~\ref{tab:tracking} reports the mean and RMS tracking difference and the variation explained by the two factor specifications. The unexplained RMS removes both fitted factors and the OLS intercept. It is therefore not equal to the original uncentered tracking RMS times $\sqrt{1-R^2}$ unless the tracking difference has zero mean. For TQQQ, FF6 plus electronic equipment explains $22.50\%$ of tracking variation. For SOXL and SOXS, it explains only $9.63\%$ and $2.60\%$. The same industry factor materially improves the fit of SOXX's total return, but that comparison concerns a different dependent variable.

These in-sample attributions distinguish broad exposure from daily implementation and benchmark differences. They do not separate fees, derivatives, financing, index mismatch, or rebalance activity as causes of the remaining variation. Daily tracking in \eqref{eq:tracking} also differs from the multi-day compounding identity in \eqref{eq:discrete}.

\begin{table}[htbp]\centering\footnotesize\setlength{\tabcolsep}{4pt}
\caption{Tracking differences from financed benchmark positions and their factor attribution. Mean, RMS, and unexplained RMS are daily basis points. $R^2$ is in percent and concerns tracking variation, not the product's total return. The unexplained RMS uses FF6 plus electronic equipment and removes the fitted intercept. All 26 products use the same 466 dates.}\label{tab:tracking}
\begin{tabular}{llrrrrrr}
\toprule
ETF & Proxy & $\beta$ & Mean bp & RMS bp & FF6 $R^2$ & +Chips $R^2$ & Unexplained RMS bp \\
\midrule
SDOW & DIA & $-3$ & -0.73 & 7.17 & 5.41 & 5.59 & 6.93 \\
DXD & DIA & $-2$ & -0.72 & 7.77 & 3.76 & 4.23 & 7.58 \\
DOG & DIA & $-1$ & -0.60 & 4.49 & 1.20 & 1.57 & 4.41 \\
DDM & DIA & $+2$ & -0.61 & 5.65 & 1.97 & 2.06 & 5.56 \\
UDOW & DIA & $+3$ & -0.84 & 6.19 & 10.95 & 10.96 & 5.79 \\
SRTY & IWM & $-3$ & -1.08 & 15.85 & 4.20 & 4.26 & 15.47 \\
TZA & IWM & $-3$ & -0.89 & 17.28 & 3.79 & 3.83 & 16.92 \\
TWM & IWM & $-2$ & -0.96 & 12.81 & 1.60 & 1.68 & 12.67 \\
RWM & IWM & $-1$ & -0.74 & 7.48 & 1.30 & 1.69 & 7.38 \\
UWM & IWM & $+2$ & -0.58 & 12.62 & 1.36 & 1.72 & 12.50 \\
TNA & IWM & $+3$ & -0.78 & 15.38 & 3.98 & 3.98 & 15.05 \\
URTY & IWM & $+3$ & -0.87 & 17.90 & 2.75 & 2.77 & 17.63 \\
SQQQ & QQQ & $-3$ & -0.99 & 11.60 & 11.67 & 12.15 & 10.83 \\
QID & QQQ & $-2$ & -0.77 & 7.45 & 4.48 & 4.74 & 7.23 \\
PSQ & QQQ & $-1$ & -0.60 & 4.48 & 4.51 & 5.32 & 4.32 \\
QLD & QQQ & $+2$ & -0.59 & 5.85 & 16.81 & 16.93 & 5.30 \\
TQQQ & QQQ & $+3$ & -1.00 & 9.68 & 21.79 & 22.50 & 8.47 \\
SOXS & SOXX & $-3$ & -1.74 & 30.52 & 1.33 & 2.60 & 30.07 \\
SOXL & SOXX & $+3$ & -2.66 & 36.55 & 7.80 & 9.63 & 34.65 \\
SPXS & SPY & $-3$ & -0.41 & 24.82 & 19.60 & 20.75 & 22.09 \\
SPXU & SPY & $-3$ & -0.62 & 23.61 & 24.06 & 25.64 & 20.36 \\
SDS & SPY & $-2$ & -0.58 & 15.16 & 23.48 & 24.47 & 13.17 \\
SH & SPY & $-1$ & -0.49 & 8.06 & 21.43 & 22.53 & 7.08 \\
SSO & SPY & $+2$ & -0.68 & 14.36 & 25.75 & 26.94 & 12.26 \\
SPXL & SPY & $+3$ & -0.97 & 24.49 & 27.70 & 28.93 & 20.63 \\
UPRO & SPY & $+3$ & -1.00 & 22.28 & 29.88 & 30.69 & 18.53 \\
\bottomrule
\end{tabular}

\end{table}

\end{document}